\documentclass[reprint,superscriptaddress,amsmath,amssymb,aps,nofootinbib,pra]{revtex4-2}

\usepackage[utf8]{inputenc}
\usepackage[english]{babel}
\usepackage{amsmath,amssymb,amsfonts,amsthm}
\usepackage{graphicx}
\usepackage{float}
\usepackage{color,xcolor}
\usepackage{bm}
\usepackage{bbold}
\usepackage{multirow, array, booktabs, dcolumn}
\usepackage{enumerate}
\usepackage[shortlabels]{enumitem}
\usepackage{cases}
\usepackage{verbatim}
\usepackage{setspace}
\usepackage{mathrsfs}
\usepackage{tikz}
\usetikzlibrary{decorations.pathmorphing}

\usepackage[colorlinks=true,urlcolor=blue,citecolor=blue,linkcolor=blue]{hyperref}
\usepackage[nameinlink]{cleveref}
\crefname{figure}{Fig.}{Figures}
\crefname{table}{Table}{Tables}
\crefname{equation}{}{}
\crefname{section}{Section}{Sections}

\usepackage[labelfont=bf]{caption}
\usepackage[labelfont=bf]{subcaption}

\begin{document}

\title{Classical-quantum study of confinement in the chaotic $x^{2}y^{2}$ Yang-Mills Hamiltonian}

\author{Mario A. Quiroz-Juarez}%
\affiliation{Universidad Nacional Aut\'onoma de M\'exico, Centro de F\'{i}sica Aplicada y Tecnolog\'{i}a Avanzada, Boulevard Juriquilla 3001, 76230 Quer\'{e}taro, M\'exico}

\author{Marco A. Zurita}%
\email{cbi2231801480@xanum.uam.mx}
\affiliation{Departamento de F\'{i}sica, Universidad Aut\'onoma Metropolitana Unidad Iztapalapa, San Rafael Atlixco 186, 09340 Cd. Mx., M\'exico}

\author{Horacio Olivares-Pilon}%
\email{horop@xanum.uam.mx}
\affiliation{Departamento de F\'{i}sica, Universidad Aut\'onoma Metropolitana Unidad Iztapalapa, San Rafael Atlixco 186, 09340 Cd. Mx., M\'exico}

\author{Adrian M. Escobar Ruiz}%
\email{admau@xanum.uam.mx}
\affiliation{Departamento de F\'{i}sica, Universidad Aut\'onoma Metropolitana Unidad Iztapalapa, San Rafael Atlixco 186, 09340 Cd. Mx., M\'exico}

\begin{abstract}
We analyze how quantum mechanics reinstates confinement in Hamiltonian systems that are classically unstable and exhibit chaotic dynamics. Specifically, we consider two paradigmatic models: the Contopoulos Hamiltonian, an isotropic oscillator perturbed by the quartic coupling $\alpha\, x^{2}y^{2}$, and the purely quartic Yang--Mills Hamiltonian $H=\tfrac{1}{2}(p_{x}^{2}+p_{y}^{2})+\alpha\, x^{2}y^{2}$. Classical dynamics, characterized through Poincar\'e sections, Lyapunov exponents, and periodic orbits, reveals distinct escape mechanisms: in the Contopoulos system, trajectories destabilize along the diagonal valleys $x=\pm y$ for $\alpha<0$, whereas in the Yang--Mills case with $\alpha>0$, escape occurs along the coordinate axes $x=0$ or $y=0$. In sharp contrast, the quantum Yang--Mills Hamiltonian with $\alpha>0$ admits only discrete, normalizable eigenstates. Semiclassical WKB and full two--dimensional analyses further show that these quantum states are localized along the classical escape channels, illustrating how transverse zero--point motion generates an effective confining barrier. Our study combines global Lyapunov--exponent heat maps with high--precision quantum spectra obtained via variational and Lagrange--mesh methods, providing quantitatively controlled results across regimes. In addition, we corroborate the classical predictions through analog electronic simulations based on operational--amplifier circuit models, offering an experimentally inspired validation of the theoretical framework.
\end{abstract}

\maketitle

\section{Introduction}

The interplay between classical chaos and quantum confinement is a central theme in nonlinear dynamics and quantum physics. 
Quartic Hamiltonians provide the minimal extension of the harmonic oscillator capable of generating nonlinear behavior: 
while quadratic terms yield integrable motion, quartic couplings introduce the simplest nonlinear resonances, saddle structures, and routes to chaos. 
At the same time, quartic interactions are remarkably universal, arising in reduced models of gauge fields~\cite{matinyan1986classical}, galactic dynamics~\cite{contopoulos_structure_1994,GContopoulos_2003}, molecular and vibrational systems~\cite{reichl2004transition}, and matrix models relevant to high--energy physics~\cite{asplund2016quantum,bastianello2019thermalization}. 
These models therefore provide a natural testbed for studying how classical instability is reconciled with quantum confinement.

This raises a fundamental paradox: classical trajectories in quartic systems may destabilize and even escape along open channels, yet the corresponding quantum Hamiltonians admit only discrete, normalizable eigenstates. 
In quantum mechanics, motion along classically open directions is inseparably coupled to transverse zero--point fluctuations. 
These fluctuations generate an \emph{effective confining potential} that prevents escape even where the classical potential is flat.  
A rigorous proof of this phenomenon was given in~\cite{Simon1983AnnPhys, Simon2009MFAT}, where it was shown that the operator
\begin{equation}
    \hat{H} \;=\; -\tfrac{1}{2}\nabla^2\ + \ \alpha\, x^2\, y^2\,, 
    \qquad \alpha > 0 ,
\end{equation}
has a purely discrete spectrum. 

A paradigmatic example is provided by the generalized Yang--Mills Hamiltonian, which combines quadratic confinement with quartic interactions~\cite{matinyan1986classical,jimenezlara_periodic_2011,kasperczuk1994integrability},  
\begin{equation}
    H_{\textrm{GYM}} = \tfrac{1}{2}(p_x^2 + p_y^2 + x^2 + y^2) \;+\; \tfrac{a}{4}x^4 \;+\; \tfrac{b}{2}x^2y^2 ,
    \label{HYM}
\end{equation}
where $a,b\in\mathbb{R}$. 
The mixed quartic term $x^2y^2$ defines the so--called Yang--Mills potential, obtained from a homogeneous reduction of the $SU(2)$ Yang--Mills field~\cite{matinyan1986classical}, and has been widely studied as the \emph{mechanical Yang--Mills Hamiltonian}~\cite{jimenezlara_periodic_2011,kasperczuk1994integrability}.  

Classical dynamics of such systems has been extensively analyzed using averaging and bifurcation theory~\cite{jimenezlara_periodic_2011,BUICA20047,maciejewski_periodic_2005}, symmetry--line constructions~\cite{pina1987symmetry}, Lyapunov spectra~\cite{benettin1980lyapunov,dubeibe2014lyapunov,arnold2006mathematical}, and Poincar\'e sections~\cite{korsch2008chaos,lichtenberg2013regular,burden2016numerical}. These approaches connect to the broader theory of Hamiltonian chaos~\cite{zaslavsky2007chaos,gutzwiller1990chaos}, semiclassical quantization~\cite{reichl2004transition}, and universal spectral statistics~\cite{bohigas1984characterization,haake2010quantum}.  

A particularly important reduction arises for $a=0$, yielding the \emph{Contopoulos Hamiltonian}~\cite{contopoulos_structure_1994}, originally motivated by galactic dynamics~\cite{GContopoulos_2003,caranicolas2011order} and closely related to the paradigmatic H\'enon--Heiles system~\cite{henon1964,bertrand2006chaos}. Both exhibit chaotic scattering, resonance overlap, and destabilization of periodic orbits. It is well established that the Yang--Mills Hamiltonian with $b\neq 0$ is non--integrable and strongly chaotic~\cite{maciejewski_periodic_2005,casetti2000classical}. More recent studies extend these ideas to Chern--Simons terms~\cite{baskan2021chaos}, Matinyan--Yang--Mills--Higgs systems~\cite{kandiran2020analysis}, and bosonic matrix models relevant to M--theory~\cite{asplund2016quantum,bastianello2019thermalization}.  

The special case of a pure quartic interaction, $V(x,y)=\alpha\, x^2y^2$ where $a=0,b=2\,\alpha>0$, is of particular interest. Classically, trajectories destabilize and may escape along the diagonals $x=\pm y$ in the Contopoulos Hamiltonian when $\alpha<0$~\cite{bolotin1997chaos}, while in the pure quartic Yang-Mills Hamiltonian escape occurs along the coordinate axes. Quantum mechanically, however, all Yang-Mills eigenstates remain normalizable and the spectrum discrete~\cite{simon1983semiclassical,cohen2019quantum}, illustrating how confinement is restored even when classical motion is unstable.  

In this work, following some of the methodology presented in \cite{ESCOBARRUIZ2025116771},  we undertake a combined classical and quantum analysis of the $x^{2}y^{2}$ Hamiltonian with and without quadratic confinement. On the classical side, we employ Poincar\'e sections, Lyapunov exponents, and symmetry-line constructions to characterize the escape mechanisms and chaotic regimes. On the quantum side, we compute spectra using variational~\cite{simon1983semiclassical,cohen2019quantum} and Lagrange-mesh methods~\cite{baye2015lagrange}, providing high--precision results across parameter ranges. For the Contopoulos Hamiltonian, we show that positive quartic coupling partially lifts the isotropic-oscillator degeneracies. For the pure quartic Yang-Mills, the spectrum is discrete and degenerate, with the ground-state energy increasing monotonically with $\alpha$.

Taken together, these results establish a sharp classical-quantum dichotomy: while trajectories destabilize and may escape at the classical level, quantum mechanics restores confinement.

\vspace{-0.6cm}

\section{Contopoulos classical Hamiltonian}

\begin{figure}[t]
    \centering
    \includegraphics[width=4.9cm]{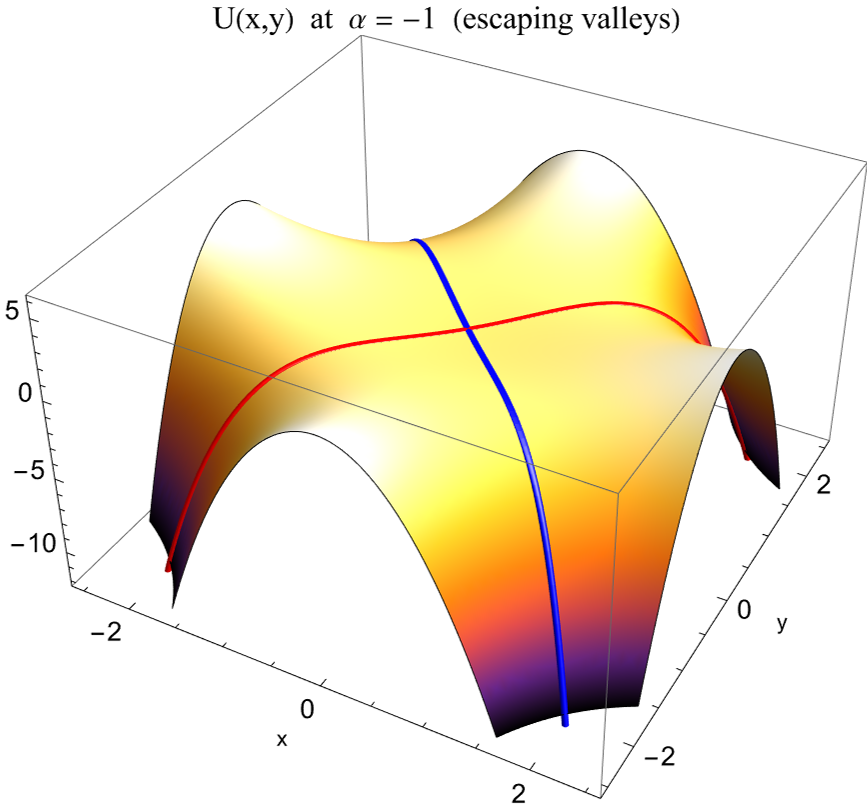}
\vspace{0.05cm}\includegraphics[width=4.9cm]{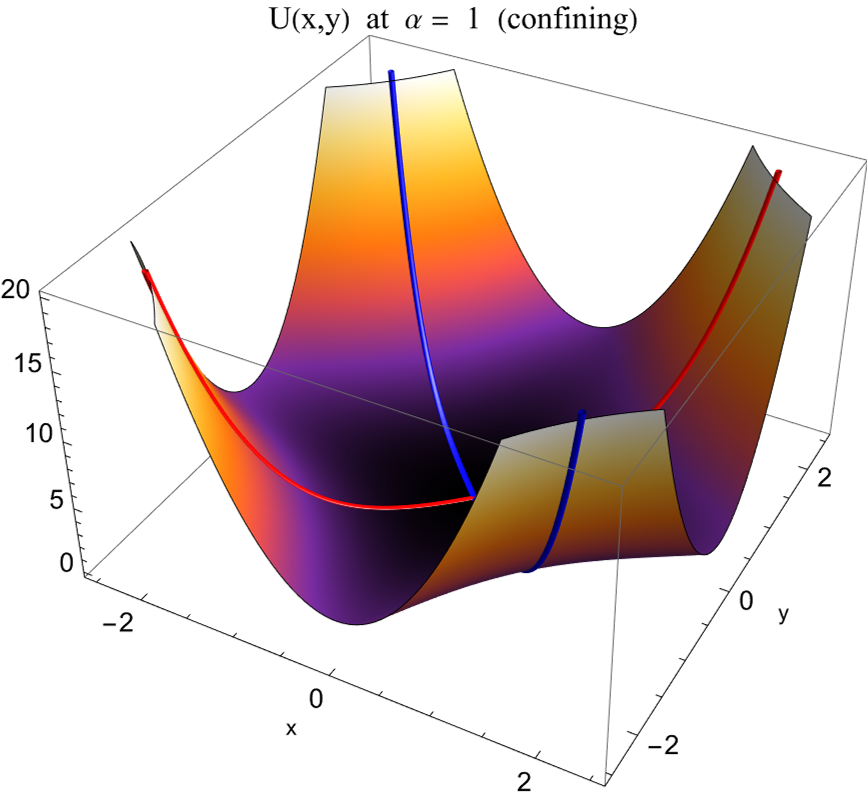}
    \caption{\small Potential landscapes of  Hamiltonian~\cref{hamiltoniano}.}
    \label{fig:potentials}
\end{figure}

We first analyze the Hamiltonian
\begin{equation}
    H \ = \ \tfrac{1}{2}\left(p_x^2 + p_y^2 + x^2 + y^2 \right) \;+\; \alpha\, x^2 y^2 ,
    \label{hamiltoniano}
\end{equation}
which describes an isotropic oscillator perturbed by a quartic coupling. This model, introduced by Contopoulos~\cite{contopoulos_structure_1994}, provides a minimal framework for exploring the interplay of regularity, instability, and chaos in two degrees of freedom. The equations of motion are
\begin{equation}
\begin{aligned}
   \dot{x} &= p_x, \qquad & \dot{y} &= p_y, \\
   \dot{p}_x &= -x\,(1+2\,\alpha\, y^2), \qquad & 
   \dot{p}_y &= -y\,(1+2\,\alpha\, x^2).
\end{aligned}
\label{eqsmo}
\end{equation}
The dynamics depends crucially on the sign of $\alpha$. For $\alpha>0$, the potential $U(x,y)=\tfrac{1}{2}(x^2 + y^2)+\alpha\,x^2 y^2$ grows without bound in all directions and is therefore globally confining. Trajectories near the origin are regular; as the energy increases, extended chaotic layers develop, yet motion remains bounded. For $\alpha<0$, additional real saddle points appear and the potential becomes unbounded along the diagonals $x=\pm y$, opening genuine escape channels and enabling chaotic scattering. The fixed points and their stability properties are summarized in Table~\ref{tab:fixedpoints}, whereas the potential landscapes in Fig.~\ref{fig:potentials} highlight the associated structural change.

\begin{table}[b]
\centering
\caption{\small Fixed points $\mathbf{r}_0=(x_0,y_0,p_{x_0}=0,p_{y_0}=0)$ of ~\cref{hamiltoniano}.}
\begin{tabular}{cccc}
\toprule
$\alpha$ & $(x_0,y_0)$ & $U(x_0,y_0)$ & Stability \\
\midrule
$>0$ & $(0,0)$ & $0$ & Center \\
$>0$ & $\left(\pm\dfrac{i}{\sqrt{2\alpha}},\pm\dfrac{i}{\sqrt{2\alpha}}\right)$ & $-\tfrac{1}{4\alpha}$ & Complex saddle \\
$<0$ & $(0,0)$ & $0$ & Center \\
$<0$ & $\left(\pm\dfrac{1}{\sqrt{2|\alpha|}},\pm\dfrac{1}{\sqrt{2|\alpha|}}\right)$ & $\tfrac{1}{4|\alpha|}$ & Real saddle \\
\bottomrule
\end{tabular}
\label{tab:fixedpoints}
\end{table}

\subsection{Poincaré Section}

A Poincaré map is obtained by recording successive intersections of the flow~\cref{eqsmo} with a transverse plane, thereby reducing the 4–dimensional phase space to a 2–dimensional representation. Ordered point sets indicate regular motion, whereas scattered distributions reveal the onset of chaos~\cite{arnold2006mathematical}. For numerical integration we employed an adaptive Runge–Kutta–Fehlberg scheme~\cite{burden2016numerical}, selecting $x=0$ as the section and sampling 300 distinct initial conditions, following Ref.~\cite{korsch2008chaos}.  

\begin{figure*}[h!]
    \centering    
    \begin{subfigure}[h!]{0.35\textwidth}
        \centering
        \includegraphics[width=\textwidth]{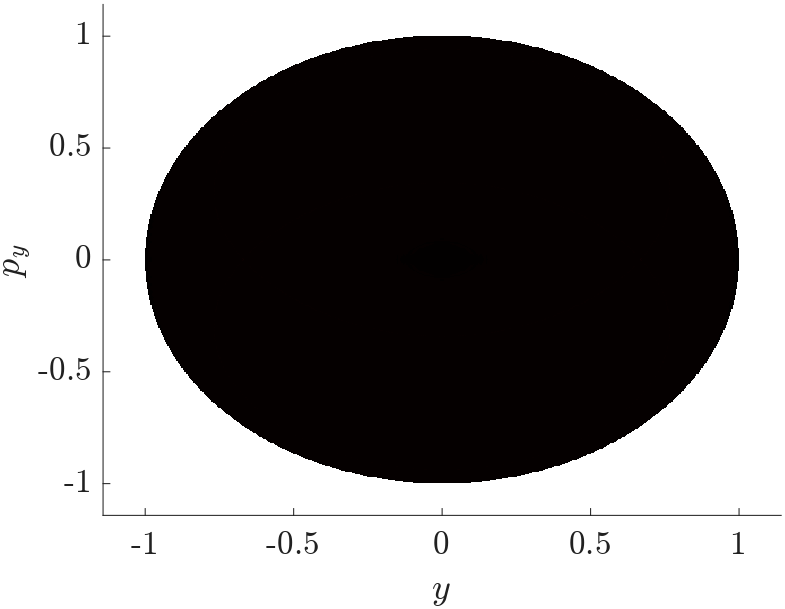} 
        \caption{$(\alpha=1/2,E=1/2)$}
    \end{subfigure}
    \begin{subfigure}[h!]{0.35\textwidth}
        \centering
        \includegraphics[width=\textwidth]{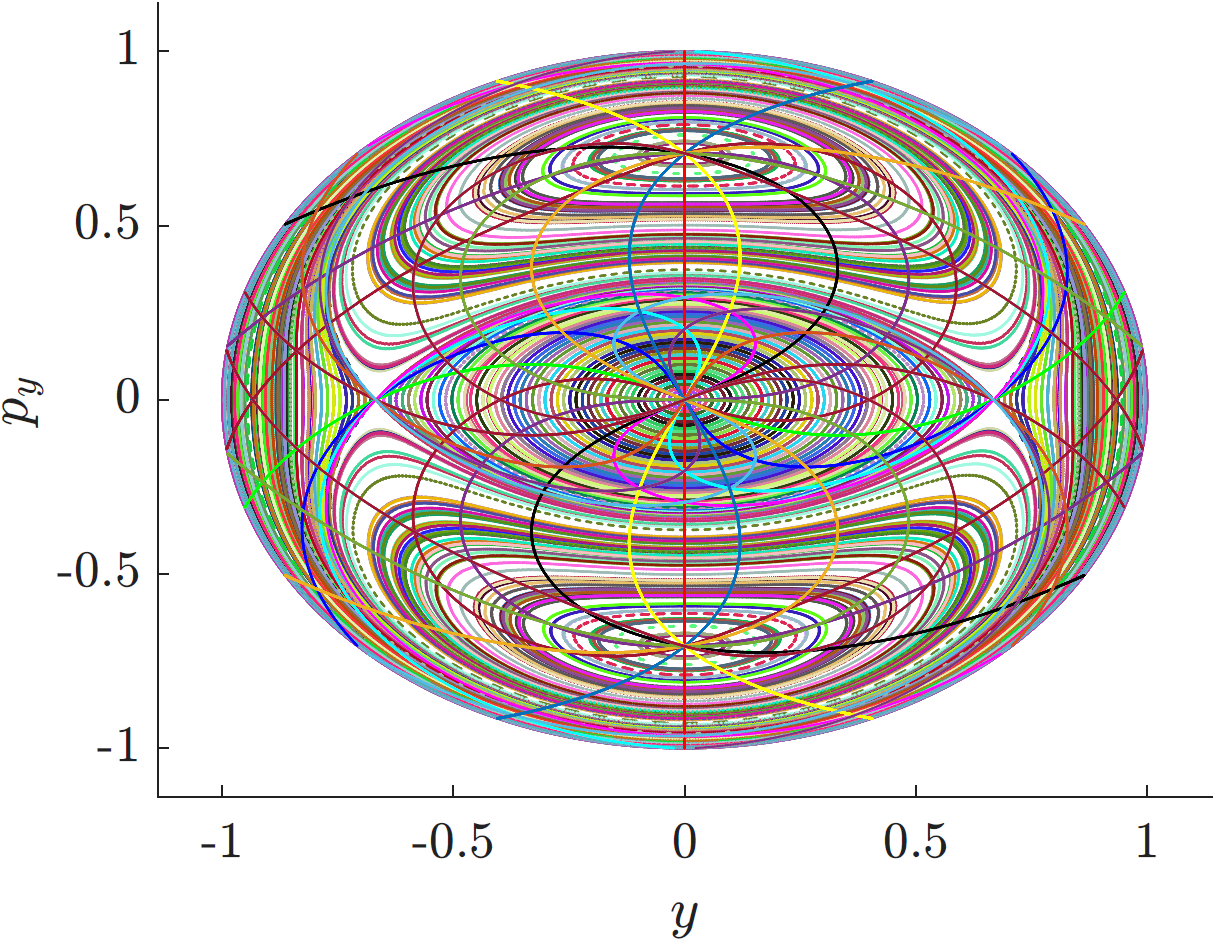} 
        \caption{$(\alpha=1/2,E=1/2)$}
    \end{subfigure}\\
    \begin{subfigure}[h!]{0.35\textwidth}
        \centering
        \includegraphics[width=\textwidth]{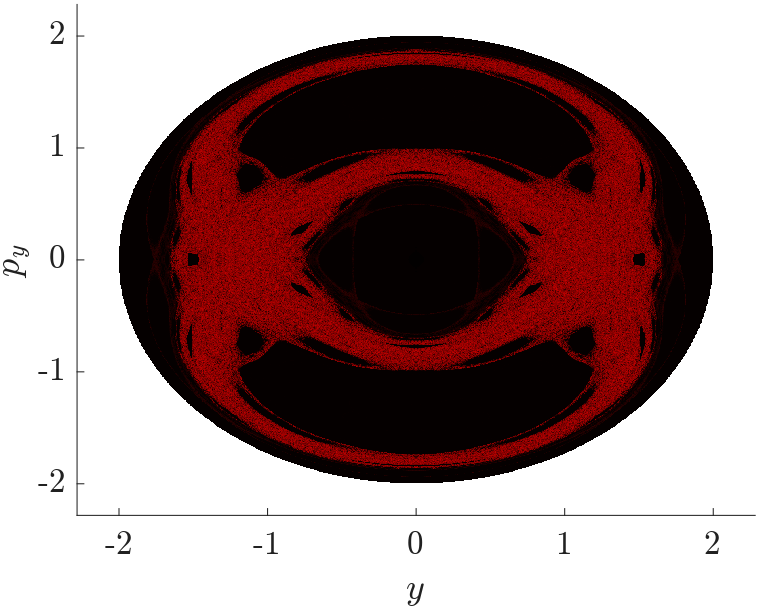}
        \caption{$(\alpha=1/2,E=2)$}
    \end{subfigure}\begin{subfigure}[h!]{0.35\textwidth}
        \centering
        \includegraphics[width=\textwidth]{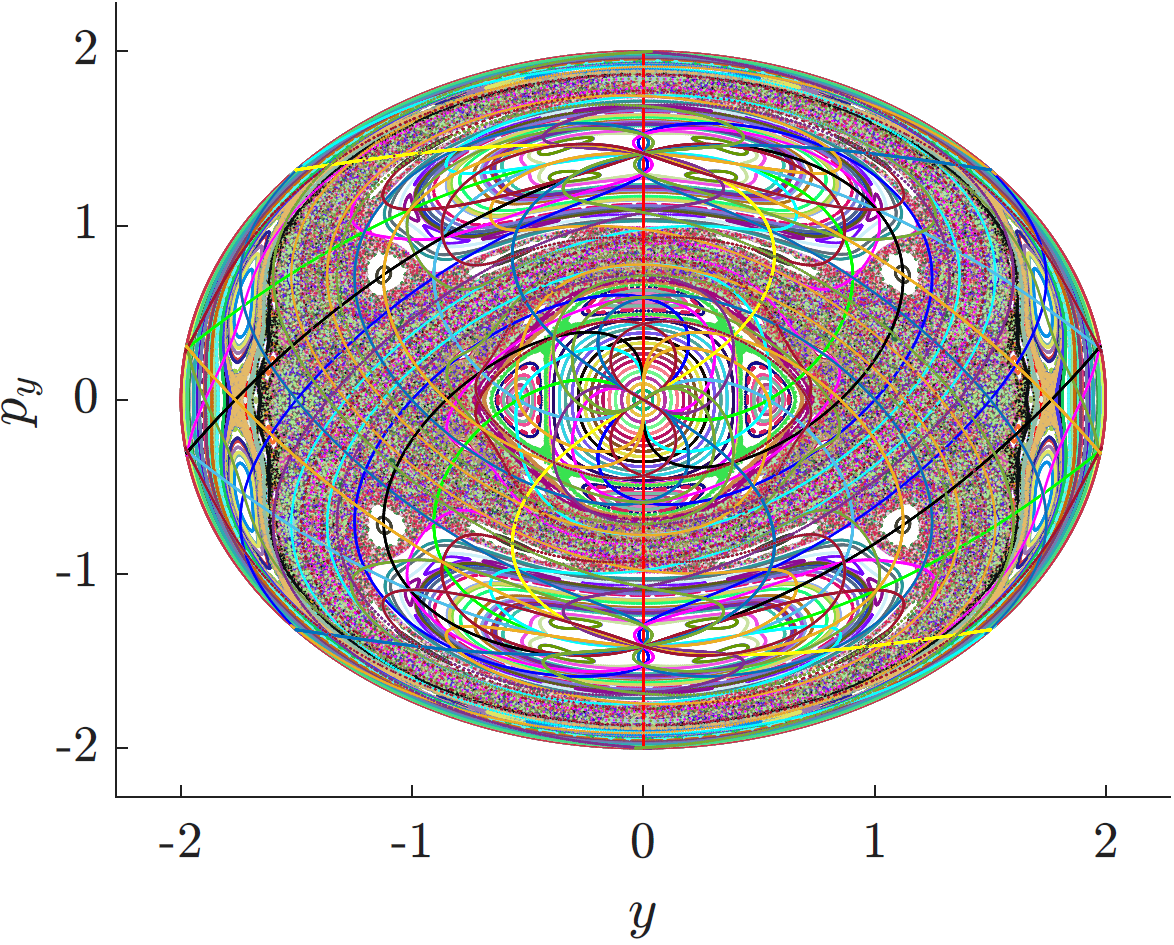} 
        \caption{$(\alpha=1/2,E=2)$}
        \label{mapa_rojo}
    \end{subfigure}\\
    \begin{subfigure}[h!]{0.35\textwidth}
        \centering
        \includegraphics[width=\textwidth]{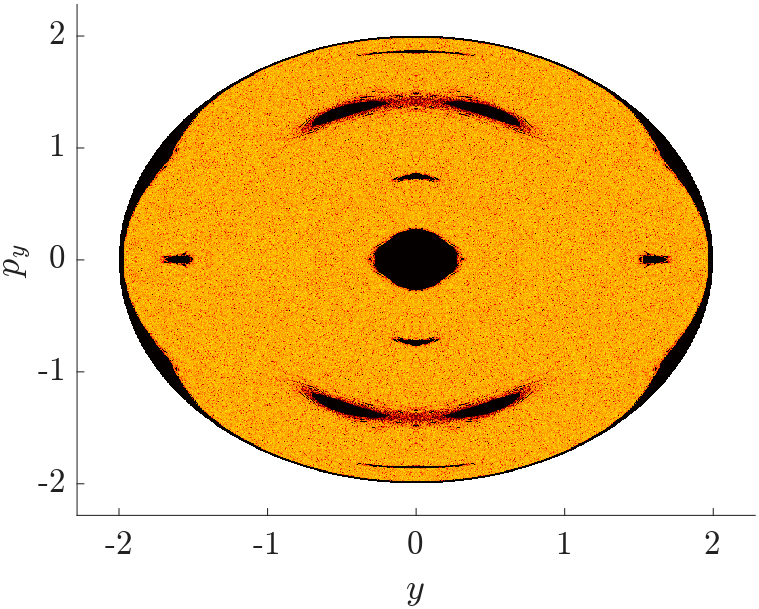} 
        \caption{$(\alpha=3/2,E=2)$}
    \end{subfigure}\begin{subfigure}[h!]{0.35\textwidth}
        \centering
        \includegraphics[width=\textwidth]{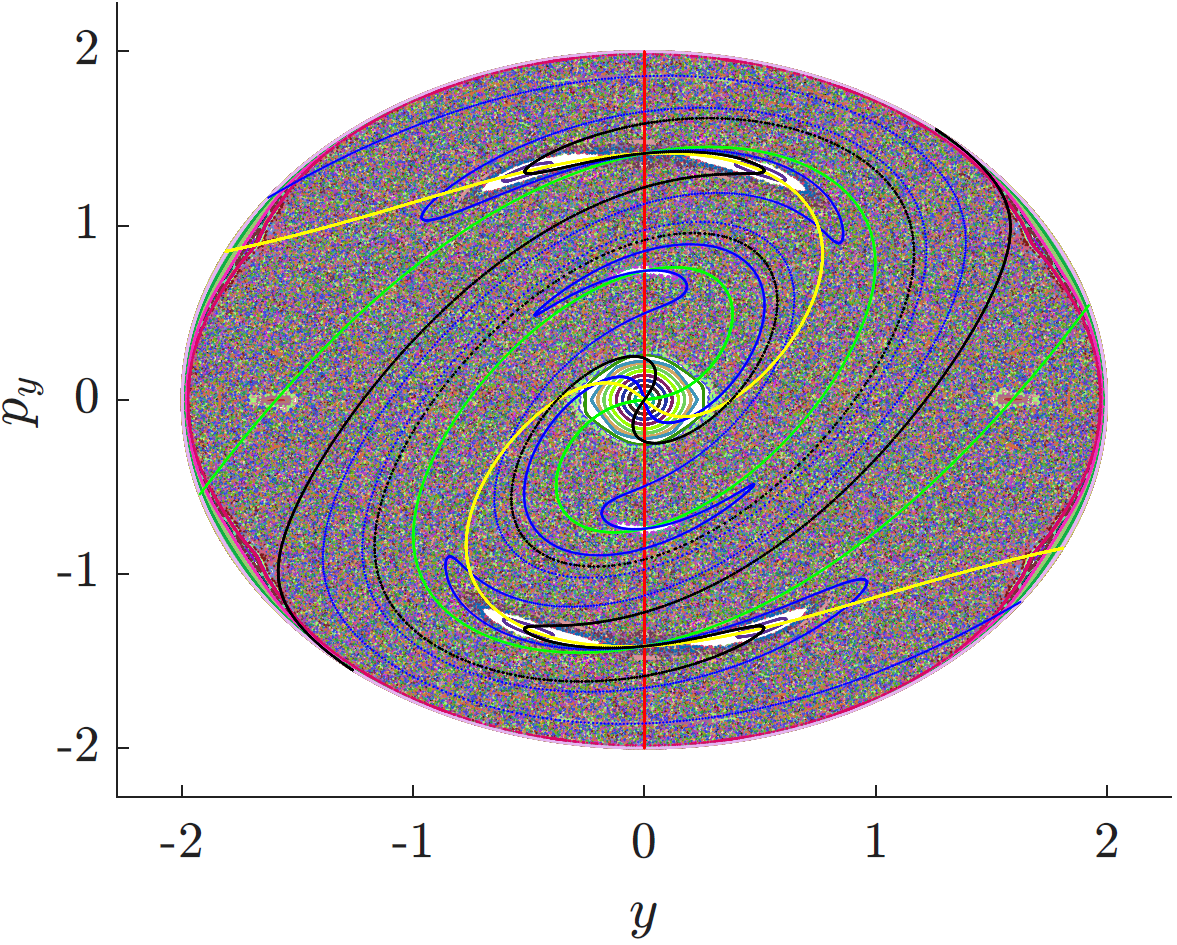} 
        \caption{$(\alpha=3/2,E=2)$}
        \label{mapa_naranja}
    \end{subfigure}\\
    \begin{subfigure}[h!]{0.35\textwidth}
        \centering
        \includegraphics[width=\textwidth]{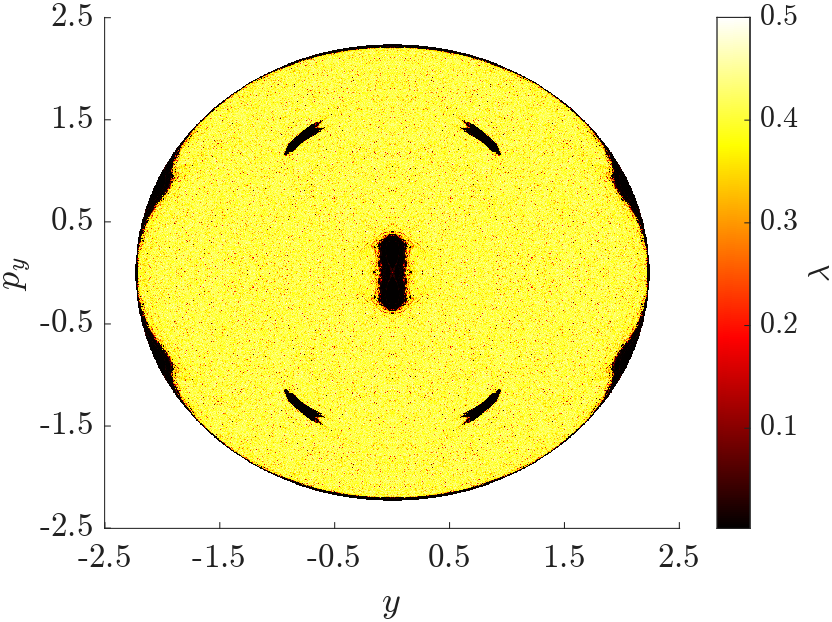} 
        \caption{$(\alpha=7/4,E=5/2)$}
        \label{exp_amarillo}
    \end{subfigure}
    \
    \begin{subfigure}[h!]{0.35\textwidth}
        \centering
        \includegraphics[width=\textwidth]{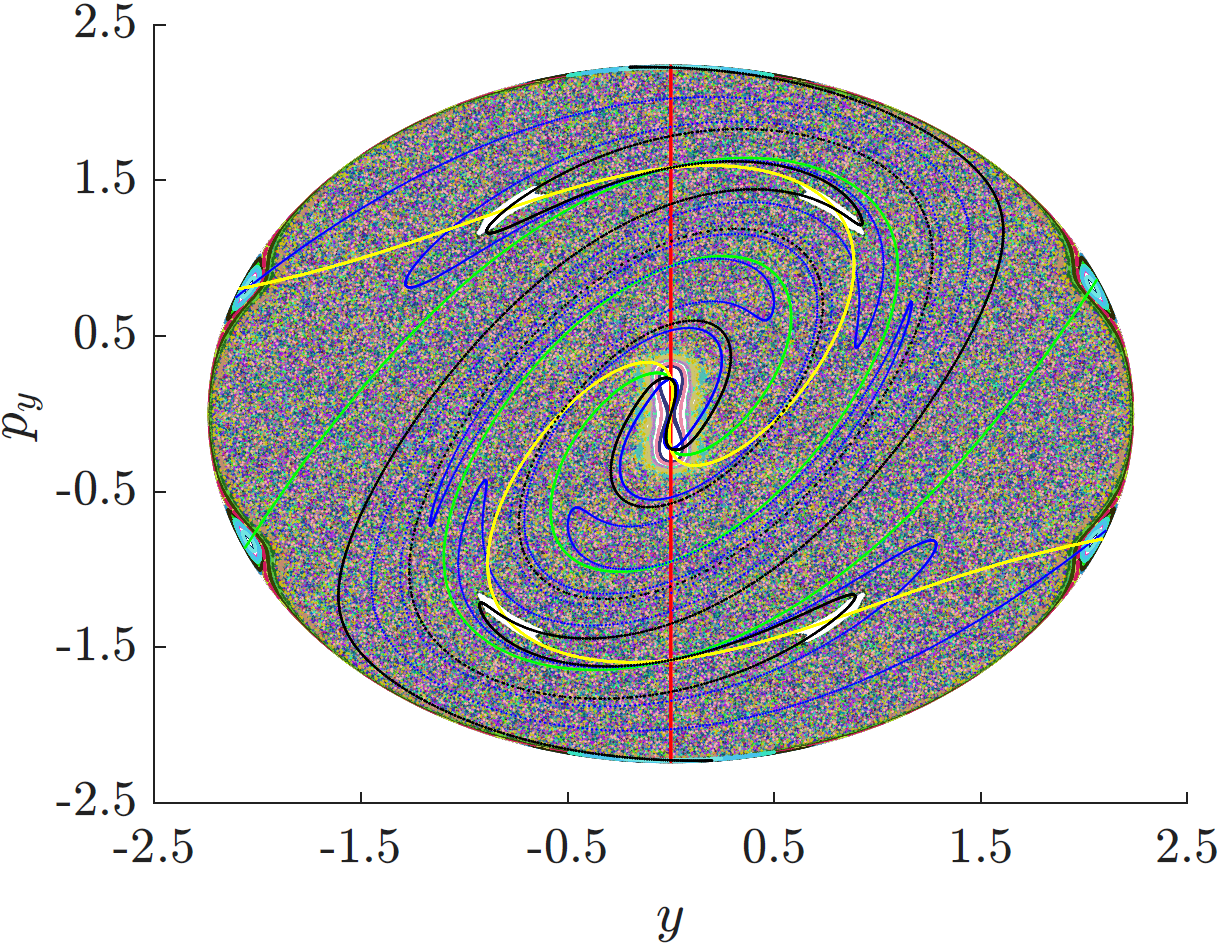} 
        \caption{$(\alpha=7/4,E=5/2)$}   
        \label{mapa_amarillo}
    \end{subfigure}
    \caption{Poincaré sections (right) and largest Lyapunov exponents (left) 
on the $(y,p_y)$ plane for the Contopoulos Hamiltonian with $\alpha>0$ 
at different energies $E$. 
Symmetry lines are indicated in each plot. 
Subfigures \textbf{(a)}–\textbf{(b)}, \textbf{(c)}–\textbf{(d)}, 
\textbf{(e)}–\textbf{(f)}, and \textbf{(g)}–\textbf{(h)} correspond to regions 
\textbf{(I)}, \textbf{(II)}, \textbf{(III)}, and \textbf{(IV)} in \cref{positivo}. 
As the energy increases, motion remains bounded but chaotic regions grow, 
illustrating the progressive destruction of invariant tori.}
    \label{alfa_positivo}
\end{figure*}

\begin{figure*}[h!]
    \centering    
    \begin{subfigure}[h!]{0.47\textwidth}
        \centering
        \includegraphics[width=\textwidth]{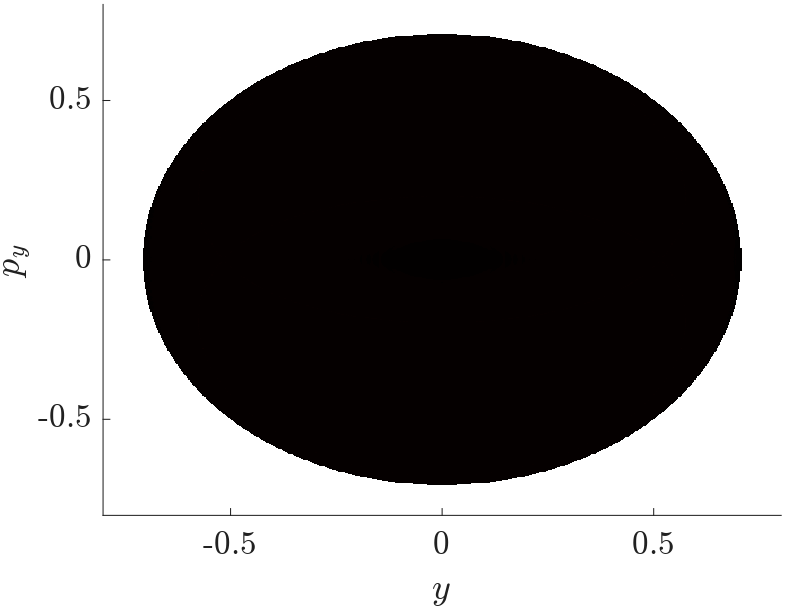} 
        \caption{$(\alpha=-1/2,E=1/4)$}
    \end{subfigure}
    \quad
    \begin{subfigure}[h!]{0.47\textwidth}
        \centering
        \includegraphics[width=\textwidth]{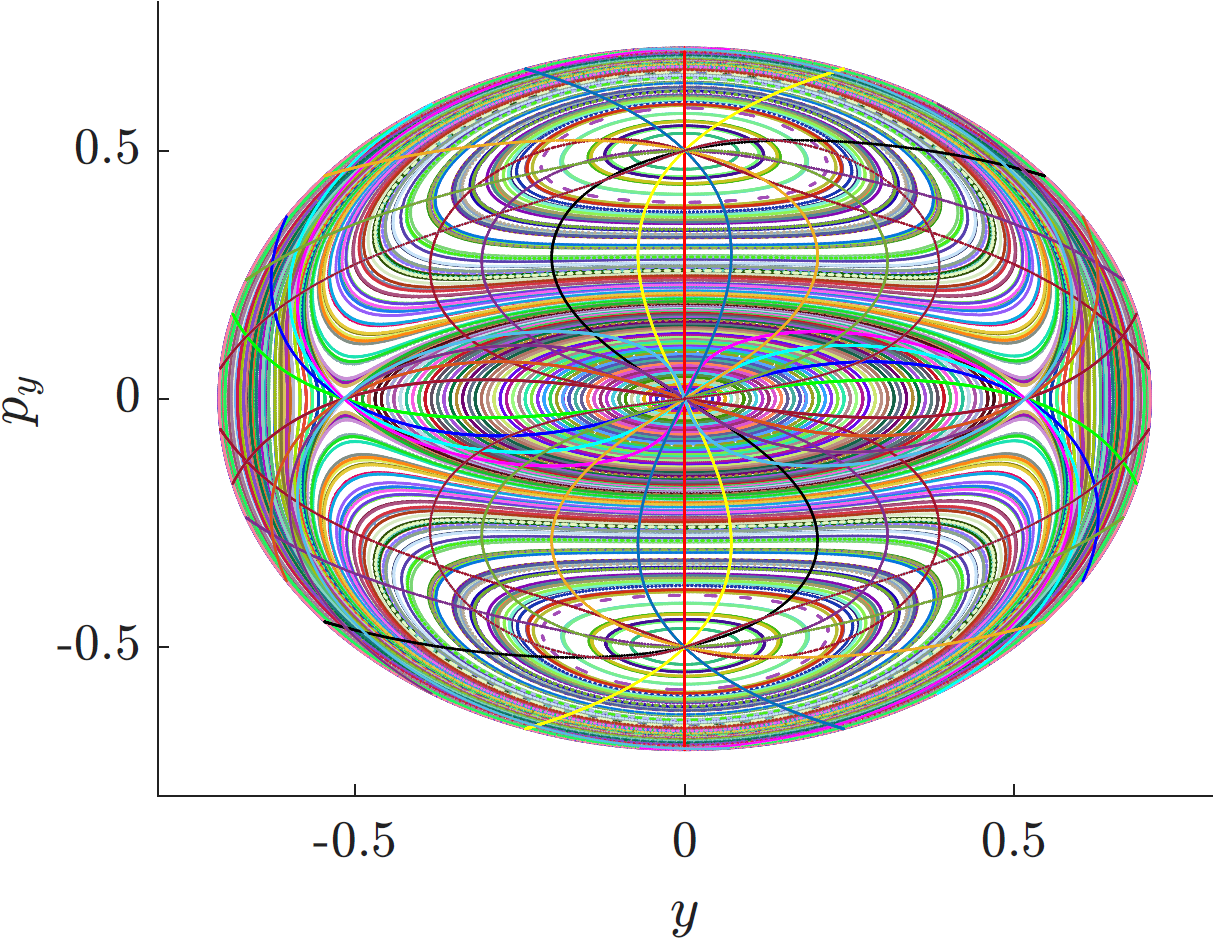}
        \caption{$(\alpha=-1/2,E=1/4)$}
    \end{subfigure}\\
    \begin{subfigure}[h!]{0.47\textwidth}
        \centering
        \includegraphics[width=\textwidth]{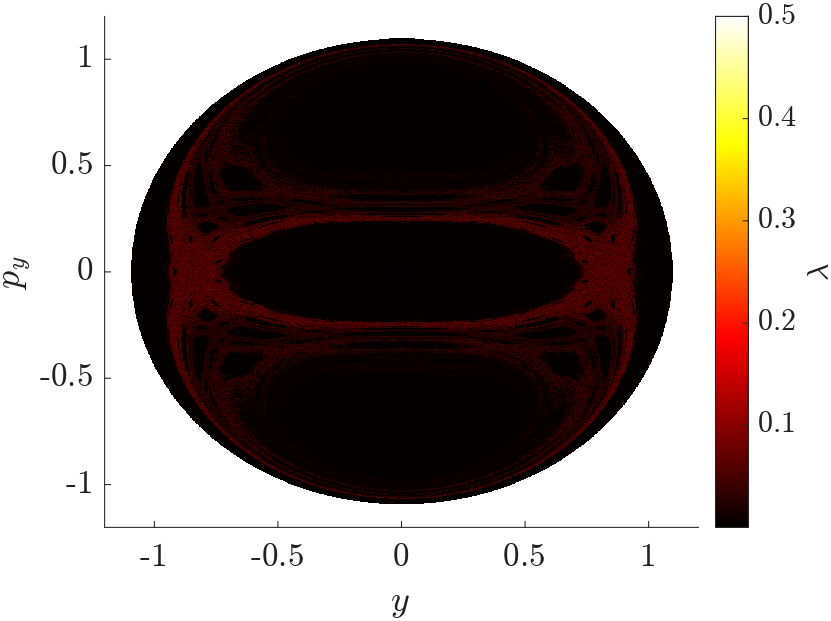}
        \caption{$(\alpha=-1/2,E=3/5)$}
    \end{subfigure}
    \quad
    \begin{subfigure}[h!]{0.47\textwidth}
        \centering
        \includegraphics[width=\textwidth]{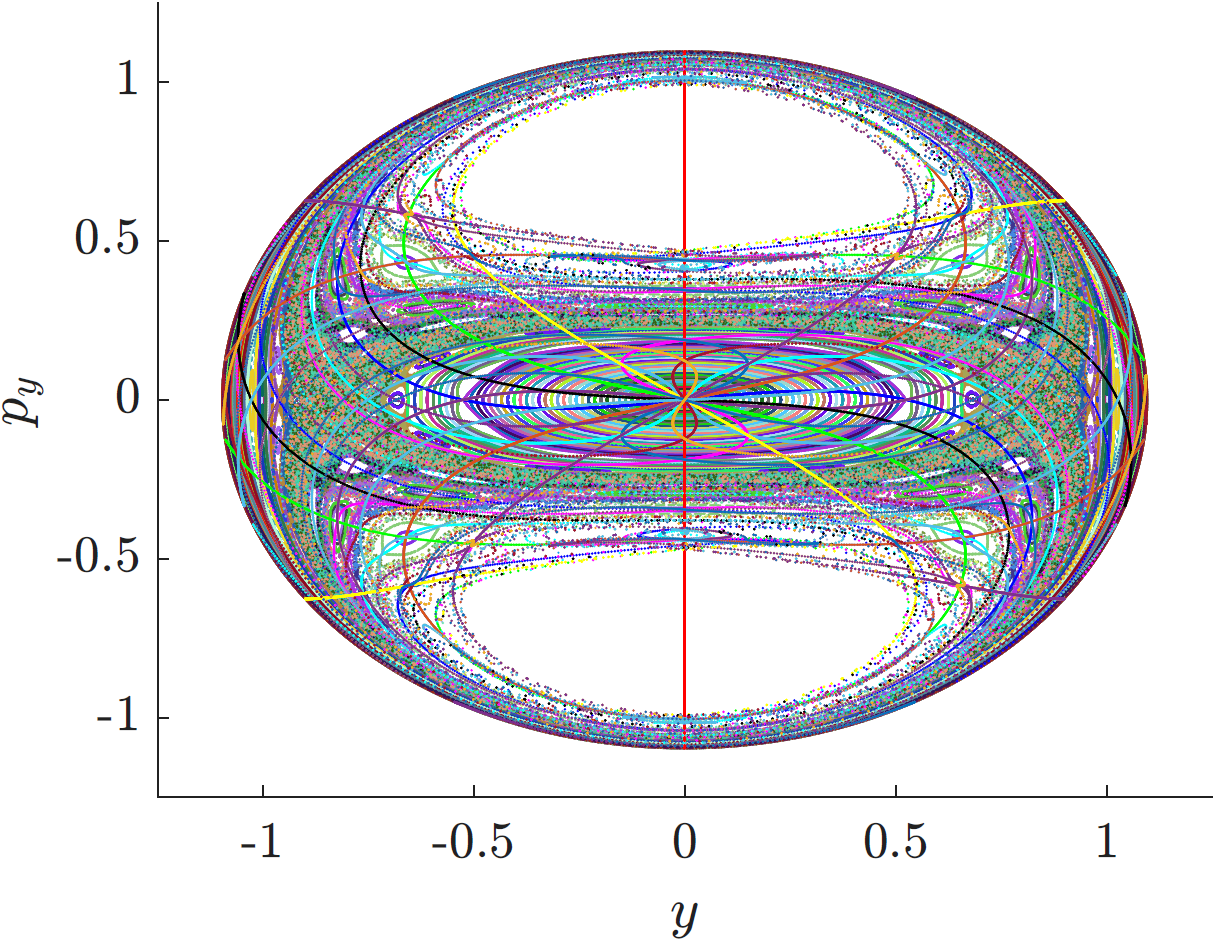}
        \caption{$(\alpha=-1/2,E=3/5)$}
        \label{mapa_neg_rojo}
    \end{subfigure} 
    \caption{Poincaré sections (right) and Lyapunov exponents (left) 
on the $(y,p_y)$ plane for the Contopoulos Hamiltonian with $\alpha<0$ 
at different energies $E$. 
Symmetry lines are displayed. 
Subfigures \textbf{(a)}–\textbf{(b)} and \textbf{(c)}–\textbf{(d)} correspond to regions 
\textbf{(A)} and \textbf{(B)} in \cref{negativo}. 
For $\alpha<0$, real saddles open escape channels along $x=\pm y$: 
at low energy trajectories remain confined, while above the threshold chaotic escape dominates.}
    \label{alfa_negativo}
\end{figure*}

Representative maps in the $(y,p_y)$ plane are shown in~\cref{alfa_positivo} ($\alpha>0$) and ~\cref{alfa_negativo} ($\alpha<0$), with symmetry lines included for reference. At fixed $\alpha=\frac{1}{2}$ and energy $E=\frac{1}{2}$, the intersections of symmetry lines form ordered curves characteristic of near–integrable motion, consistent with a small Lyapunov exponent $\lambda\simeq 3\times 10^{-3}$. By contrast, at the same coupling but higher energy $E=2$, the section displays widespread stochasticity with only small islands of regular motion, and the Lyapunov exponent increases up to $\lambda \gtrsim 0.2$. These results provide direct evidence of the transition from regular to chaotic dynamics as $E$ increases.

\clearpage

\subsection{Lyapunov heat maps of chaos}

The largest Lyapunov exponent (LLE) provides a quantitative measure of chaos by characterizing the exponential divergence of nearby trajectories: regular motion corresponds to $\lambda=0$, while a positive LLE signals sensitive dependence on initial conditions.  

We compute the LLE using the variational method, evolving the equations of motion~\cref{eqsmo} simultaneously with the associated variational equations,
\begin{equation}
   \dot{\phi}^t(\Vec{x}_0) = D_{x}F(f^t(\Vec{x}_0))\, \phi^t(\Vec{x}_0),
   \label{varia}
\end{equation}
where $\phi^t$ is the state–transition matrix and $D_xF$ the Jacobian of the flow. The exponent is extracted from the long–time growth of perturbations,
\begin{equation}
   \lambda = \lim_{t\to\infty}\,\frac{1}{t}\ln\frac{\|\phi^t(\Vec{x}_0)\cdot \delta \Vec{x}_0\|}{\|\delta \Vec{x}_0\|},
   \label{lamda}
\end{equation}
averaged over an ensemble of random perturbations. Numerical integration was performed with an adaptive Runge–Kutta–Fehlberg scheme up to $t=8000$. 

\begin{figure}[h!]
    \centering
    \includegraphics[width=9cm]{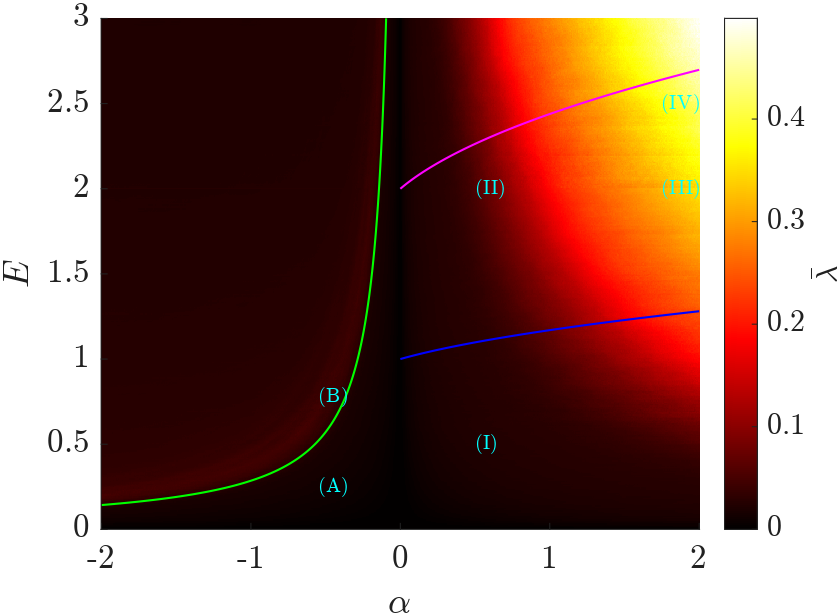}
    \caption{Global Lyapunov heat map of chaos. The average LLE in the $(\alpha,E)$ plane. 
The green curve marks the classical escape threshold. 
For $\alpha<0$, trajectories above this line escape along $x=\pm y$, 
while for $\alpha>0$ motion remains bounded but increasingly chaotic. Lines in blue and violet denote the ground- and first excited energies of the quantum Contopoulos system, respectively.}
    \label{grande}
\end{figure}

To construct Fig.~\ref{grande}, for each fixed pair $(\alpha, E)$ we compute the LLE for a large ensemble of trajectories initialized on a dense grid in the $(p_y, y)$ plane. The LLEs of these trajectories are then averaged, and this average value $\Bar{\lambda}$ corresponds to a single point in the $(\alpha, E)$ heat map. This provides a global classification of the dynamics across parameter space. 
For $\alpha>0$ four regimes (\textbf{I}–\textbf{IV}) emerge, 
while for $\alpha<0$ two distinct regimes of instability are observed 
(regions \textbf{A}, \textbf{B}), as summarized in~\cref{positivo,negativo}, respectively. 

Typical values illustrate the contrast: nearly regular orbits yield $\lambda\simeq 3\times 10^{-3}$, whereas chaotic trajectories reach $\lambda\gtrsim 0.3$. This quantitative distinction complements the Poincaré sections, linking the visual breakdown of invariant tori to the metric growth rates of nearby trajectories.

\begin{table}[h!]
\centering
\caption{Average $\Bar{\lambda}$ of the LLE for different regions in the plane $(\alpha,E)$ for $\alpha>0$ by \cref{grande}}
\begin{tabular}{cc} \noalign{\hrule height 0.8pt} \noalign{\vskip 3pt} 
 Region $(\alpha, E)$ & \quad $\Bar{\lambda}$ 
 \vspace{3pt}
 \\  \hline \noalign{\vskip 5pt}
 \textbf{(I)}\,=\,$(1/2,1/2)$ & \quad  0.00273\\[10pt] 
 \textbf{(II)}\,=\,$(1/2,2)$ &  \quad  0.18960\\[10pt]
 \textbf{(III)}\,=\,$(7/4,2)$ & \quad  0.28029\\[10pt]
  \textbf{(IV)}\,=\,$(7/4,5/2)$ & \quad  0.38311\\[5pt]
\noalign{\hrule height 1.2pt}
\end{tabular}
\label{positivo}
\end{table}

\begin{table}[h!]
\centering
\caption{Average $\Bar{\lambda}$ of the largest Lyapunov exponent for different regions in the plane $(\alpha,E)$ for $\alpha<0$ by \cref{grande}}
\begin{tabular}{cc} \noalign{\hrule height 0.8pt} \noalign{\vskip 3pt} 
 Region $(\alpha, E)$ & \quad $\Bar{\lambda}$ 
 \vspace{3pt}
 \\  \hline \noalign{\vskip 5pt}
 \textbf{(A)}\, = \, $(-1/2,1/4)$ & \quad  0.00274\\[10pt] 
 \textbf{(B)}\, = \, $(-1/2,3/5)$ & \quad 0.08270\\[5pt]
\noalign{\hrule height 1.0pt}
\end{tabular}
\label{negativo}
\end{table}

\subsection{Symmetry lines and periodic orbits}

Periodic orbits in area-preserving maps can be efficiently located using symmetry lines~\cite{pina1987symmetry}. Let $T$ denote the Poincaré map defined by successive intersections of the flow with a chosen section. If $n$ is the smallest integer such that $T^n x = x$, the set $\{x,Tx,\dots,T^{n-1}x\}$ defines a periodic orbit of period $n$. In reversible systems $T$ can be expressed as the product of two involutions, $T=I_1I_0$, and the corresponding fixed-point sets define the fundamental symmetry lines $\Gamma_0$ and $\Gamma_1$.  

The Hamiltonian~\cref{hamiltoniano} possesses reflection symmetry in $p\rightarrow -p$ and invariance under time and spatial inversion. As a result, the Poincaré sections display reflection symmetry with respect to the $y$ and $p_y$ axes, corresponding to the fundamental lines $\Gamma_0$ and $\Gamma_1$. Additional symmetry lines $\Gamma_k$ are generated by repeated application of $T$ or $I$, with $\Gamma_0$ producing all even lines and $\Gamma_1$ all odd ones. Intersections of symmetry lines yield the initial conditions of symmetric periodic orbits, with the orbit period determined by $|i-j|$ or a divisor thereof.  

In practice, we initialize trajectories along the vertical line $\Gamma_0$ (red in \cref{alfa_positivo,alfa_negativo}), discretized into $8000$ points, and iterate under $T$ using the Runge–Kutta–Fehlberg method to generate higher-order lines $\pm \Gamma_1,\dots,\pm \Gamma_8$. The resulting intersections provide initial conditions of periodic orbits, whose stability is directly visible in the Poincaré maps: stable orbits correspond to island centers, while unstable ones coincide with chaotic scattering regions. A representative periodic orbit $\gamma_1(t)$ is shown in \cref{p1a,p1b}, with initial conditions in Table~\ref{B}. Further examples are provided in Appendix~\ref{appx:AA} (see~\cref{p2,p3,p4,p5}), and the corresponding initial conditions are collected in~\cref{II,III,IV}.

\begin{figure}[H]
    \centering    
    \begin{subfigure}[h!]{0.47\textwidth}
        \centering
        \includegraphics[width=\textwidth]{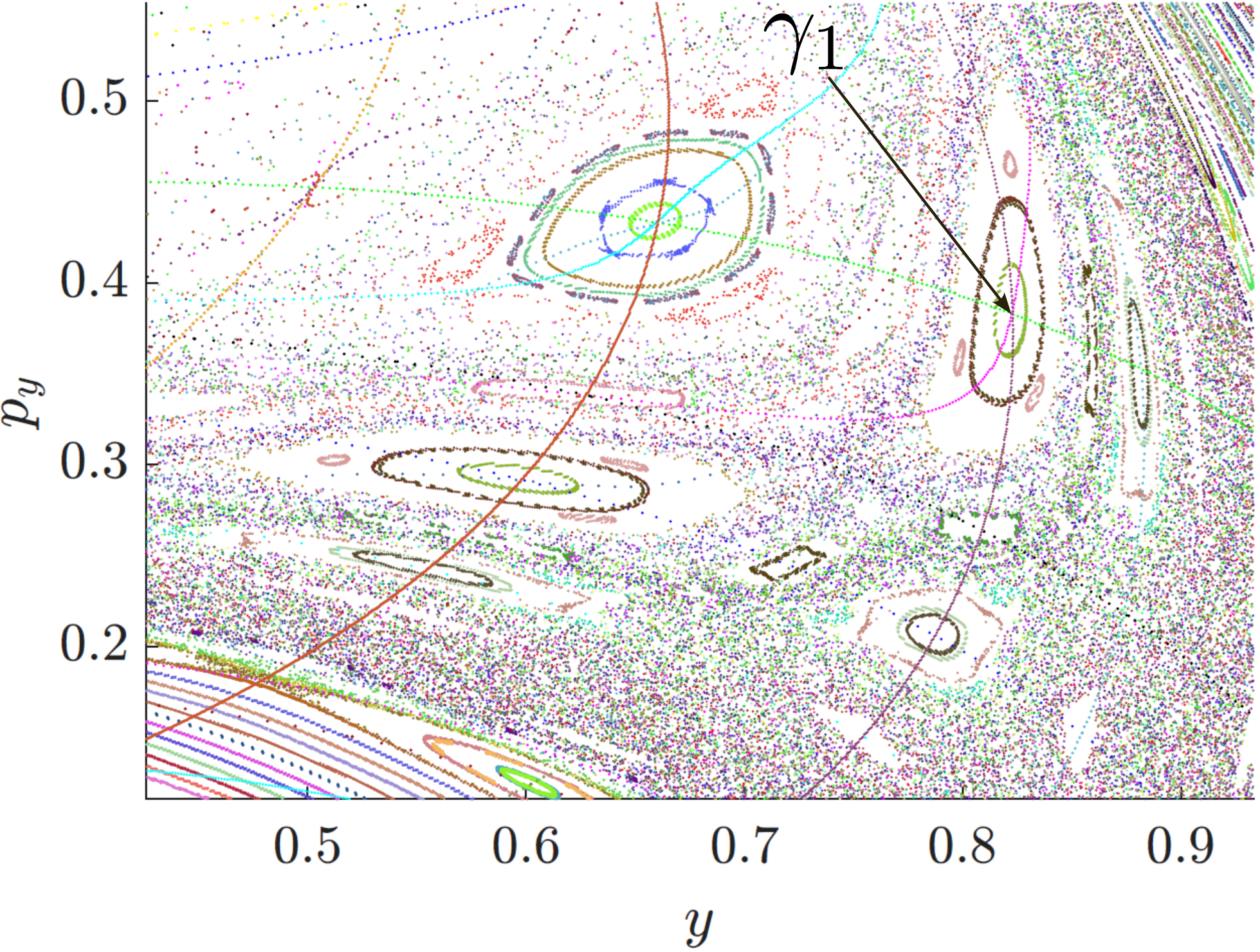} 
        \caption{}
    \end{subfigure}
    \quad
    \begin{subfigure}[h!]{0.47\textwidth}
        \centering
        \includegraphics[width=\textwidth]{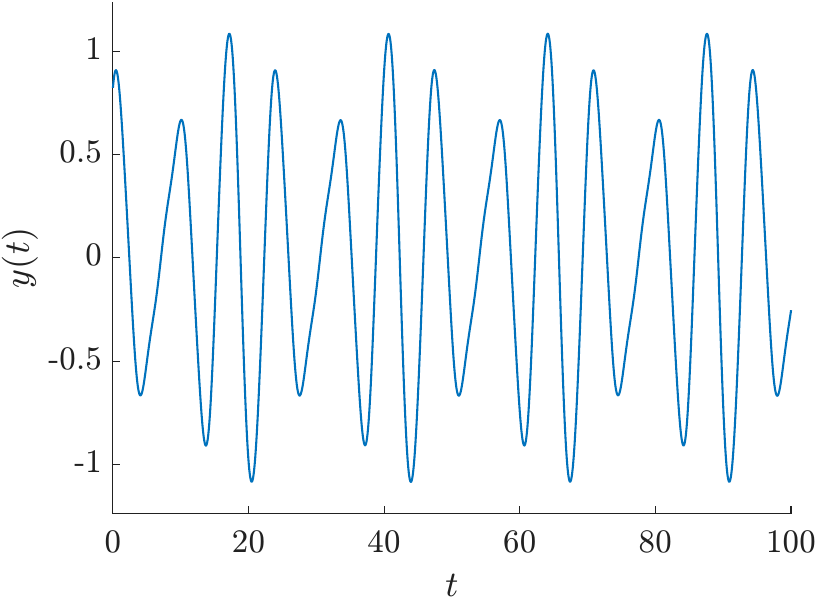}
        \caption{}
    \end{subfigure}
    \caption{Representative periodic orbit $\gamma_1$ at $(\alpha=-1/2,E=3/5)$ obtained from symmetry–line intersections. 
Panels show (a) a phase–space magnification of \cref{mapa_neg_rojo} highlighting the orbit, 
and (b) the time series $y(t)$. The corresponding initial conditions are given in \cref{B}.}
    \label{p1a}
\end{figure}

\begin{figure}[H]
    \centering    
    \begin{subfigure}[h!]{0.5\textwidth}
        \centering
        \includegraphics[width=\textwidth]{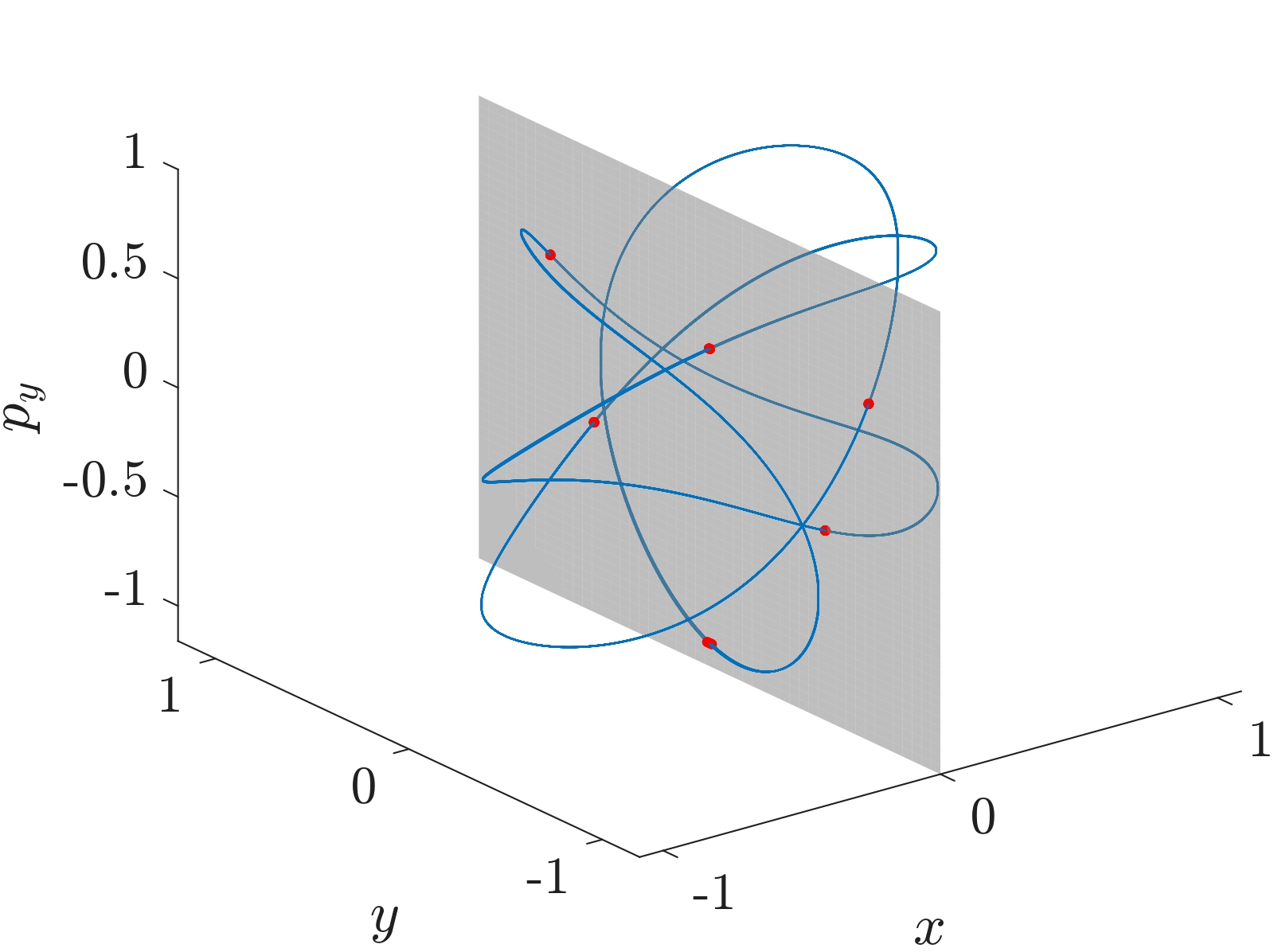}
        \caption{}
    \end{subfigure}
    \quad
    \begin{subfigure}[h!]{0.5\textwidth}
        \centering
        \includegraphics[width=\textwidth]{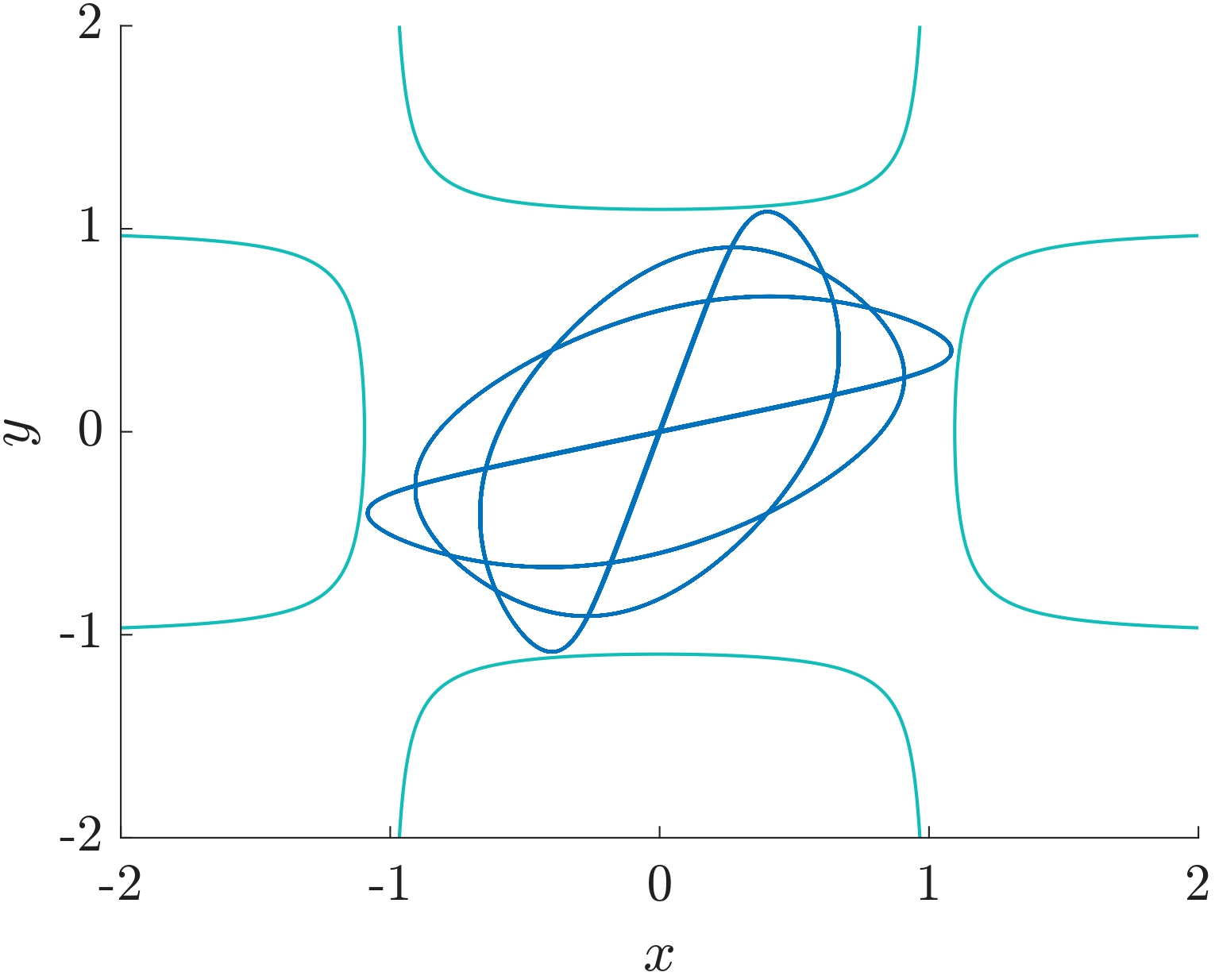}
        \caption{}
    \end{subfigure}         
    \caption{Representative periodic orbit $\gamma_1$ at $(\alpha=-1/2,E=3/5)$ obtained from symmetry–line intersections. Panels show (a) the trajectory on the $x=0$ Poincaré section, 
and (b) the projection on the $(x,y)$ plane. The corresponding initial conditions are given in \cref{B}.}
    \label{p1b}
\end{figure}

\begin{table*}[htp]
\centering
\caption{\small Initial conditions obtained numerically through the intersection of symmetry lines for region \textbf{(B)} by \cref{alfa_negativo}}
\begin{tabular}{cc} \noalign{\hrule height 0.8pt} \noalign{\vskip 3pt} 
 Orbit & $(x,p_x,y,p_y)$ 
 \vspace{3pt}
 \\  \hline \noalign{\vskip 5pt}
 Periodic $\gamma_1$ & $(0,\pm 0.614026743631817,\pm 0.822393150292746,\pm 0.382936893569267)$ \\[10pt] 
 Quasiperiodic & $(0,\pm 0.974679434480896,\pm 1/2,0)$ \\[10pt] 
 Chaotic & $(0,\pm 0.748331477354788,\pm 4/5,0)$\\[5pt]
\noalign{\hrule height 0.8pt}
\end{tabular}
\label{B}
\end{table*}

\subsection{Periodic Orbits from Averaging Theory}

Periodic orbits can also be obtained analytically using averaging methods~\cite{BUICA20047}. Following Jim\'enez-Lara and Llibre~\cite{jimenezlara_periodic_2011}, we consider the rescaling $x=\sqrt{\varepsilon}X$ with parameters $a=0$, $\varepsilon=\alpha$, and $b=2\alpha$. By locating the zeros of the averaged functions $f_{11}$ and $f_{12}$, the first family of periodic solutions arises for $(\alpha^*,\theta)=(0,\pi)$, leading to the initial conditions
\begin{equation}
   x(0)=\sqrt{E}, \quad p_x(0)=0, \quad 
   y(0)=\sqrt{E}, \quad p_y(0)=0.
   \label{lidia}
\end{equation}
A representative example of such orbits, derived directly from the averaging approach, is shown in~\cref{analitico1}, with a further example provided in Appendix~\ref{appx:AA} (see~\cref{analitico2}).

\begin{figure}[h!]
    \centering    
    \begin{subfigure}[h!]{0.5\textwidth}
        \centering
        \includegraphics[width=\textwidth]{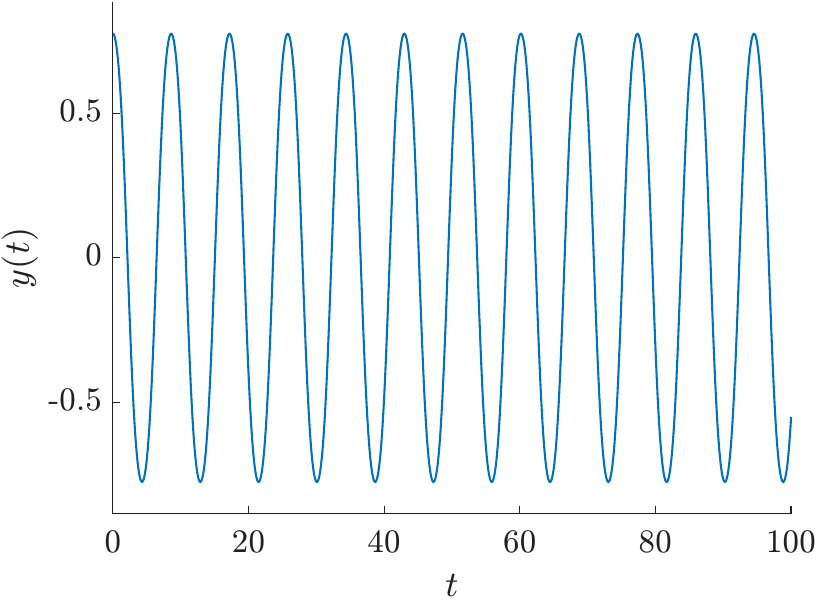} 
    \end{subfigure}\\
    \begin{subfigure}[h!]{0.5\textwidth}
        \centering
        \includegraphics[width=\textwidth]{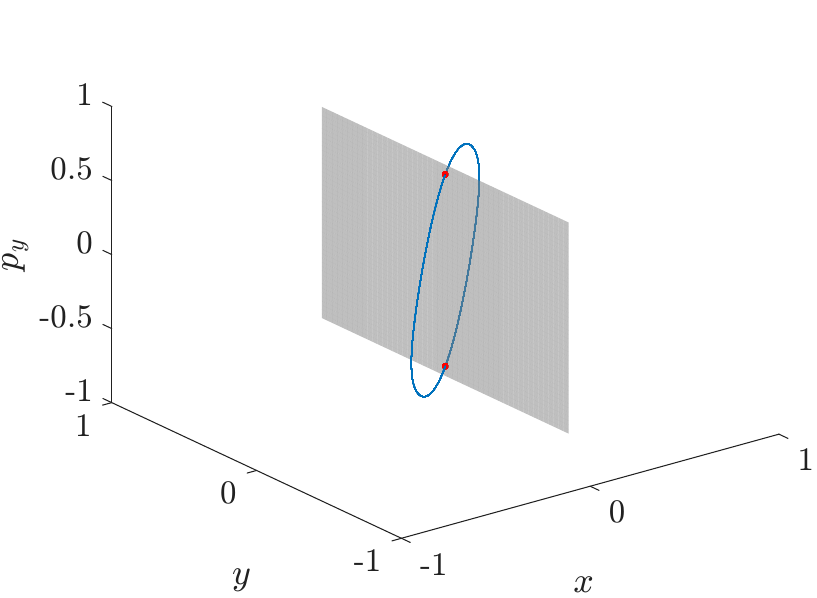}
    \end{subfigure}
    \caption{\small Periodic orbit obtained from averaging theory for $(\alpha=-1/2, E=3/5)$. Panel~\textbf{(a)} shows the trajectory $y(t)$ and its phase–space representation, and panel~\textbf{(b)} the corresponding Poincaré section at $x=0$ (gray plane). Initial conditions follow from~\cref{lidia}.}
    \label{analitico1}
\end{figure}

\paragraph*{Diagonal periodic orbits.}
Averaging theory predicts a family of periodic solutions along the diagonal
subspace $x=y$, corresponding to $(r^*,\theta^*)=(\sqrt{h},0)$ in the
rescaled coordinates of Jiménez--Lara and Llibre~\cite{jimenezlara_periodic_2011}.
Mapping back to our original variables, and using $E\approx \alpha h$, the
initial conditions reduce to (\ref{lidia}). Along this invariant subspace the dynamics reduces to the Duffing oscillator
\begin{equation}
\ddot q + q + \alpha\, q^3 = 0 ,
\qquad
q(t)\equiv x(t)=y(t) ,
\label{eq:duffing}
\end{equation}
with energy
\begin{equation}
E=\tfrac12 A^2+\tfrac{\alpha}{4}A^4 ,
\end{equation}
where $A$ is the turning amplitude. The period reads
\begin{equation}
T(E,\alpha)=
\begin{cases}
\dfrac{2\sqrt{2}}{(1+4\alpha E)^{1/4}}\;
K\!\left(\dfrac{\sqrt{1+4\alpha E}-1}{2\sqrt{1+4\alpha E}}\right),
& \alpha>0 , \\[2.5ex]
\dfrac{4}{\sqrt{\,1+\sqrt{1-4|\alpha|E}\,}}\;
K\!\left(\dfrac{1-\sqrt{1-4|\alpha|E}}{1+\sqrt{1-4|\alpha|E}}\right),
& \alpha<0 ,
\end{cases}
\label{eq:period}
\end{equation}
where $K(m)$ is the complete elliptic integral of the first kind. 
For small nonlinearity this reduces to
\begin{equation}
T(E,\alpha)\;\approx\;2\pi\Big(1-\tfrac{3}{4}\,\alpha E\Big),
\end{equation}
in agreement with the time series of Fig.~\ref{analitico1}.

The existence of diagonal periodic solutions was established by
Jim\'enez--Lara and Llibre~\cite{jimenezlara_periodic_2011} using
averaging theory. Our contribution here is to make the reduction
explicit in the original coordinates, showing that the symmetric
Duffing dynamics yields a closed--form period written directly in terms
of the physical parameters $(E,\alpha)$, see Eq.~\eqref{eq:period}.
To the best of our knowledge, this form of the period has not been
presented before in the context of the quartic Yang--Mills/Contopoulos
Hamiltonians.

%%%%%%%%%%%%%%%%%%%%%%%%%%%%%%%%%%%%%%%%%%%%%%%%%%%%%

\clearpage
\section{Yang-Mills classical Hamiltonian}

Removing the quadratic confinement from~\cref{hamiltoniano} yields the Yang--Mills mechanical Hamiltonian,
\begin{equation}
    H_{\textrm{YM}} \ = \ \tfrac{1}{2}\left(p_x^2 + p_y^2\right) \;+\; \alpha \, x^2 y^2 ,
\end{equation}
where the quartic interaction defines the Yang--Mills potential. For any $\alpha \neq 0$, this system is non--integrable and with $\alpha>0$ exhibits strongly chaotic dynamics.  

\begin{figure}[h!]
    \centering
    \includegraphics[width=9cm]{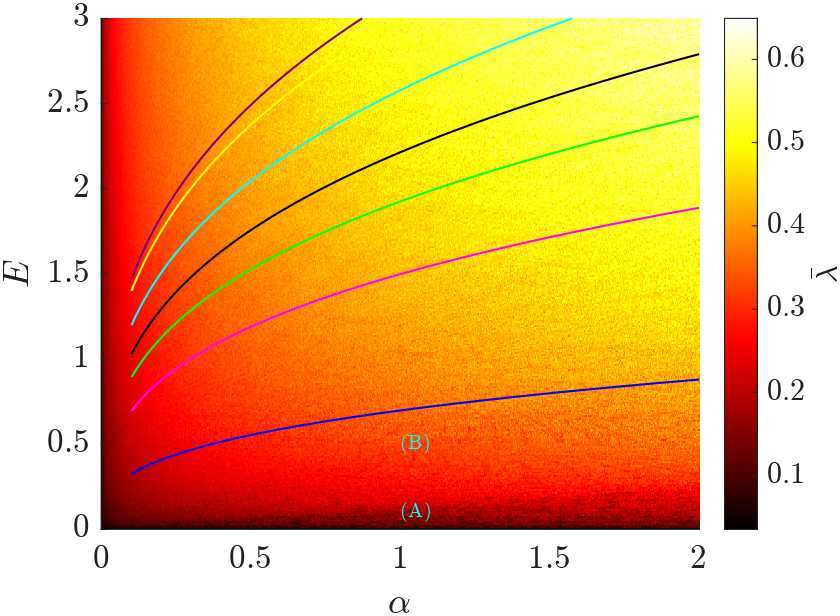}
    \caption{The average LLE in the $(\alpha,E)$ plane for the pure Yang–Mills Hamiltonian. 
Instability appears across the full $(\alpha,E)$ plane, 
reflecting the absence of quadratic confinement. 
Escape occurs along the coordinate axes. Quantum energies are displayed as well.}
    \label{grande2}
\end{figure}

As in the previous section, we characterize the dynamics through the largest Lyapunov exponent $\lambda$. A global heatmap of the averaged values $\Bar{\lambda}$, shown in Figure~\ref{grande2}, reveals the dependence on both $\alpha$ and the energy. Two representative chaotic regimes, labeled \textbf{A} and \textbf{B}, are identified for detailed analysis (see Table~\ref{exponentes_v1}). Local Poincaré sections (not shown) confirm the global picture of Figure~\ref{grande2}. Therefore, even in the absence of quadratic confinement, quantum mechanics restores discrete normalizable states in a system that is classically dominated by chaotic escape.

\begin{table}[h!]
\centering
\caption{Average $\Bar{\lambda}$ of the largest Lyapunov exponent for different regions in the plane $(\alpha,E)$ by Figure~\ref{grande2}.}
\begin{tabular}{cc} \noalign{\hrule height 0.8pt} \noalign{\vskip 3pt} 
 Region $(\alpha, E)$ & \quad $\Bar{\lambda}$ 
 \vspace{3pt}
 \\  \hline \noalign{\vskip 5pt}
 \textbf{(A)}\,=\,$(1,1/10)$ & \quad 0.09031\\[10pt] 
 \textbf{(B)}\,=\,$(1,1/2)$ & \quad 0.29689\\[5pt]
\noalign{\hrule height 0.8pt}
\end{tabular}
\label{exponentes_v1}
\end{table}

\begin{figure}[h!]
    \centering    
    \begin{subfigure}[h!]{0.45\textwidth}
        \centering
        \includegraphics[width=\textwidth]{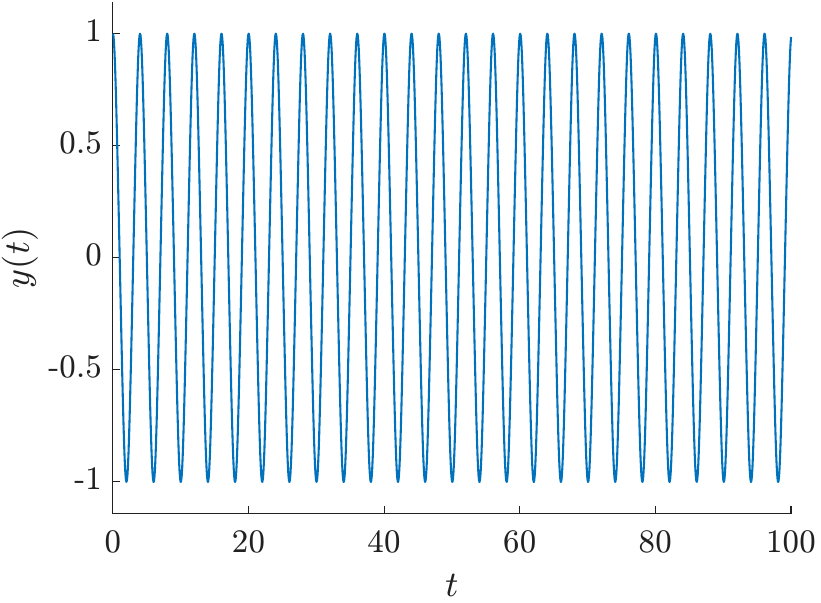} 
        \caption{}
    \end{subfigure}
    \begin{subfigure}[h!]{0.45\textwidth}
        \centering
        \includegraphics[width=\textwidth]{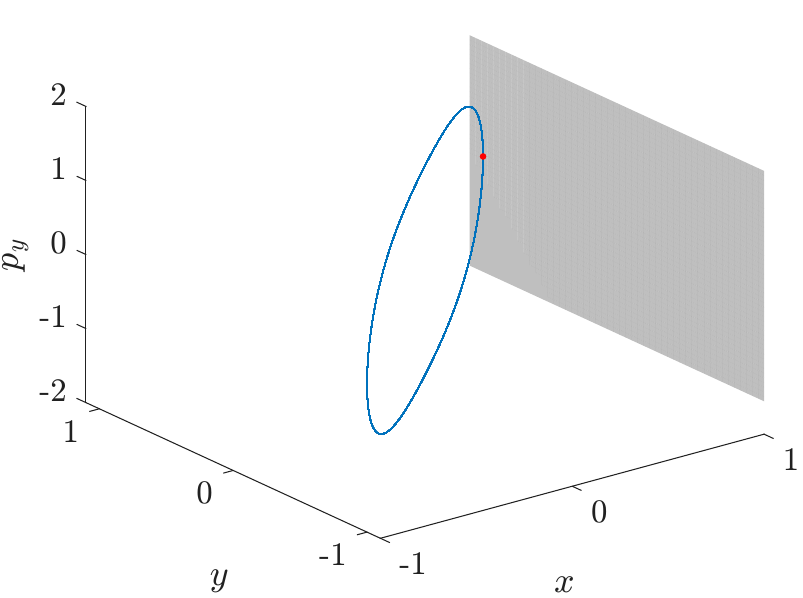}
        \caption{}
    \end{subfigure}
    \caption{\small (Top) projections of the periodic orbit $\gamma_6=(x,y,p_x,p_y)$ with values $(\alpha=1,E=1/2)$. It was found through the intersection of symmetry lines. The corresponding initial conditions are given in \cref{V}. In \textbf{(a)} the time series $y(t)$ and \textbf{(b)} the orbit in phase space as the Poincaré section $x=1$ (gray plane).}
    \label{billar}
\end{figure}

\begin{table}[h]
\centering
\caption{\small Initial conditions obtained numerically through the intersection of symmetry lines 
for region \textbf{(B)} by Figure~\ref{billar}.}
\begin{tabular}{cc} \noalign{\hrule height 0.8pt} \noalign{\vskip 3pt} 
 Orbit & $(x,p_x,y,p_y)$ 
 \vspace{3pt}
 \\  \hline \noalign{\vskip 5pt}
 Periodic $\gamma_6$ & $(1, 0, \pm 1, 0)$ \\[10pt] 
 Chaotic & $(1,\pm 1,\pm \sqrt{2},0)$\\[5pt]
\noalign{\hrule height 0.8pt}
\end{tabular}
\label{V}
\end{table}

Unlike the Contopoulos system, chaos is present at all energies and escape occurs along the axes rather than diagonals.

In contrast to the Contopoulos Hamiltonian, where bounded motion and stability islands coexist at low energies, 
the pure Yang--Mills system exhibits global chaos from the outset, with no persistent regular islands 
and classical confinement lost through escape along the coordinate axes.

\clearpage

\section{Electronic Simulations for the Classical Contopoulos Hamiltonian}

To complement the analytical and numerical analysis, we present an analog electronic implementation of the dynamical equations~\cref{eqsmo} for the Contopoulos Hamiltonian. This provides an experimentally inspired platform where classical chaotic dynamics can be reproduced and controlled using operational–amplifier circuits (see \cite{3bodyexp, ESCOBARRUIZ2025116771}), thereby enabling the construction of experimental systems that offer implementation advantages—such as reduced cost, lower complexity, and simplified measurement protocols—over the original setups \cite{quiroz2017emergence, quiroz2021reconfigurable, jimenez2021experimental,ulmann2022analog, quiroz2022demand, xu2023dynamical}. The circuit design relies on two functional building blocks: adders and integrators, realized with standard op--amp configurations. Nonlinear terms are introduced through integrated analog multipliers.

The functional blocks obey the relations
\begin{equation}
    V_o(t) = -\sum_i \frac{R_f}{R_i}\, V_i(t) ,
\end{equation}
for the adder, and
\begin{equation}
    V_o(t) = -\int \frac{1}{C_i R_i}\, V_i(t)\, dt ,
\end{equation}
for the integrator, where $V_o$ and $V_i$ denote output and input voltages, respectively. Here, $R_j$ are resistors and $C_j$ are capacitors. The analog multipliers generate the nonlinear products,
\begin{equation}
    V_o(t) = \frac{[x_1(t)-x_2(t)] [y_1(t)-y_2(t)]}{10} + z(t) ,
\end{equation}
where $x_1(t)$, $x_2(t)$, $y_1(t)$, $y_2(t)$, and $z(t)$ are input voltages. The prefactor $1/10$ represents the intrinsic attenuation of the device.

\begin{figure*}[htp]
    \centering    
    \includegraphics[width=18.0cm]{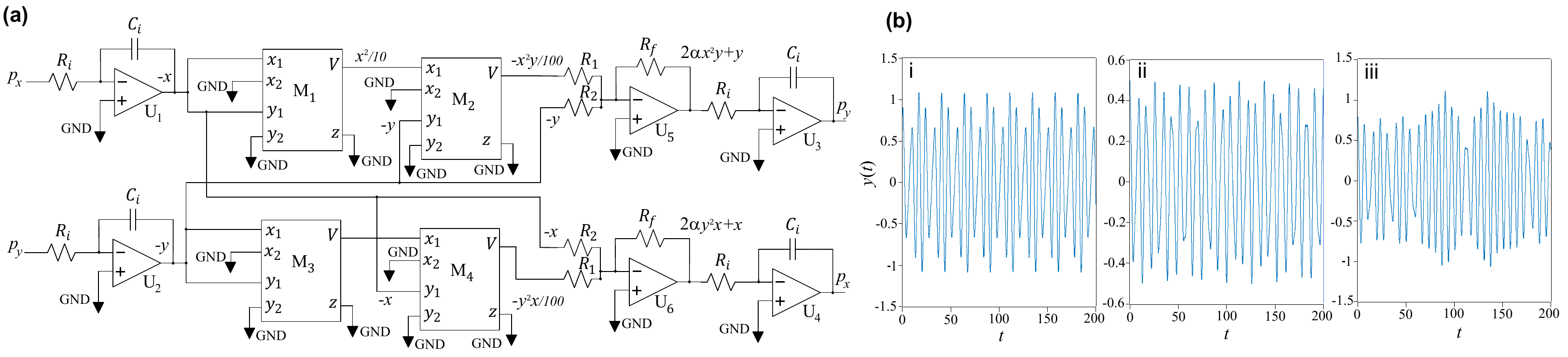} 
    \caption{\small (a) Electronic circuit implementing the dynamical equations~\cref{eqsmo}, built from operational amplifier integrators ($U_1$–$U_4$) and adders ($U_5$, $U_6$), together with analog multipliers ($M_1$–$M_4$) that generate the nonlinear coupling. $R_i$, $R_1$, $R_2$, and $R_f$ denote resistors; $C_i$ denote capacitors. Output voltages for each block are indicated after their respective symbol. (b) Time series of the $y$--variable for energy $E=3/5$ and parameter $\alpha=-1/2$, showing (i) periodic, (ii) quasiperiodic, and (iii) chaotic regimes. Initial conditions are given in~\cref{B}.}
    \label{fig:circuit}
\end{figure*}

The synthesized circuit corresponding to the equations of motion~\cref{eqsmo} is shown in Fig.~\ref{fig:circuit}(a). General--purpose operational amplifiers ($U_j$) implement the integrators ($U_1$–$U_4$) and adders ($U_5$, $U_6$), while the multipliers ($M_j$) generate the quartic coupling. In this configuration the effective coupling constant is tuned by resistor ratios, with
\begin{equation}
    \frac{R_f}{1000 R_1} = 2\alpha .
\end{equation}
To satisfy the normalization of~\cref{eqsmo}, the integrators and the second input terminals of the adders must operate with unit amplification, requiring
\begin{equation}
    \frac{R_f}{R_2} = 1, 
    \qquad \frac{1}{R_i C_i} = 1 .
\end{equation}

The full circuit was simulated in MATLAB using the Simscape toolbox, which enables accurate modeling of the dynamical response and direct validation against theoretical predictions. This approach provides a controlled test of the circuit design before potential hardware realization.

Figure~\ref{fig:circuit}(b) displays representative time series of the $y$--variable for three different dynamical regimes: (i) periodic cf. with Fig.\ref{p1a}, (ii) quasiperiodic, and (iii) chaotic motion. Simulations were performed with energy $E=3/5$ and coupling $\alpha=-1/2$, using the initial conditions listed in~\cref{B}. These results confirm that the circuit faithfully reproduces the diverse behaviors of the Contopoulos Hamiltonian, from regular oscillations to irregular trajectories, under controlled variation of the initial conditions.

Importantly, these analog simulations demonstrate how complex chaotic behavior and escape channels of classical Hamiltonians can be emulated in a tangible platform, thereby reinforcing the central theme of this work: although classical trajectories destabilize, the quantum theory restores confinement.

\section{Quantum systems}

Now, let us consider the quantum counterpart of the models introduced above. 
We first focus on the \emph{Contopoulos Hamiltonian}, which combines the isotropic 
harmonic oscillator with the quartic coupling, and then we analyze the purely quartic 
case corresponding to the \emph{Yang-Mills} mechanical Hamiltonian. Both cases illustrate 
the contrast between bounded quantum spectra and the underlying classical dynamics.  

\subsection{Contopoulos Hamiltonian: variational method for the ground state}

The Hamiltonian operator for the Contopoulos system reads
\begin{equation}
    \hat{\mathcal{H}} \ = \ -\tfrac{1}{2}\left(\tfrac{\partial^2}{\partial x^2} + \tfrac{\partial^2}{\partial y^2}\right)
    + \tfrac{1}{2}(x^2+y^2) + \alpha \, x^2 y^2 ,
    \label{H_cuantico}
\end{equation}
which reduces to the two--dimensional isotropic oscillator when $\alpha=0$. 
For $\alpha \neq 0$, the problem is no longer exactly solvable.  
The exact energy of the ground state $E_0$ can be approximated through the use of the 
variational method, where the variational energy $E^{var}_0$ is given by
\begin{equation}
    E^{var}_0 = \underset{\{\xi_1, \dots\}}{min}\, \langle \psi(\xi_1,\dots) | \hat{\mathcal{H}} | \psi(\xi_1,\dots) \rangle \ \geq \ E_0\,,
\end{equation}
%with $E[\psi] = \langle \psi | \hat{\mathcal{H}} | \psi \rangle$.  
and the trial function $\psi(\xi_1,\dots)$ depends of some parameters $\{\xi_1,\dots\}$.
First, let us consider the normalized Gaussian trial function
\begin{equation}
    \psi_1(x,y) \ = \ \sqrt{\tfrac{2\tau}{\pi}}\, e^{-\tau(x^2+y^2)} ,
\label{psi1C}
\end{equation}
with a single variational parameter $\tau$ and leading to a fully analytical expression 
for the energy $E_0^{\rm var,1}(\alpha)$ (see Appendix~\ref{app:variational}).  
A more general choice is
\begin{equation}
\psi_2(x,y) \ \propto \ e^{-\tau_1(x^2+y^2)-\tau_2 x^2y^2}\ ,
\label{psi2C}
\end{equation}  
with two variational parameters $(\tau_1,\tau_2)$. For this trial function, the 
variational energy $E_0^{\rm var,2}(\alpha)$ is given in terms of the modified Bessel 
function of the second kind $K_q(z)$ (see Appendix~\ref{app:variational}).

\begin{figure}[h!]
    \centering
    \includegraphics[width=8.5cm]{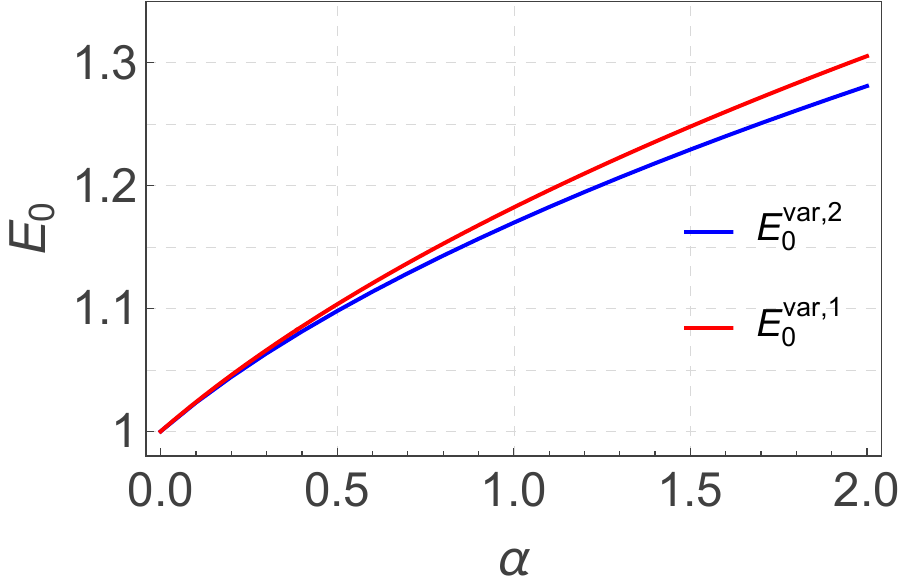}
    \caption{Ground--state energy of the Contopoulos Hamiltonian. 
            Variational energies $E_{0}^{\text{var},1}$ and $E_{0}^{\text{var},2}$ 
            obtained with the trial functions with one--parameter $\psi_1$~(\ref{psi1C}) and 
            two--parameters $\psi_2$~(\ref{psi2C}), respectively.}
    \label{fig:variational_energy}
\end{figure}

\begin{figure}[h!]
    \centering
    \includegraphics[width=8.5cm]{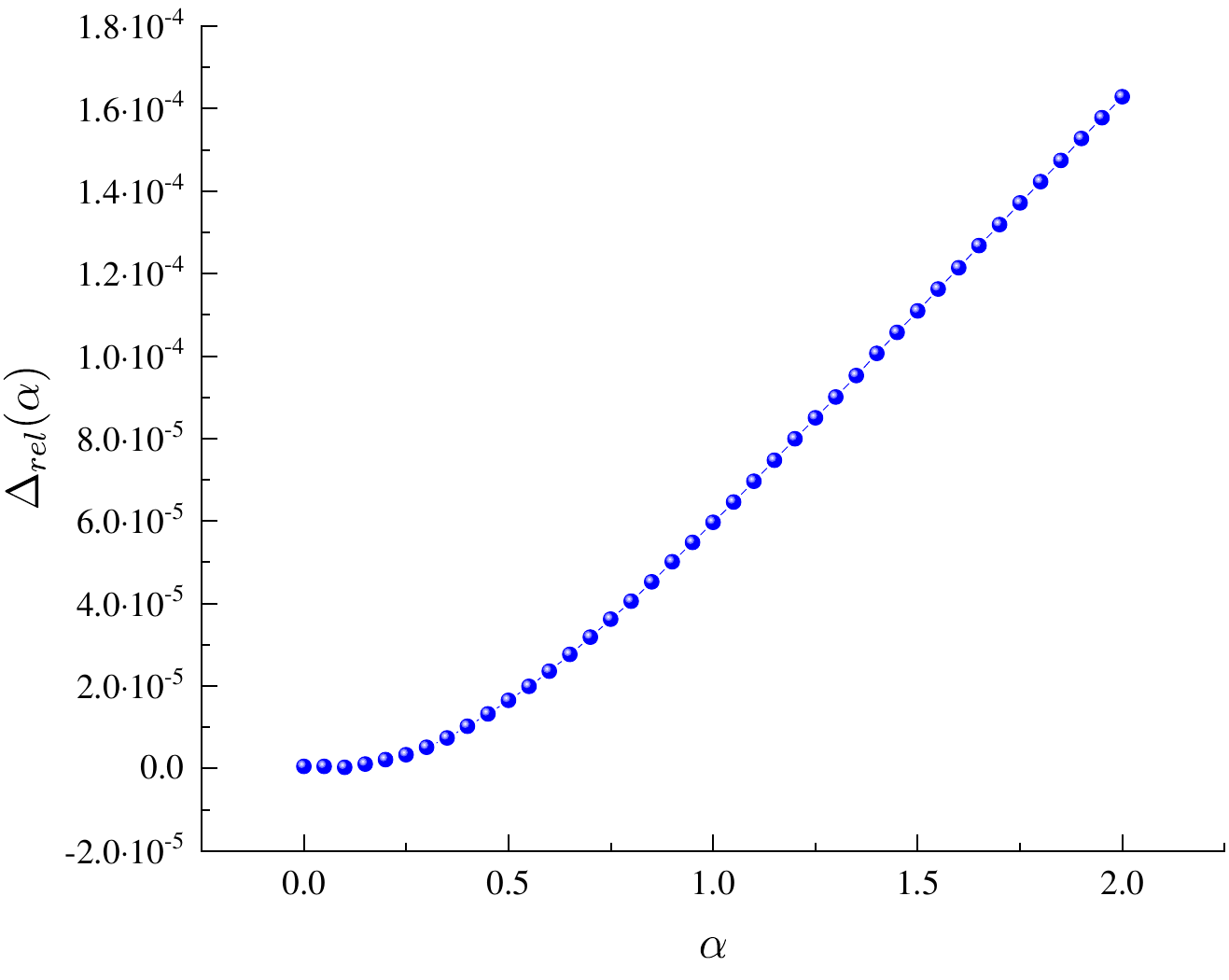}
    \caption{Relative error $\Delta_{\mathrm{rel}}(\alpha)$ between variational energies $E_{0}^{\mathrm{var},2}$ and Lagrange--mesh results $E_{0}^{LM}$.}
    \label{FErel}
\end{figure}

The resulting ground--state energies are shown in Figure~\ref{fig:variational_energy}. 
The relative error $\Delta_{\mathrm{rel}}(\alpha)$ in comparison with the most accurate 
results $E_{0}^{LM}(\alpha)$ using the Lagrange-Mesh method (see below)  for $E_{0}^{\mathrm{var},2}(\alpha)$
\begin{equation}
\Delta_{\mathrm{rel}}(\alpha) \ = \  
\frac{   E_{0}^{\mathrm{var},2}(\alpha)\,-\,E_{0}^{LM}(\alpha) }
     {E_{0}^{LM}(\alpha)} \,,   \label{Erel}
\end{equation}
remains below $10^{-4}$ across the studied range of~$\alpha$ (see Figure~\ref{FErel}).

Further details of the analytic derivation, the full expression for $E_0^{\rm var,1}(\alpha)$, and the behavior of the optimal parameters $\{\tau,\tau_1,\tau_2\}$ are given in Appendix~\ref{app:variational}.

\subsection{Contopoulos Hamiltonian: Lagrange--mesh method}

Now, the previous variational analysis is extended by using the Lagrange-Mesh 
method~\cite{baye2015lagrange}, a highly accurate method for solving eigenvalue problems. 
The method combines Lagrange interpolating 
functions with Gaussian quadrature, leading to sparse Hamiltonian matrices that can be diagonalized 
efficiently, even with relatively large bases. The Lagrange–mesh provides benchmark-quality eigenvalues to be used for spectral statistics. 

In the Lagrange-mesh method, the wave function $\Psi(x,y)$ for the 
Hamiltonian~(\ref{H_cuantico}) can be expressed as
\begin{equation}
\Psi(x,y)=\sum_{i=1}^{N_x} \sum_{j=1}^{N_y} c_{i,j}\,f_i(x)\,g_j(y)\,.  
\end{equation}
The set of $N_x$ Lagrange functions $\{f_i(x)\}$ are given by~\cite{dbaye2006}
\begin{equation}
\label{lfhp}
f_i(x) = \dfrac{(-1)^{N_x+i}}{\sqrt{\pi^{1/2}\,2^{N_x +1}\,N_x!}}\dfrac{H_{N_x}(x)}{x-x_i}e^{-\frac{x^2}{2}}\,,  
\end{equation}
where $H_{N_x}$ is the Hermite polynomial of degree $N_x$ and $x_i$ are
the zeros $H_{N_x}(x_i)=0$. In the same way, the $N_y$ functions 
$\{g_j(y)\}$ have the same structure as that for the $f_i(x)$ 
functions~(\ref{lfhp}) but with  $y$ as the independent variable. 

The energy spectra were computed using a mesh with $N_x = N_y = 40-100$, 
implemented in \textit{Mathematica} with a working precision of 20.

In the unperturbed oscillator ($\alpha=0$), except for the ground state, all states with $n_x+n_y=n$ are degenerate at 
energy $E^{(\alpha=0)}_n=n+1$. The quartic perturbation $V=\alpha \,x^2 y^2$ systematically lifts this 
degeneracy as can be seen in Figure~\ref{enerLMCP}. Numerical results are reported in Table~\ref{tLMC}
of Appendix~\ref{appx:LMM}. The ground state ($n=0$) increases monotonically 
as a function of parameter $\alpha$, while the first excited state ($n=1$) remains a degenerate doublet. 
At $n=2$, the three states $\{\lvert 2,0\rangle,\lvert 1,1\rangle,\lvert 0,2\rangle\}$ split into three distinct levels, with $\lvert 1,1\rangle$ decoupled and the other two forming symmetric/antisymmetric combinations. For $n=3$, the four states split into two doublets. This progression illustrates how the quartic interaction breaks the high degeneracy of the isotropic oscillator.

\begin{figure}[h!]
    \centering
    \includegraphics[width=8.0cm]{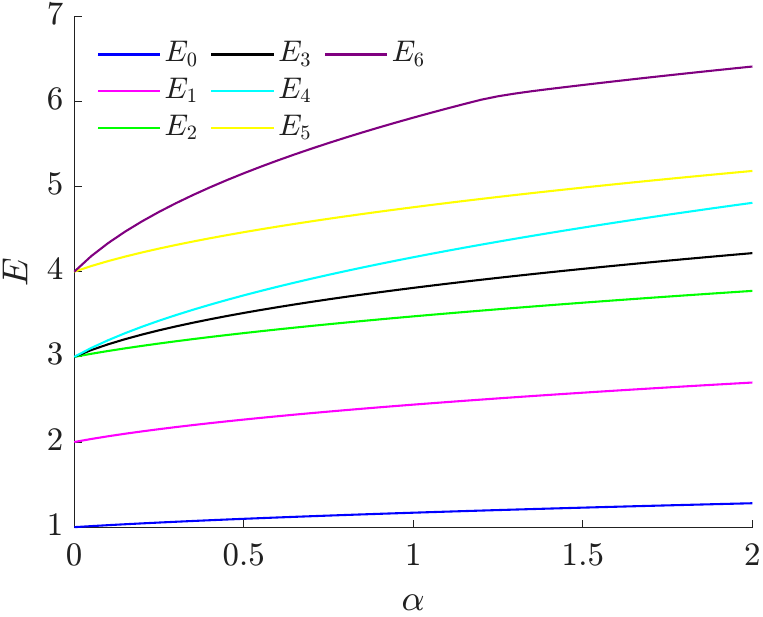}
    \caption{Energies $E_n$ of the lowest states $n=0,1,\ldots,6$ as a function of the parameter $\alpha$ 
    for the Contopoulos Hamiltonian when implementing the Lagrange-Mesh method. Quartic coupling lifts 
    oscillator degeneracies, with splittings increasing at higher energies.}
    \label{enerLMCP}
\end{figure}

\begin{figure}[h]
  \centering
  \subfloat[]{\includegraphics[width=.49\linewidth]{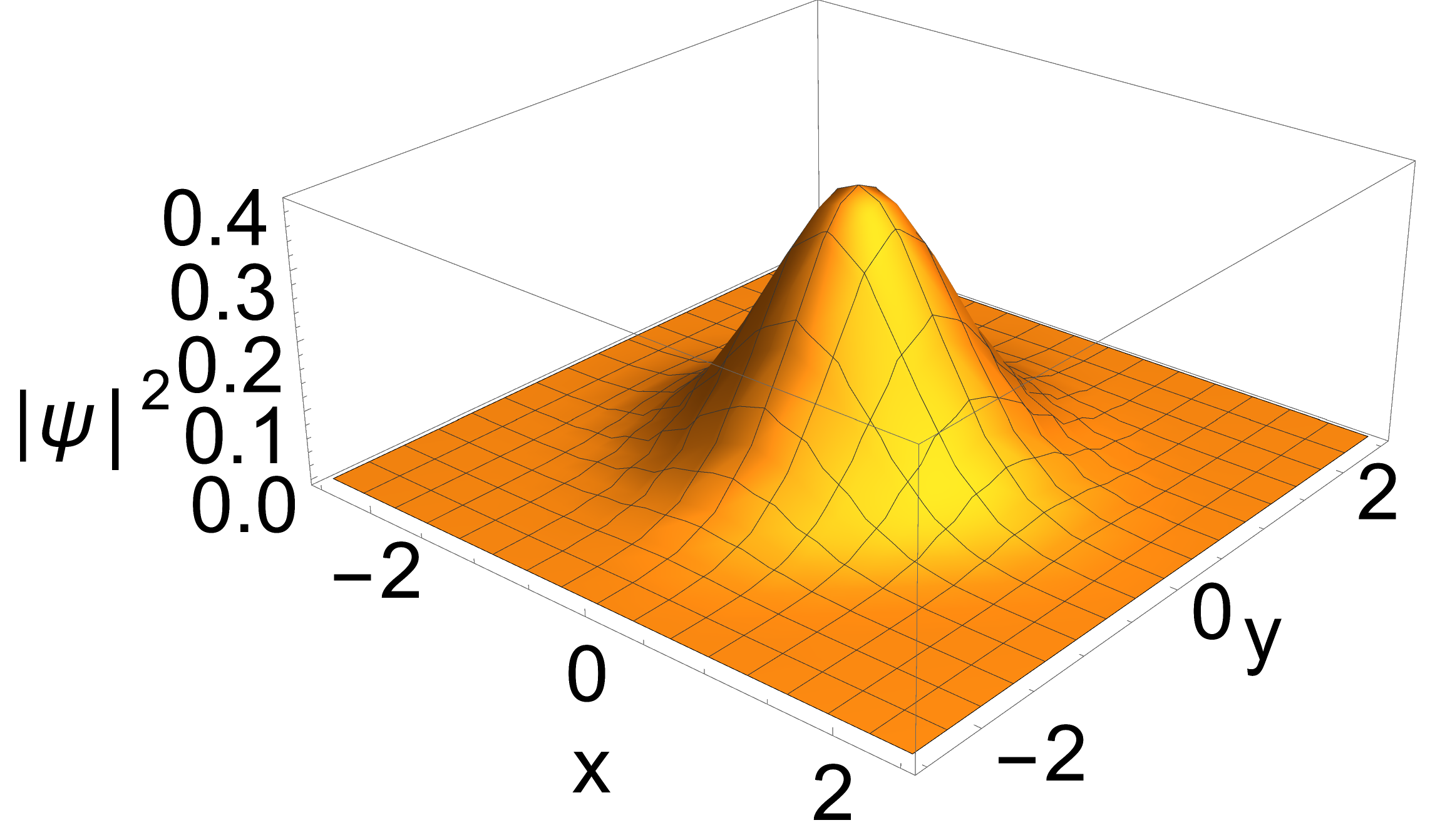}}\hfill
  \subfloat[] {\includegraphics[width=.49\linewidth]{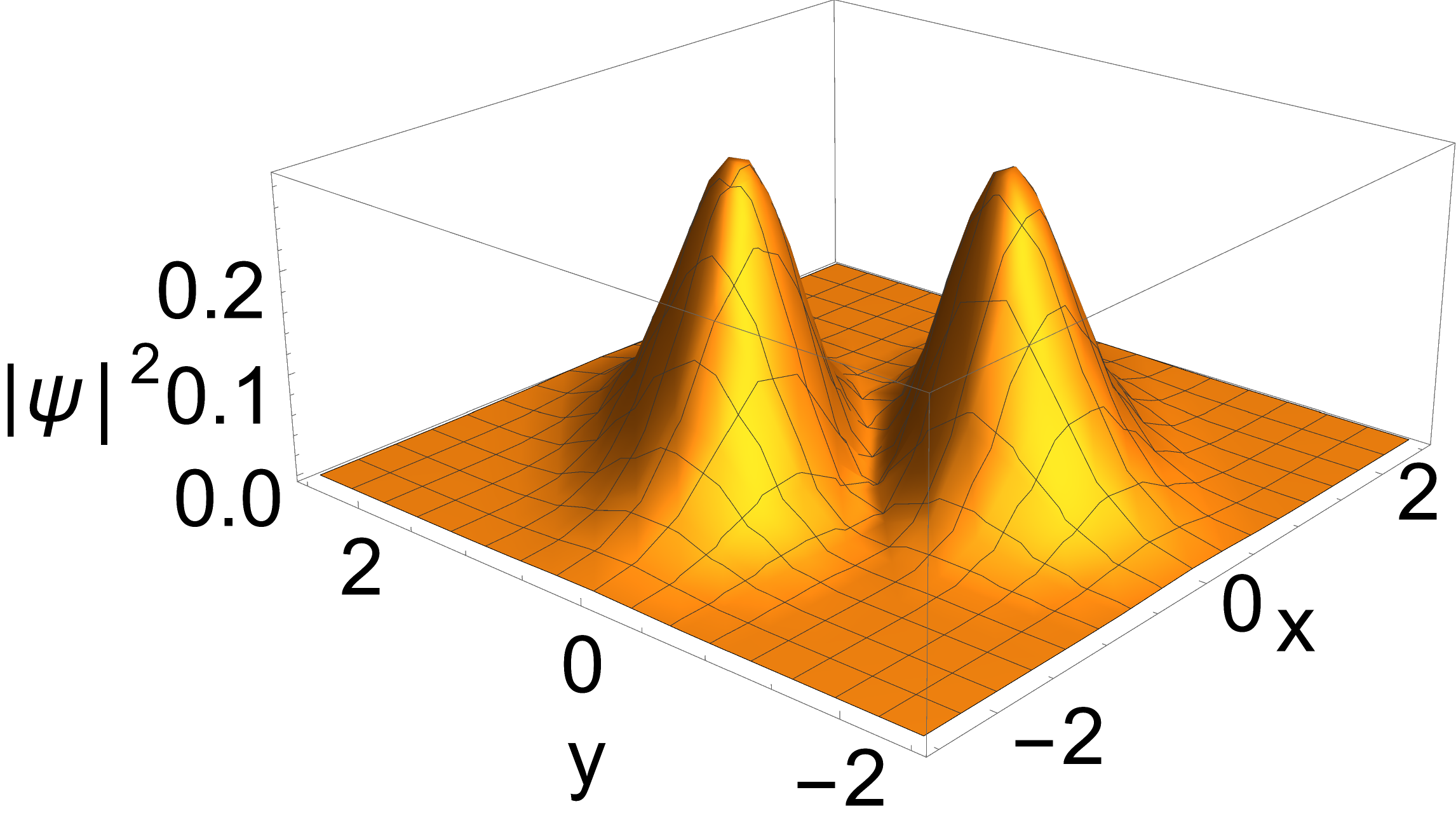}}
  \caption{$a$) $|\psi|^2$ of the ground state ($n=0$) and $b$) the first excited state ($n=1$)
  for the Contopoulos potential with $\alpha=1$.}
  \label{psi2CP}
\end{figure}

The probability density $|\psi|^2$ for the ground state ($n=0$) and the first excited state ($n=1$)
are depicted in Figure~\ref{psi2CP} for $\alpha=1$. The $|\psi|^2$ profiles are similar for different 
values of the $\alpha$ parameter.

\subsection{Pure Yang--Mills Hamiltonian}

Finally, we turn to the purely quartic quantum system,
\begin{equation}
\label{YMV}
    \hat{\mathcal{H}}_{\rm YM} \ = \ -\dfrac{1}{2}\left(\dfrac{\partial^2}{\partial x^2} + \dfrac{\partial^2}{\partial y^2}\right) + \alpha \, x^2 y^2 ,
\end{equation}
($\alpha>0$) which corresponds to the Yang--Mills quantum Hamiltonian. Classically, this system is non--integrable, has escaping channels, and exhibits strongly chaotic dynamics.  

In the quantum regime, however, the behavior is markedly different. For $\alpha > 0$, the eigenfunctions remain normalizable and the spectrum is discrete, despite the absence of the harmonic term. Unlike the Contopoulos Hamiltonian, the pure Yang--Mills system lacks the continuous oscillator symmetry and therefore has no rotational degeneracy structure. Nevertheless, the quartic potential is invariant under independent sign reversals of each coordinate and under the exchange $x \leftrightarrow y$, forming the discrete symmetry group $D_{4}$. Consequently, the spectrum exhibits systematic degeneracies corresponding to the one-- and two--dimensional irreducible representations of $D_{4}$, resulting in pairs of degenerate energy levels.

\begin{figure}[h!]
    \centering
    \includegraphics[width=8.5cm]{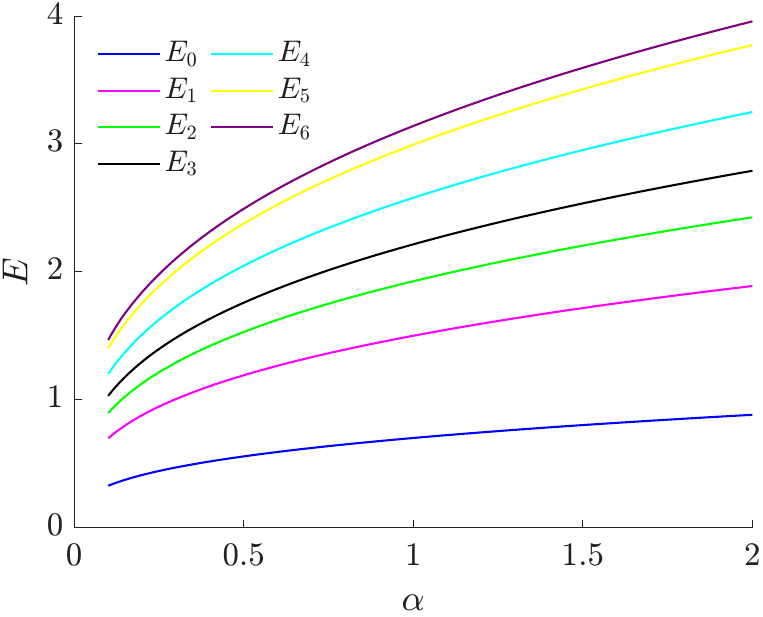}
    \caption{The Lagrange-Mesh energies $E_n$ of the lowest states $n=0,1,\ldots,6$ as a function of the parameter $\alpha$ for the pure Yang--Mills Hamiltonian. }
\label{estados_sin_oscilador}
\end{figure}

The Lagrange--mesh method provides accurate numerical results for this case as well. The resulting spectrum, shown in 
Figure~\ref{estados_sin_oscilador}, illustrates the qualitative difference with the Contopoulos Hamiltonian: the levels are more irregularly spaced and lack the systematic degeneracies of the oscillator-based model.  
Numerical results are reported in Table~\ref{tLMYM} of Appendix~\ref{appx:LMM}.

\begin{figure}[h]
  \centering
  \subfloat[]{\includegraphics[width=.49\linewidth]{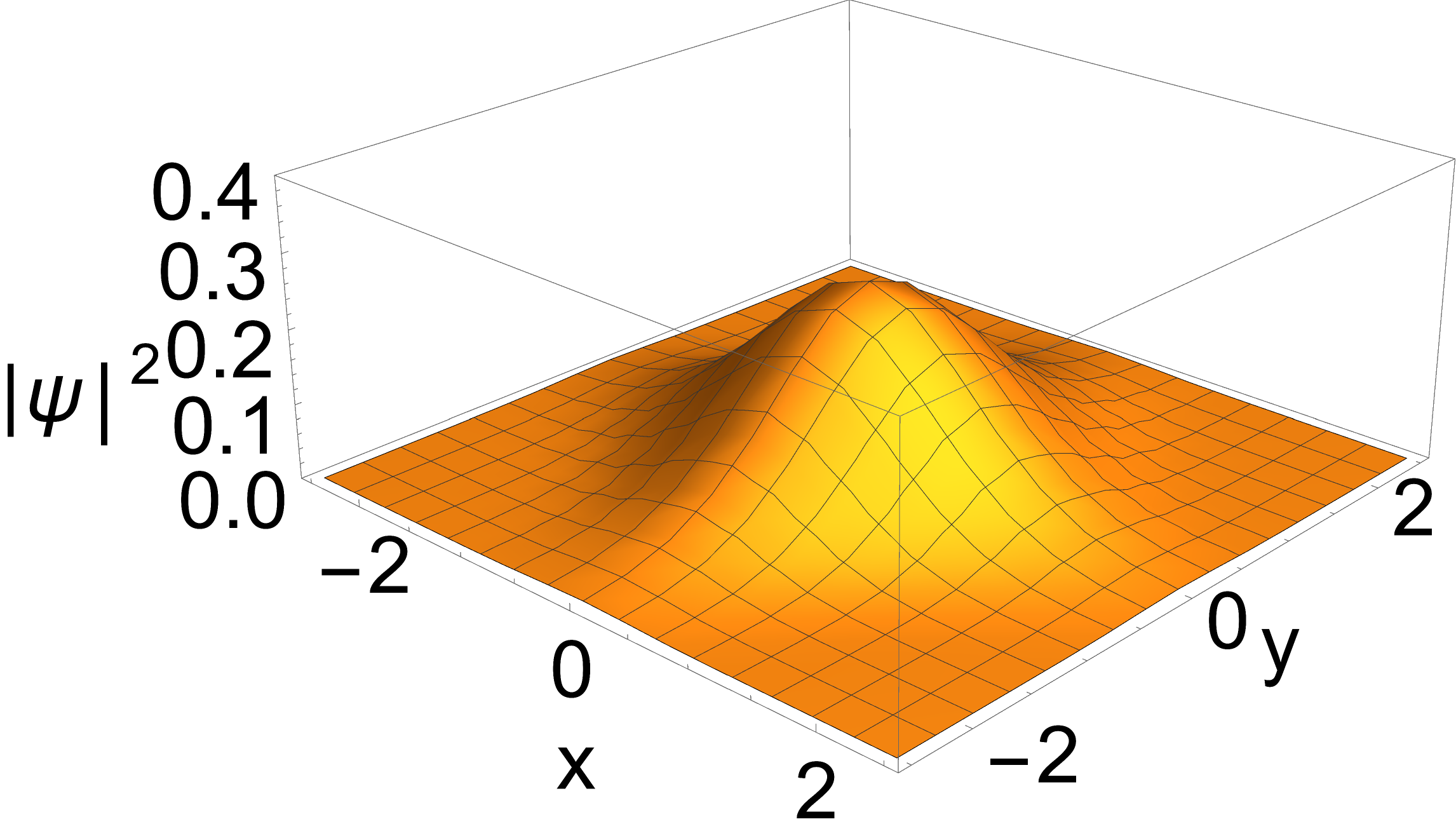}}\hfill
  \subfloat[] {\includegraphics[width=.49\linewidth]{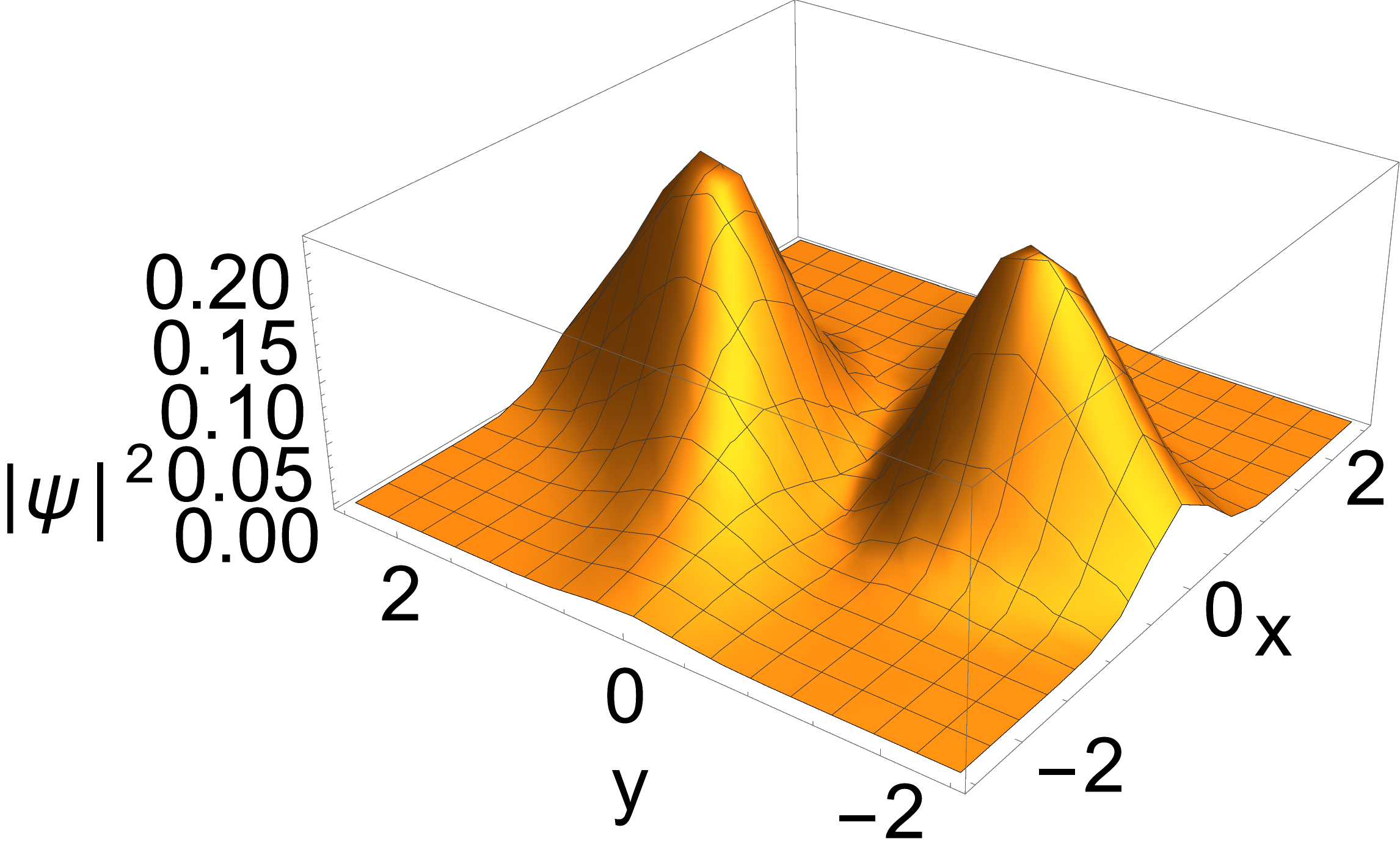}}
  \caption{$a$) $|\psi|^2$ of the ground state ($n=0$) and $b$) the first excited state ($n=1$)
  for the Yang-Mills potential~(\ref{YMV}) with $\alpha=1$.}
  \label{psi2YMP}
\end{figure}

In Figure~\ref{psi2YMP} the probability density $|\psi|^2$ for the ground state ($n=0$) and the 
first excited state ($n=1$) are depicted  for $\alpha=1$. Similar profiles for $|\psi|^2$ 
are obtained for different values of the parameter $\alpha$. 

\section{Semiclassical mechanism of restoration of confinement}
\label{sec:semi+num}

To clarify \emph{how} quantum mechanics restores confinement in systems with 
classical escape channels let us examine again the two planar 
Hamiltonians (\ref{H_cuantico}) and (\ref{YMV}).
Throughout this section we set $\hbar=1$.
%\begin{equation}
%H_{\mathrm{C}}=\tfrac12(p_x^2+p_y^2)+\tfrac12(x^2+y^2)+\alpha\,x^2y^2,
%\end{equation}
%\begin{equation}
%H_{\mathrm{YM}}=\tfrac12(p_x^2+p_y^2)+\alpha\,x^2y^2,
%\end{equation}
%to clarify \emph{how} quantum mechanics restores confinement in systems with classical escape channels. Throughout this section we set $\hbar=1$.

\subsection{Effective confinement from transverse quantization}
\label{subsec:BOlinear}

Again, take the quartic Yang--Mills Hamiltonian~(\ref{YMV})
\begin{equation}
  {\hat H}_{\rm YM} = \tfrac12(p_x^2+p_y^2)+\alpha\,x^2y^2 .
  \label{YMVn}
\end{equation}
For fixed value of $x$, the motion in the $y$-coordinate experiences a harmonic potential
$V_y(y)= (\alpha x^2)\, y^2$ with frequency
\begin{equation}
  \omega_y(x)=\sqrt{2\alpha}\,|x| .
\end{equation}
Projecting onto the $n$-th transverse eigenstate $\chi_n(y;x)$ (Born--Oppenheimer
approximation, neglecting derivative couplings) gives the ansatz
$\Psi(x,y)\approx \phi_n(x)\chi_n(y;x)$, where $\phi_n(x)$ satisfies
\begin{equation}
  \left[-\frac{1}{2}\frac{d^2}{dx^2} + V_{\rm eff}^{(n)}(x)\right]\phi_n(x)=E\,\phi_n(x) ,
  \label{eq:BOeq}
\end{equation}
with the effective potential
\begin{equation}
  V_{\rm eff}^{(n)}(x)=\Bigl(n+\tfrac12\Bigr)\,\omega_y(x)
  =\Bigl(n+\tfrac12\Bigr)\sqrt{2\alpha}\,|x| .
  \label{eq:VeffLinear}
\end{equation}
Thus, \emph{transverse zero--point motion produces a linearly rising barrier}
$V_{\rm eff}^{(n)}(x)\propto|x|$ along the classically open direction.
Since $|x|$--type tails yield purely discrete spectra in one dimension,
each adiabatic channel is confining, and the full two--dimensional spectrum is discrete.
This physical picture agrees with rigorous results proving that
$-\tfrac12\Delta+\alpha x^2y^2$ has a purely discrete spectrum for $\alpha>0$
despite the existence of classical escape channels
\cite{Simon1983AnnPhys,Simon2009MFAT}.

\subsection{Channel quantization and zero--point barriers}

\paragraph{Yang--Mills potential.}
For $n=0$, the effective one--dimensional Hamiltonian in Eq.~\eqref{eq:BOeq} reduces to
\begin{equation}
H_x = -\frac12\frac{d^2}{dx^2} + \frac{\Omega}{2}|x|, 
\qquad \Omega=\sqrt{2\alpha},
\label{Hred}
\end{equation}
which corresponds to a particle in a linear potential. 
Its semiclassical WKB quantization yields an Airy--type spectrum with the characteristic scaling
$E_m \propto \alpha^{1/3}$ 
\cite{simon1983semiclassical,matinyan1986classical,cohen2019quantum} 
(see Sec.~\ref{YMComp} for a more detailed comparison with numerical data). 
This mechanism illustrates how transverse zero--point fluctuations generate an effective linear barrier along the classical escape directions, thereby ensuring confinement in $H_{\mathrm{YM}}$ at all energies.

\paragraph{Contopoulos potential.}
Consider instead the Contopoulos Hamiltonian, and rotate to channel--aligned coordinates 
\(u=(x+y)/\sqrt{2}\), \(v=(x-y)/\sqrt{2}\). 
The potential becomes
\begin{equation}
  V(u,v)=\tfrac12\,(u^2+v^2)+\frac{\alpha}{4}\,(u^2-v^2)^2 .
\end{equation}
Expanding around the escape line $v=0$ and keeping terms up to $v^2$ gives
\begin{equation}
  V(u,v) = \Big[\tfrac12-\tfrac{\alpha}{2}\,u^2\Big]v^2 + \tfrac12\,u^2 + \frac{\alpha}{4}\,u^4
          \;+\;\frac{\alpha}{4}\,v^4 ,
\end{equation}
so that the transverse mode has frequency 
\(\omega_v^2(u)=1-\alpha u^2\), contributing a zero--point correction
\begin{equation}
  V_{\rm ZP}(u)=\tfrac12\,\sqrt{\,1-\alpha u^2\,}.
\end{equation}
The corresponding adiabatic Hamiltonian along the channel reads
\begin{equation}
  H_u = \tfrac12 p_u^2 + V_{\rm eff}(u), \qquad
  V_{\rm eff}(u) = \tfrac12\,u^2 + \frac{\alpha}{4}\,u^4 + \tfrac12\,\sqrt{\,1-\alpha u^2\,}.
\end{equation}
This refines low-lying spacings near channels, but it is not needed to ensure discreteness of the spectra. For $\alpha>0$ the full two--dimensional potential $V(u,v)$ already grows 
quartically in all directions and is therefore classically confining: no 
quantum mechanism is required to restore boundedness. In this case the 
adiabatic potential $V_{\rm eff}(u)$ should be viewed as a local 
semiclassical refinement of the low--lying spectrum near the channels. 
The apparent problem that $V_{\rm ZP}(u)=\tfrac12\sqrt{1-\alpha u^2}$ 
becomes imaginary when $|u|>1/\sqrt{\alpha}$ does not signal a physical 
instability, but rather the breakdown of the harmonic approximation in the 
transverse coordinate $v$. Physically, the full 2D potential remains smooth and 
confining; only the adiabatic approximation ceases to be valid outside 
the harmonic region. Within its range of applicability, the effective one-dimensional
Hamiltonian provides accurate turning points and level spacings that match 
both direct one--dimensional numerics and the full two--dimensional spectrum.
Within its range of validity, quantization follows from
\begin{equation}
  \frac{1}{\pi}\int_{u_-}^{u_+}\!\sqrt{2\,[E-V_{\rm eff}(u)]}\,du
  \;=\; m+\tfrac12,
\end{equation}
providing semiclassical estimates for turning points and level spacings that agree with direct 1D numerics, and serving as useful benchmarks for the 2D spectrum.

For $\alpha<0$, by contrast, the quartic channel term behaves as $-|\alpha|u^4/4$ and dominates at large $|u|$. 
Although the zero--point correction grows as $V_{\rm ZP}(u)\sim \tfrac12\sqrt{|\alpha|}\,|u|$, it cannot overcome the negative quartic tail. 
The Hamiltonian is then unbounded from below, and no true bound states exist. 
In this regime the WKB integral only characterizes \emph{local} quasi--bound motion near the origin (resonances), but cannot describe global confinement.

%%---------------------------------------------------------------------
%%---------------------------------------------------------------------
\subsection{WKB, 1D numerics and 2D levels (Yang--Mills)}
\label{YMComp}

For the Yang-Mills Hamiltonian ${\hat H}_{\mathrm{YM}}$ three data sets are compared
$i$) WKB, $ii$) 1D numerics and $ii)$ 2D levels, 
which are obtained as follows.

\subsubsection{WKB}
In accordance with the above analysis, quantizing the transverse oscillator with
\(\omega_y(x)=\Omega|x|\) yields the linear barrier \(V_{\rm eff}(x)=\frac{\Omega}{2} |x|\) with
\(\Omega=\sqrt{2\alpha}\). Semiclassical WKB quantization gives
\begin{equation}
\label{EWKB}
E^{WKB}_m=\left[\frac{3\pi}{8\sqrt{2}}\;\Omega\;\Big(m+\tfrac12\Big)\right]^{\!2/3},
\end{equation}
where $m=0,1,2,\ldots$, from which the energy behaves as 
\(E^{WKB}_m\propto \alpha^{1/3}\).
Values for the 6 lowest states $m=0,\dots,5$ are presented in the second column 
of Table~\ref{EBKvsE1D}.

\subsubsection{Direct 1D numerics}
To evaluate the semiclassical prediction~(\ref{EWKB}), the Schr\"odinger equation
\begin{equation}
\label{sch1d}
\Bigl[-\frac{1}{2}\,\frac{d^2}{dx^2}+ \frac{\Omega}{2}\,|x|\Bigr]\phi(x)=E\,\phi(x),
\qquad \Omega=\sqrt{2\alpha},
\end{equation}
for the Hamiltonian~(\ref{Hred}), is solved numerically in a domain 
$x\in[-L,L]$ with Dirichlet boundary conditions and a grid of $M$  points
with a separation \(h=2L/(M{+}1)\). A five–point \(O(h^4)\) stencil for 
\(-\tfrac12\,d^2/dx^2\) yields a sparse pentadiagonal matrix with diagonal entries 
\(5/(4h^2)+\frac{\Omega}{2}|x_i|\), first off–diagonals \(-2/(3h^2)\), and second off–diagonals 
\(1/(24h^2)\). The lowest eigenvalues were obtained by
sparse shift–invert (Arnoldi/Lanczos). The resulting spectrum \(E_m^{\mathrm{1D}}\) 
(third column of Table~\ref{EBKvsE1D}), follows the
expected \((m+\tfrac12)^{2/3}\) law and agrees with the WKB results to within a few percent for
the first levels with accuracy improving as \(m\) increases and as \(L,M\) are enlarged.
The results in Table~\ref{EBKvsE1D} correspond to $L=40$ and $M=8000$.

\begin{table}[htbp]
\centering
\caption{\small First six lowest levels for $\alpha=1$. $E_m^{\rm WKB}$
from~(\ref{EWKB}) and $E_m^{\rm 1D}$ are the results obtained by numerically 
solving the Schr\"odinger equation~(\ref{sch1d}) 
($L=40$, $M=8000$ and 5-point stencil). $E^{2D}_{m}$ is the full 2D spectra ($\delta>0.41$). }
%\label{tab:ebk_e1d_alpha1N}
\begin{tabular}{c|ccc|c}
\hline
$m$ & $E_m^{\rm WKB}$ & $E_m^{\rm 1D}$ & rel. error (\%)&$E^{2D}_{m}$\\
\hline
0 & 0.702696 & 0.641803 & 9.488&0.698137\\
1 & 1.461666 & 1.472915 & 0.764&1.498448\\
2 & 2.054695 & 2.046237 & 0.413&2.214279\\
3 & 2.571379 & 2.575247 & 0.150&2.578724\\
4 & 3.040389 & 3.036473 & 0.129&2.994059\\
5 & 3.475595 & 3.477735 & 0.062&3.464140\\
\hline
\end{tabular}
\label{EBKvsE1D}
\end{table}

\subsubsection{Projected 2D subset}
From the complete solution of the Schrodinger equation for the Yang-Mill 
Hamiltonian in two dimensions~(\ref{YMV}), one can project the eigenstates 
onto a thin tube $-y_0\le y\le y_0$ around the escape axis and retain those 
with dominant probability in the tube and compare their energies against the 1D/WKB sets.
For this purpose, we consider the \emph{TubeFraction} $\delta$ which is the fraction of 
the probability density $\lvert \psi(x,y)\rvert^2$ inside a horizontal tube of half-width $y_0$ around the $x$–axis defined as
\begin{equation}
\label{tfeq}
\delta \equiv 
\frac{ \displaystyle \int_{-\infty}^{\infty} \int_{-y_0}^{y_0} \lvert \psi(x,y)\rvert^2 \, dy \, dx }
     { \displaystyle \int_{-\infty}^{\infty} \int_{-\infty}^{\infty} \lvert \psi(x,y)\rvert^2 \, dy\,dx}.
\end{equation}
It measures how much of the wavefunction is concentrated near \(y=0\).
Typical interpretations are: $\delta \approx 1$ (strongly localized near the channel),
$\delta \approx 0.5$ (half of the probability in the tube), and $\delta \approx 0$
(probability concentrated away from the tube).

\begin{figure}[htbp]
\centering

% --------- Row 1 ---------
\begin{subfigure}[b]{0.48\linewidth}\centering
\includegraphics[width=\linewidth]{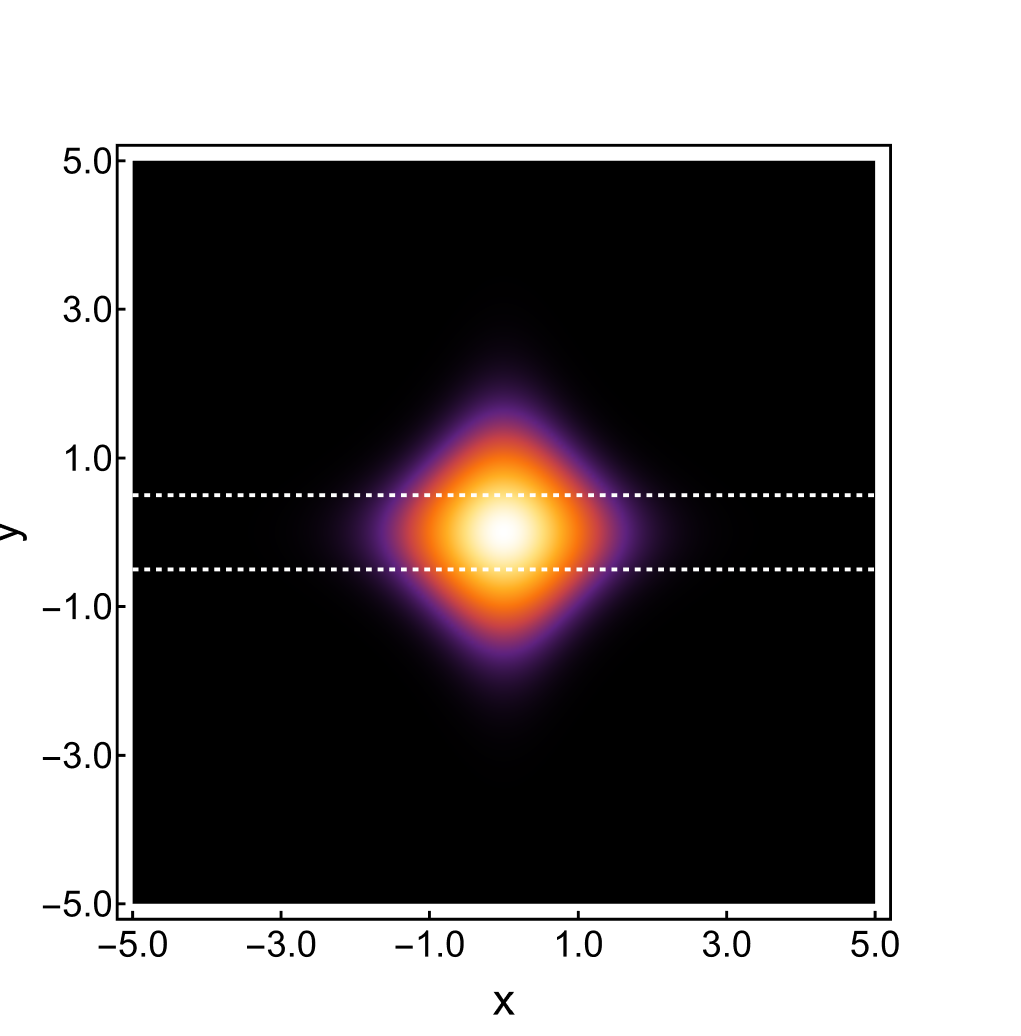}
\caption{$E_0=0.698137$, $\delta=0.5$}
\end{subfigure}\hfill
\begin{subfigure}[b]{0.48\linewidth}\centering
\includegraphics[width=\linewidth]{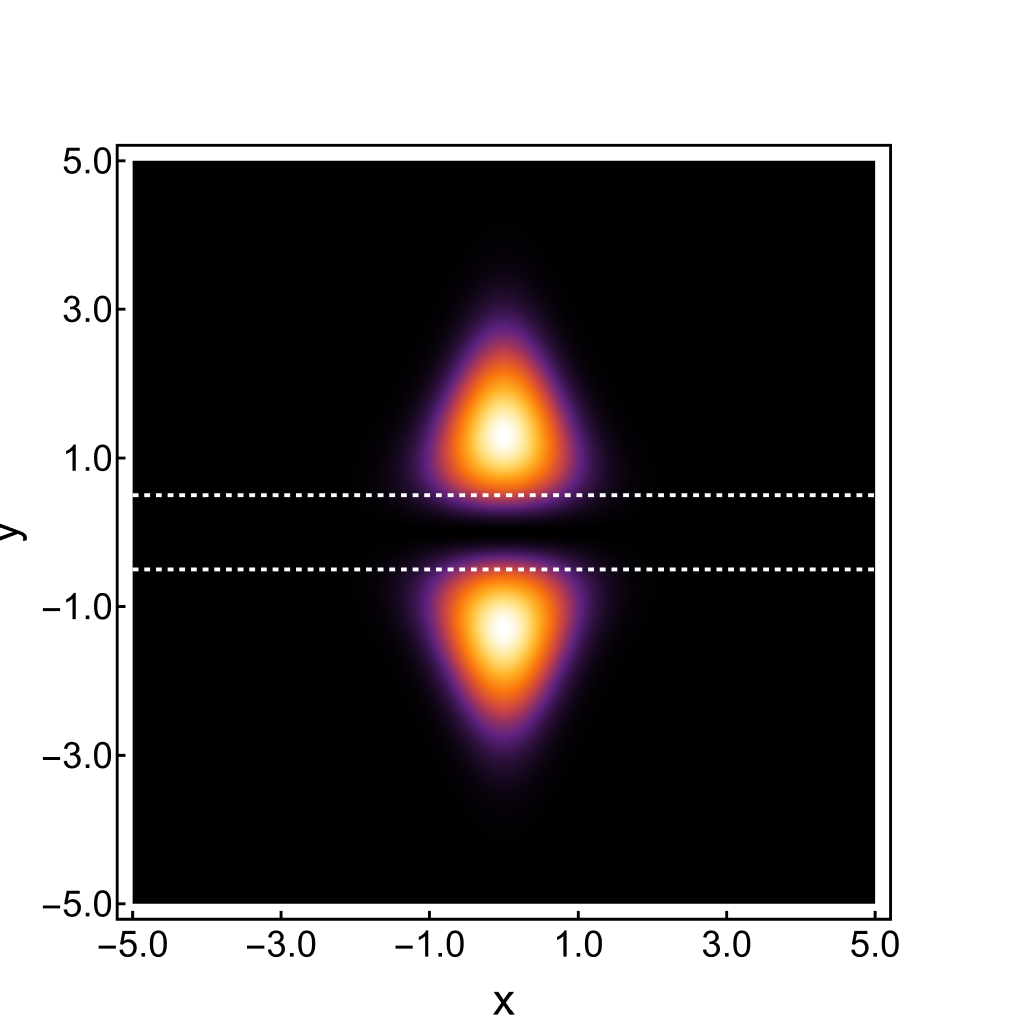}
\caption{$E_1=1.498448$, $\delta=0.05$}
\end{subfigure}
\vspace{0.8ex}
% --------- Row 2 ---------
\begin{subfigure}[b]{0.48\linewidth}\centering
\includegraphics[width=\linewidth]{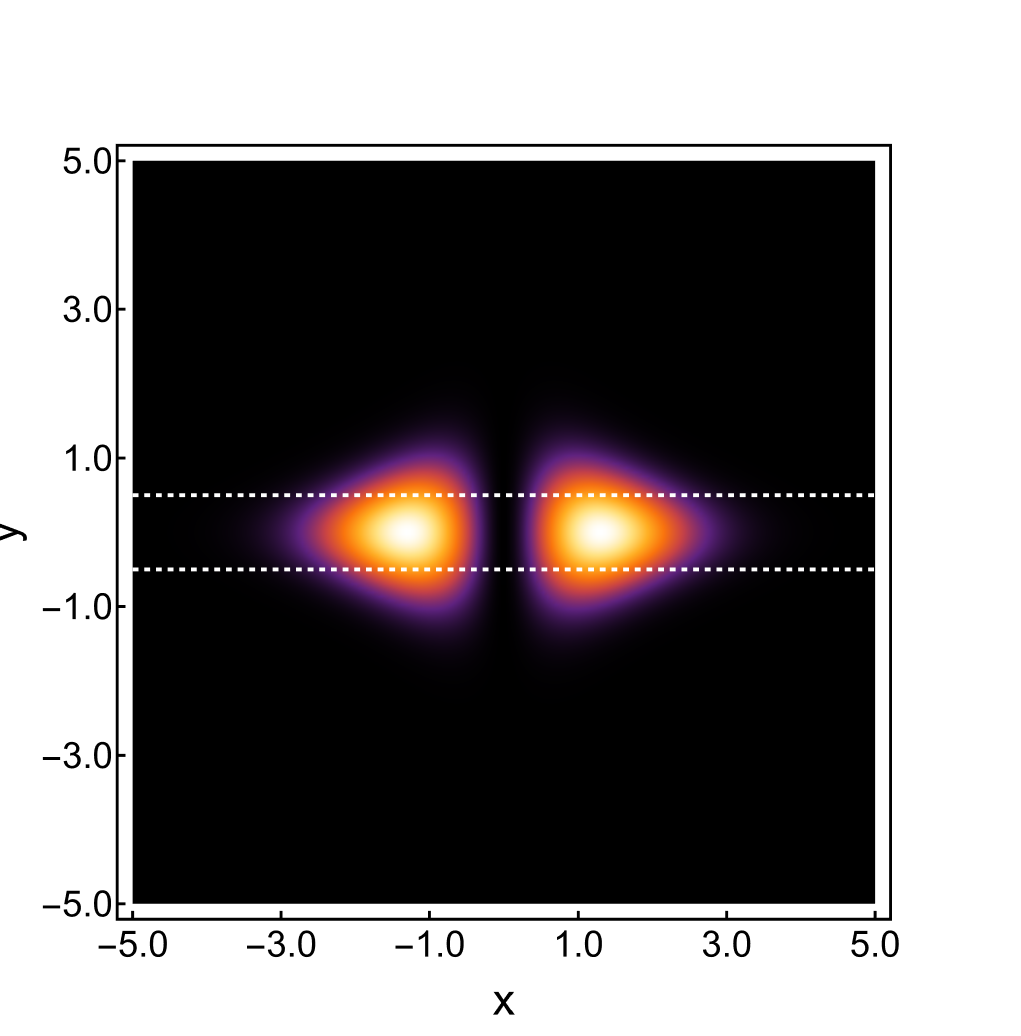}
\caption{$E_1=1.498448$, $\delta=0.68$}
\end{subfigure}\hfill
\begin{subfigure}[b]{0.48\linewidth}\centering
\includegraphics[width=\linewidth]{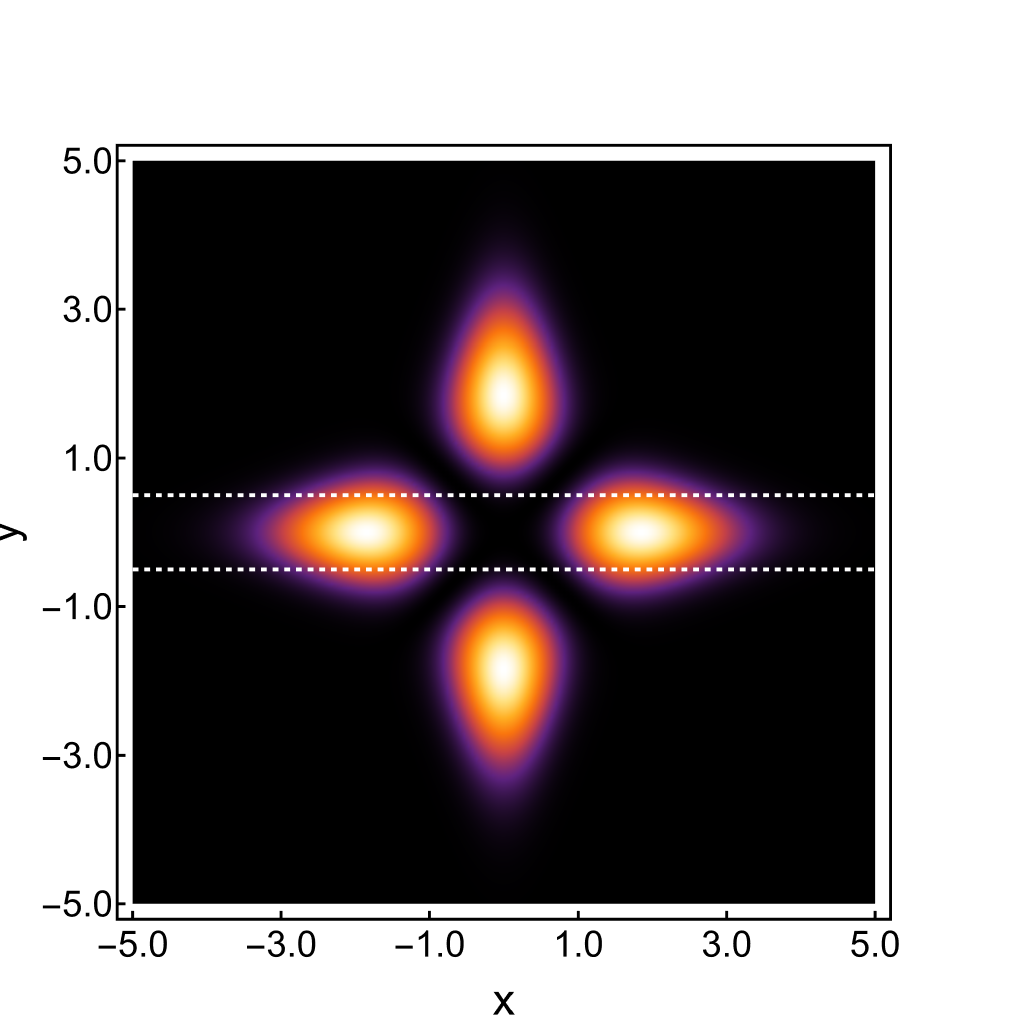}
\caption{$E_2=1.925210$, $\delta=0.4$}
\end{subfigure}
\vspace{1.2ex}
% --------- Shared colorbar ---------
\includegraphics[width=0.45\linewidth]{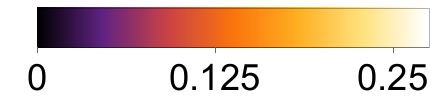}
\caption{Probability densities $|\psi(x,y)|^2$ for the lowest 
eigenstates $n=0,1,2$ of the Yang--Mills Hamiltonian
for $\alpha=1$ and $y_0=0.5$. Each panel is labeled by the energy $E_n$ and tube
fraction $\delta$~(\ref{tfeq}). Localization along symmetry--related channels
reflects the $D_4$ symmetry and agrees with the 1D semiclassical
reduction, with energies consistent from Table~\ref{EBKvsE1D}.
The shared colorbar indicates the normalized density scale.}
\label{fig:density-grid-I}
\end{figure}

\begin{figure}[htbp]
\centering
% --------- Row 3 ---------
\begin{subfigure}[b]{0.48\linewidth}\centering
\includegraphics[width=\linewidth]{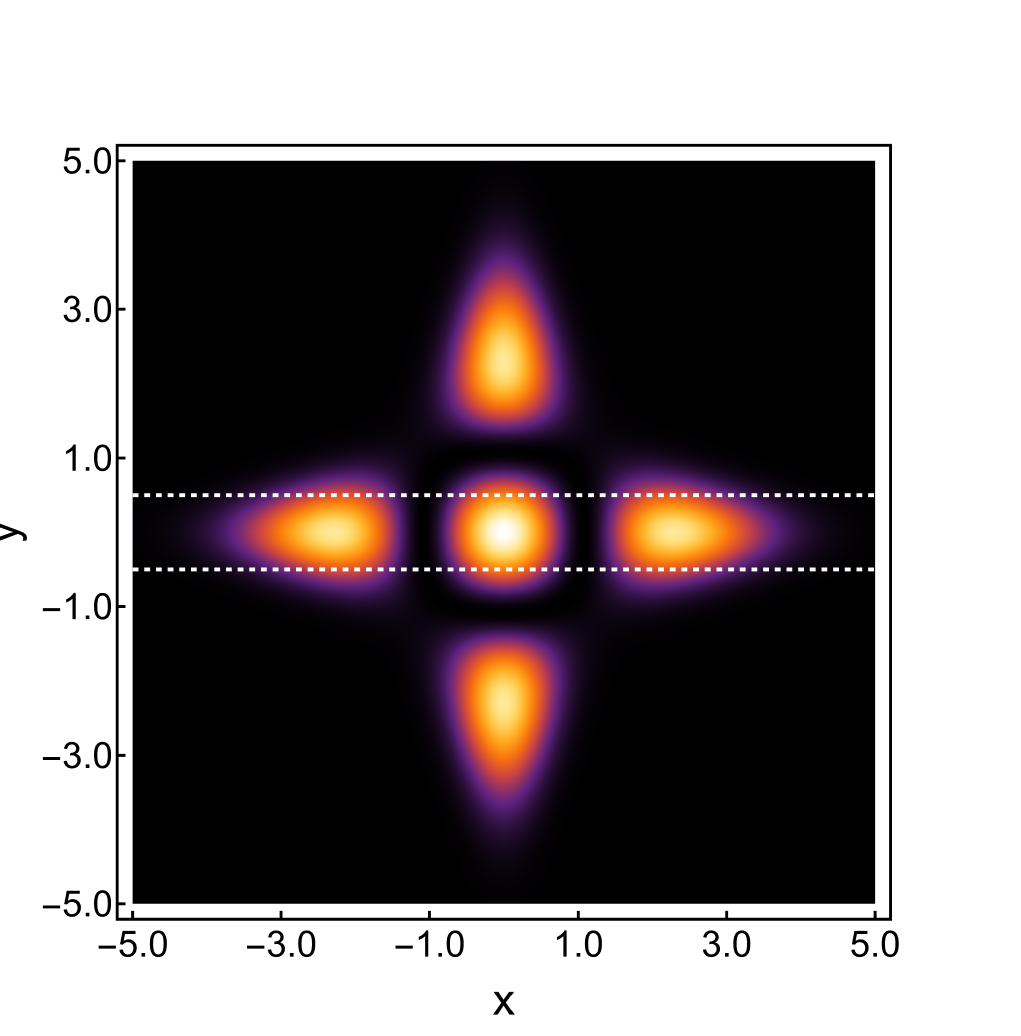}
\caption{$E_3=2.214279$, $\delta=0.47$}
\end{subfigure}\hfill
\begin{subfigure}[b]{0.48\linewidth}\centering
\includegraphics[width=\linewidth]{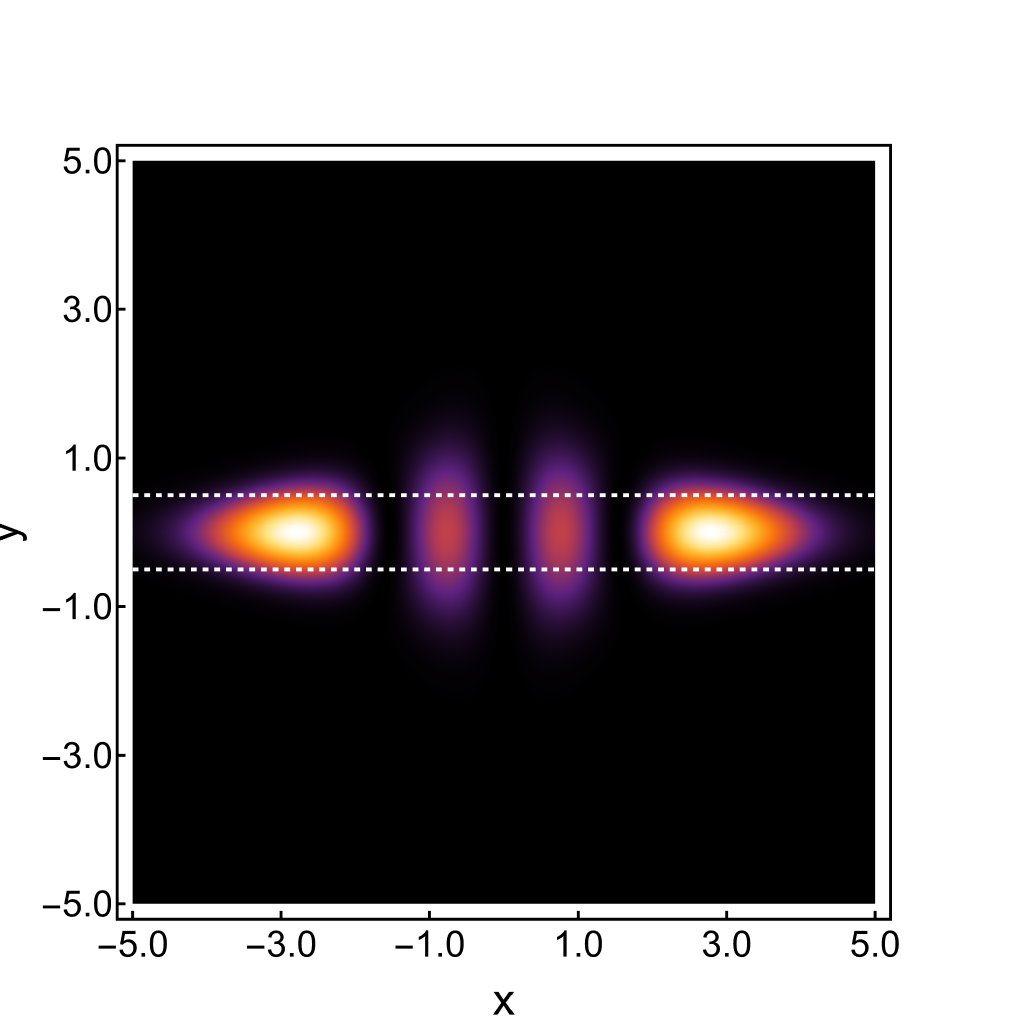}
\caption{$E_4=2.578724$, $\delta=0.77$}
\end{subfigure}
\vspace{0.8ex}
% --------- Row 4 ---------
\begin{subfigure}[b]{0.48\linewidth}\centering
\includegraphics[width=\linewidth]{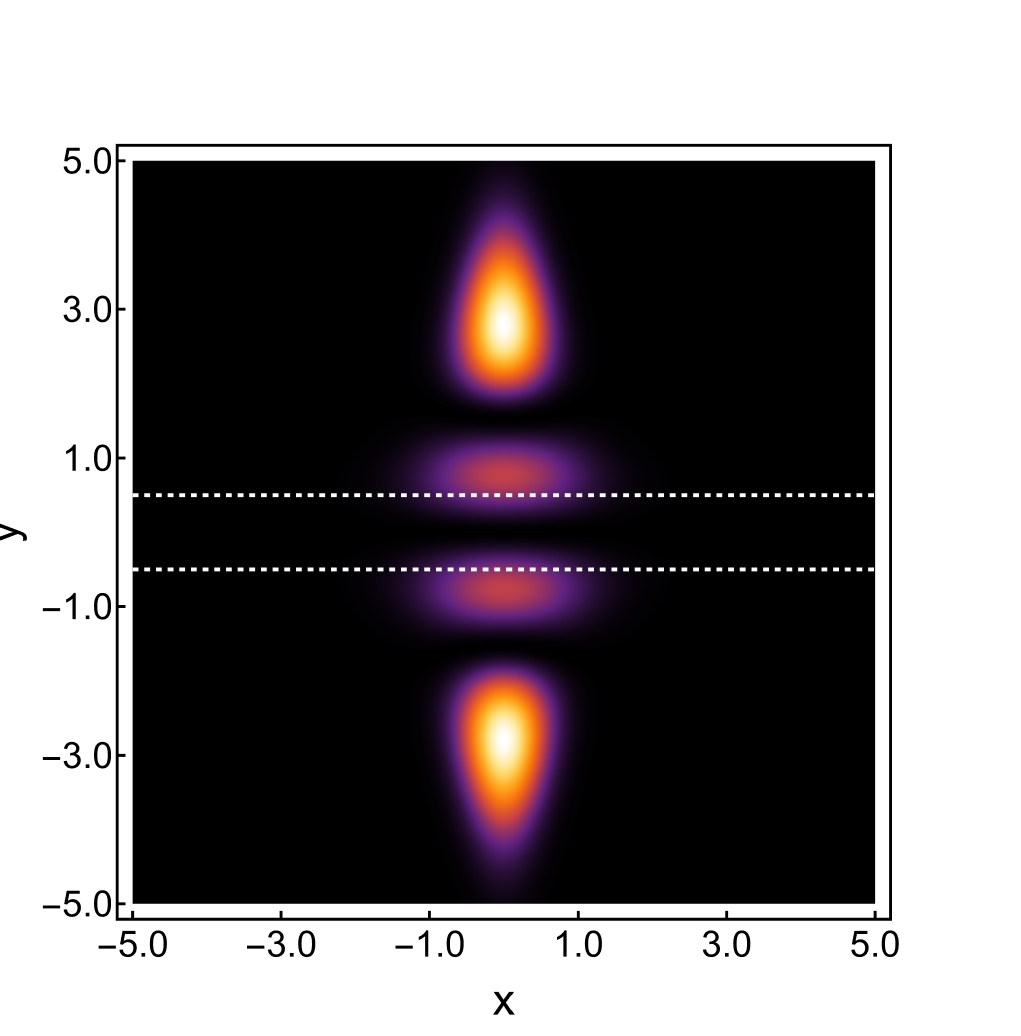}
\caption{$E_4=2.578724$, $\delta=0.05$}
\end{subfigure}\hfill
\begin{subfigure}[b]{0.48\linewidth}\centering
\includegraphics[width=\linewidth]{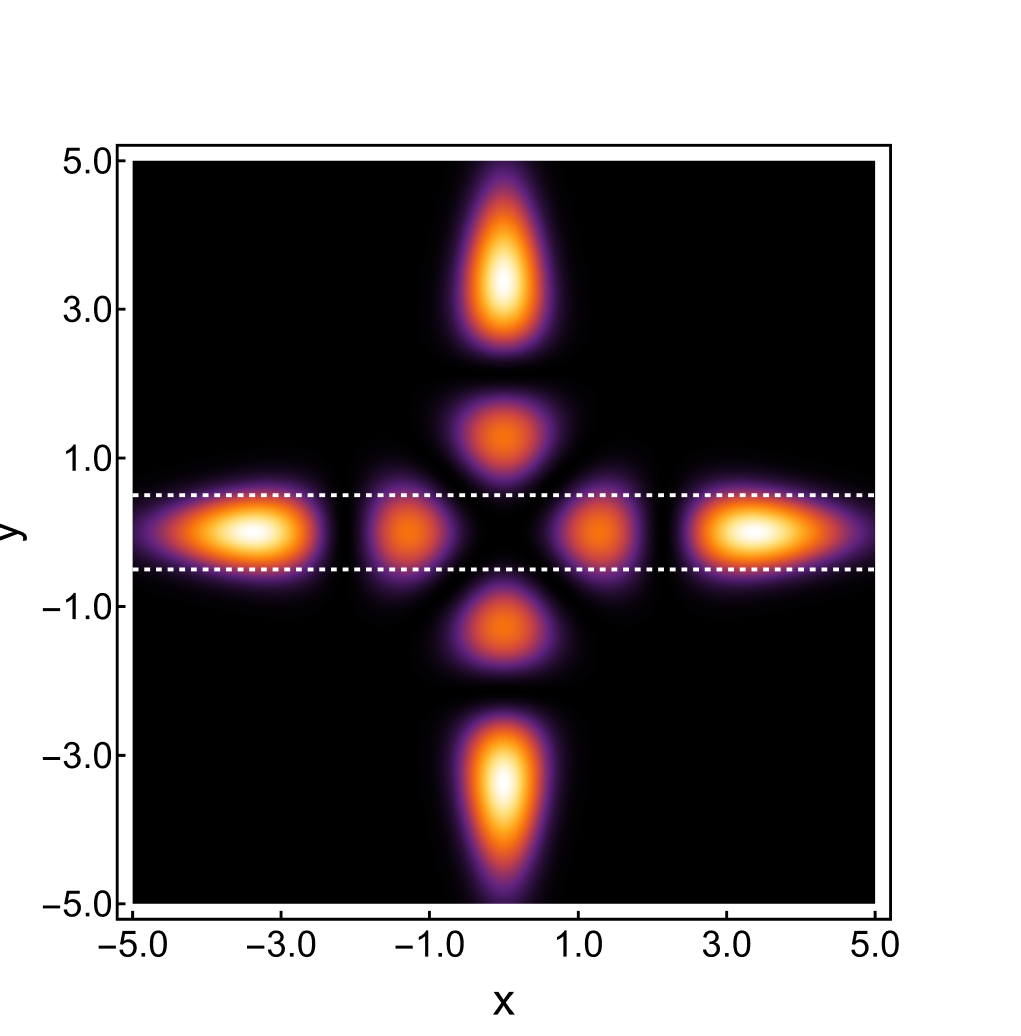}
\caption{$E_5=2.994059$, $\delta=0.43$}
\end{subfigure}
\vspace{1.2ex}
% --------- Shared colorbar ---------
\includegraphics[width=0.45\linewidth]{BARPSI.pdf}
\caption{Probability densities $|\psi(x,y)|^2$ for the excited Yang--Mills 
eigenstates $n=3,4,5$ for $\alpha=1$ and $y_0=0.5$. Energies $E_n$ and tube
fractions $\delta$~(\ref{tfeq}) are indicated. The complementary localization of
degenerate partners illustrates the $D_4$ symmetry. The patterns agree
with the 1D semiclassical reduction and the numerical
comparison in Table~\ref{EBKvsE1D}. The shared colorbar indicates the normalized
density scale.}
\label{fig:density-grid-II}
\end{figure}

To test the channel picture, we compare the closed-form WKB levels~(\ref{EWKB}) for the 
effective linear barrier $V_{\rm eff}(x)=s\,|x|$  against a subset of the exact 
2D eigenvalues selected by a large \emph{TubeFraction} $\delta$.
The tube–selected 2D levels (Figure~\ref{fig:density-grid-I} 
and~\ref{fig:density-grid-II}) are in agreement  with 
the WKB results $E_{m}^{\rm WKB}$ already for the lowest few states and follow
the predicted scaling $\propto \alpha^{1/3}(m+\tfrac12)^{2/3}$ 
(see Table~\ref{EBKvsE1D}). Residual deviations correlate
with smaller $\delta$, reflecting transverse leakage and nonadiabatic mixing. This agreement
confirms that the linear zero–point barrier captured by WKB governs the low-lying, channel–dominated
part of the 2D spectrum.

%%----
Across the sampled \(\alpha\), the lowest \(\sim\!6\) channel levels agree at the
few–percent level between WKB and 1D numerics, and the tube–projected 2D
levels fall on the same trend within visible error bars. The observed
\(E^{WKB}_m\!\propto\!\alpha^{1/3}\) behavior is consistent with Airy quantization in the
linear \(V_{\rm eff}\). This validates a simple mechanism for \emph{quantum
restoration of confinement}: transverse zero–point motion erects a linear barrier
along classical escape directions, rendering the spectrum discrete without recourse
to global spectral–statistics diagnostics.
%%----

% ==============================
% First figure (subfigures 1--4)
% ==============================

% ===============================
% Second figure (subfigures 5--8)
% ===============================

A characteristic feature of the quantum Yang--Mills Hamiltonian is
the systematic presence of degeneracies throughout the spectrum.
These arise from the quartic potential, which is symmetric under
the exchange $x\leftrightarrow y$ and under independent sign
reversals of each coordinate. As a consequence, eigenstates can be
organized into irreducible representations of the $D_4$ symmetry
group. This implies that spectral degeneracies are at most twofold:
non--degenerate states correspond to one--dimensional irreps, while
pairs of degenerate states span the two--dimensional irrep. In
practice, such identical eigenvalues are observed across the
entire spectrum, reflecting the fact that the corresponding wave
functions are localized along symmetry--related escape channels.
This connection can be quantified using the tube fraction
diagnostic: degenerate partners typically display complementary
values of $\delta$, with each state concentrating probability
density in a distinct but symmetry--equivalent channel. Although the
visibility of the degeneracies may diminish as states become more
extended at higher energies, their origin lies in the exact
discrete symmetries of the Hamiltonian, and they remain present
across the full spectrum.
%%---------------------------------------------------------------------
%%---------------------------------------------------------------------

\subsection{Toward a mobility edge in the Contopoulos system}
\label{sec:mobility-edge}

A semiclassical picture of the Contopoulos Hamiltonian can be obtained
by expanding around the $1{:}1$ resonance. Introducing oscillator
actions and angles, the quartic perturbation generates a term
$\cos 2(\theta_1-\theta_2)$, resonant when $I_1\!\approx\!I_2$.
Averaging over the fast phase reduces the dynamics to a pendulum
Hamiltonian
\begin{equation}
H_{\rm eff}(K,\psi;J)\simeq \tfrac{1}{2M}K^2 - V_0\cos(2\psi),
\end{equation}
with $M=2/\alpha$ and $V_0=\alpha J^2/8$
\cite{arnold2006mathematical,korsch2008chaos}.
The separatrix area compared with the accessible $(K,\psi)$ domain
suggests that a finite portion of phase space becomes chaotic near the
resonance. This provides a simple estimate for the fraction of chaotic
motion, though its generality beyond the normal–form approximation
remains an open question.

According to Berry–Robnik theory \cite{berry1984semiclassical},
mixed spectra decompose into independent Poisson and GOE subsequences
with weights equal to the corresponding classical phase–space
measures. Approximating the adjacent–gap ratio as
\cite{OganesyanHuse2007,Atas2013}
\begin{equation}
\langle r\rangle(E)\approx Q(E)\,\langle r\rangle_{\rm P}
+\big[1-Q(E)\big]\langle r\rangle_{\rm GOE},
\end{equation}
with $\langle r\rangle_{\rm P}=0.386$ and
$\langle r\rangle_{\rm GOE}=0.536$, one expects intermediate
statistics in agreement with numerical results.

Finally, scaling $x,y,p_x,p_y \mapsto \sqrt{E}(x',y',p_x',p_y')$
shows that the dynamics depends only on the parameter
$\kappa=\alpha E$. The onset of widespread chaos occurs at
$\kappa\sim\kappa_0$, implying a threshold
\[
E_{\rm edge}\sim \kappa_0/\alpha ,
\]
which we interpret as a classical precursor of a possible quantum
\emph{mobility edge}: below $E_{\rm edge}$, eigenstates may remain
largely localized, while above it they acquire extended, chaotic
character. Whether such a threshold survives beyond the present normal-form approximation is an open question.

\medskip
\noindent
This normal–form analysis, together with Berry–Robnik theory,
provides a parsimonious semiclassical motivation for exploring the
existence of a mobility edge in the Contopoulos system. A complete
quantum study based on
level–spacing statistics remains for future work.

\section{Conclusions}

We have investigated the interplay between classical chaos and quantum confinement in quartic Hamiltonians, focusing on the Contopoulos system and the purely quartic Yang-Mills Hamiltonian. 
Our results reveal a striking dichotomy: while classical trajectories destabilize and escape through well-defined channels, quantum mechanics restores confinement and enforces a discrete spectrum.

On the classical side, we employed Poincar\'e sections, Lyapunov exponents, and symmetry-line constructions to identify escape routes and characterize the chaotic regimes. 
In the Contopoulos Hamiltonian, escape occurs along the diagonals $x=\pm y$ for negative quartic coupling, whereas in the Yang-Mills Hamiltonian with $\alpha>0$ escape channels appear along the coordinate axes. 
These analyses underline the sensitivity of chaotic dynamics to symmetry and coupling sign.

On the quantum side, we computed high-precision spectra using variational and Lagrange-mesh methods. 
All eigenstates of the pure Yang-Mills Hamiltonian remain normalizable and the spectrum discrete, in agreement with rigorous results~\cite{Simon1983AnnPhys}. 
In the Contopoulos case, positive quartic coupling partially lifts oscillator degeneracies and separates the spectrum into localized and chaotic regimes. 
Quantum fluctuations, therefore, act as an effective restoring force, confining motion even in directions that are classically unstable.

Finally, we validated the classical dynamics through analog electronic simulations with operational-amplifier circuits. 
These experiments provide a tangible bridge between theoretical predictions and laboratory realizations, demonstrating that confinement beyond instability can be explored in accessible platforms.

\medskip
In summary, our study contributes three main insights:
\begin{itemize}
    \item Quantum fluctuations universally restore confinement in quartic Hamiltonians, even when classical dynamics is unstable and chaotic. The Contopoulos model also serves as an ideal candidate for investigating the quantum fluctuations near the classical instanton configurations \cite{ PhysRevD.92.025046, PhysRevD.92.025047} as well as the classical-quantum correspondence in chaotic systems.  
    \item Classical instability mechanisms differ qualitatively between the Contopoulos and Yang-Mills Hamiltonians, revealing how symmetry and coupling sign dictate escape channels. 
    \item Analog electronic simulations offer a novel experimental-inspired validation of classical chaos, suggesting broader applications of circuit models to nonlinear and quantum-inspired dynamics. 
\end{itemize}

These results establish a coherent picture of confinement beyond instability and highlight the $x^{2}y^{2}$ Hamiltonian as a fertile model for bridging classical chaos, quantum confinement, and experimental analogs. 
They also open avenues for further studies on semiclassical quantization, spectral statistics, and matrix-model analogues relevant to high-energy physics.

\subsection*{Declaration of competing interest}

The authors declare that they have no known competing financial interests or personal relationships that could have appeared to influence the work reported in this paper.

\subsection*{Data availability}

No data was used for the research described in the article.

\section*{Acknowledgments}
A.M.E.R. and M.A.Q.J. would like to thank the support from Consejo Nacional de Humanidades, Ciencias y Tecnologías (CONAHCyT, now SECIHTI) of Mexico under Grant CF-2023-I-1496. A.M.E.R. also thanks the support from UAM research grant DAI 2024-CPIR-0. M.A.Q.-J. thankfully acknowledges financial support by DGAPA-UNAM under the Project UNAM-PAPIIT IA103325.  

\nocite{*}
\bibliographystyle{apsrev4-2}
\bibliography{Manuscript} 

%-----------------------------------------------------------------------------
%-----------------------------------------------------------------------------
\appendix

\section{Extended results for the Contopoulos Hamiltonian}
\label{appx:AA}

This appendix collects the extended catalogue of trajectories, Poincar\'e sections, and periodic orbits 
that complement the representative examples shown in Sec.~II. 
The material presented here is not required for the main discussion but provides 
a systematic survey of parameter regimes and initial conditions, 
allowing detailed comparison and reproducibility of the results.  

\begin{figure*}[h!]
    \centering    
    \begin{subfigure}[h!]{0.47\textwidth}
        \centering
        \includegraphics[width=\textwidth]{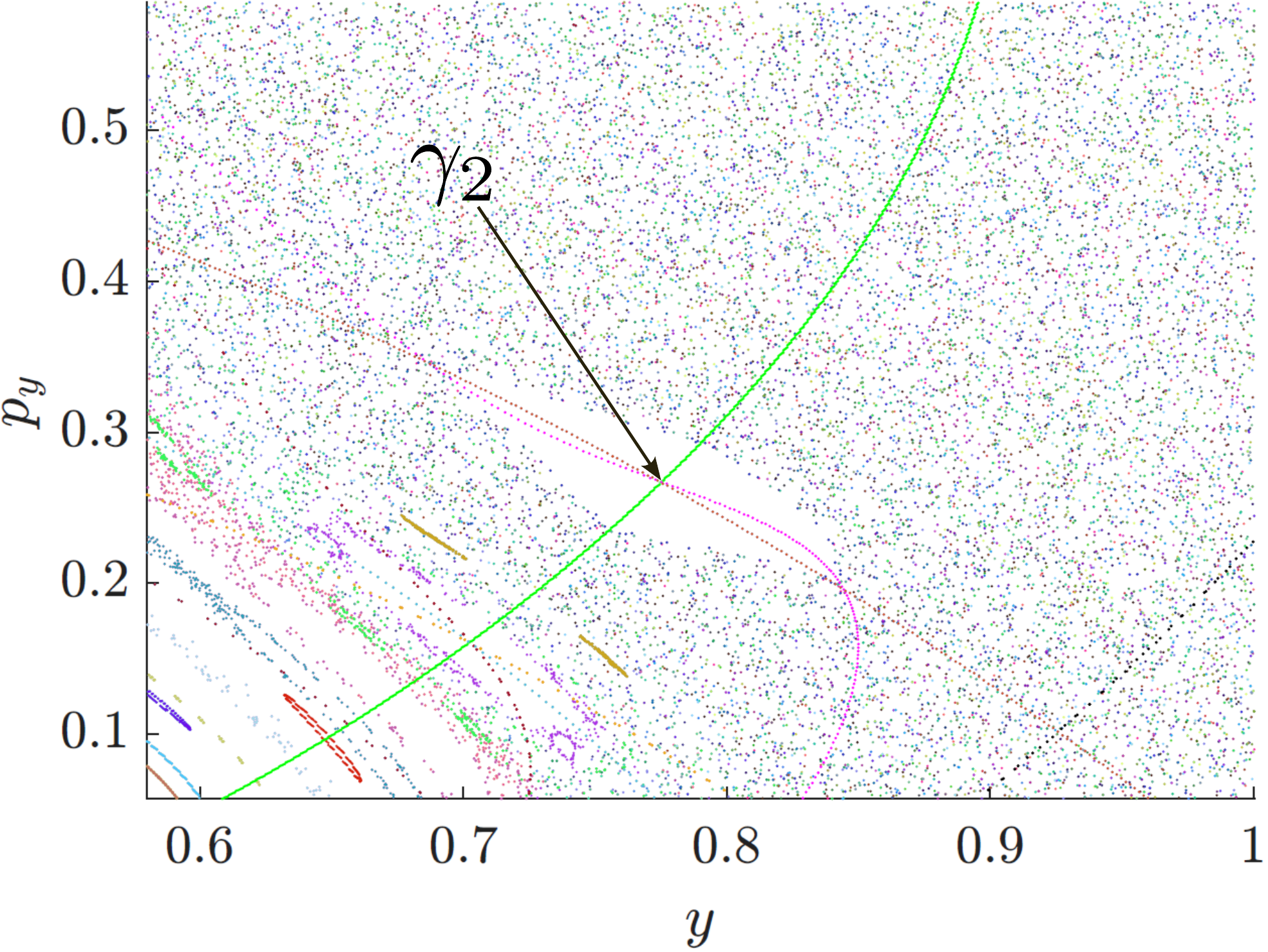} 
        \caption{}
    \end{subfigure}
    \quad
    \begin{subfigure}[h!]{0.47\textwidth}
        \centering
        \includegraphics[width=\textwidth]{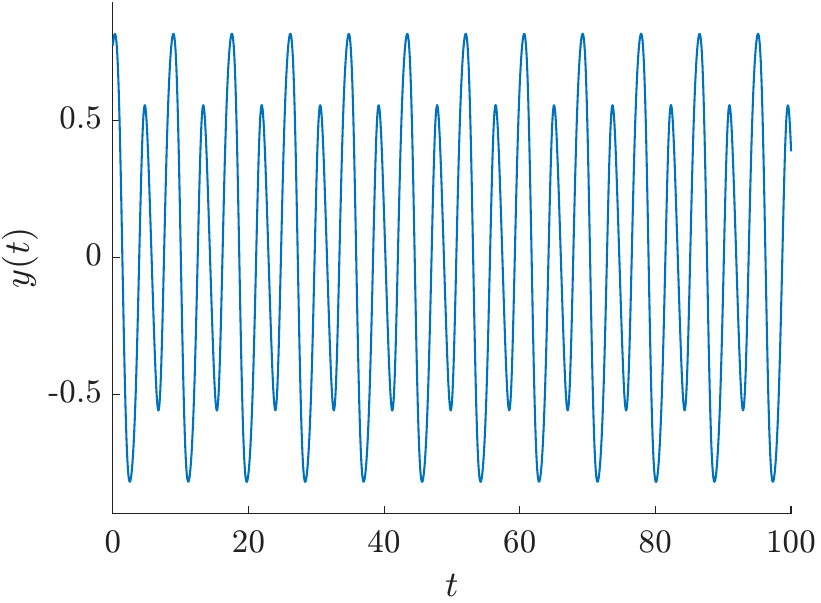}
        \caption{}
    \end{subfigure}
    \quad
    \begin{subfigure}[h!]{0.47\textwidth}
        \centering
        \includegraphics[width=\textwidth]{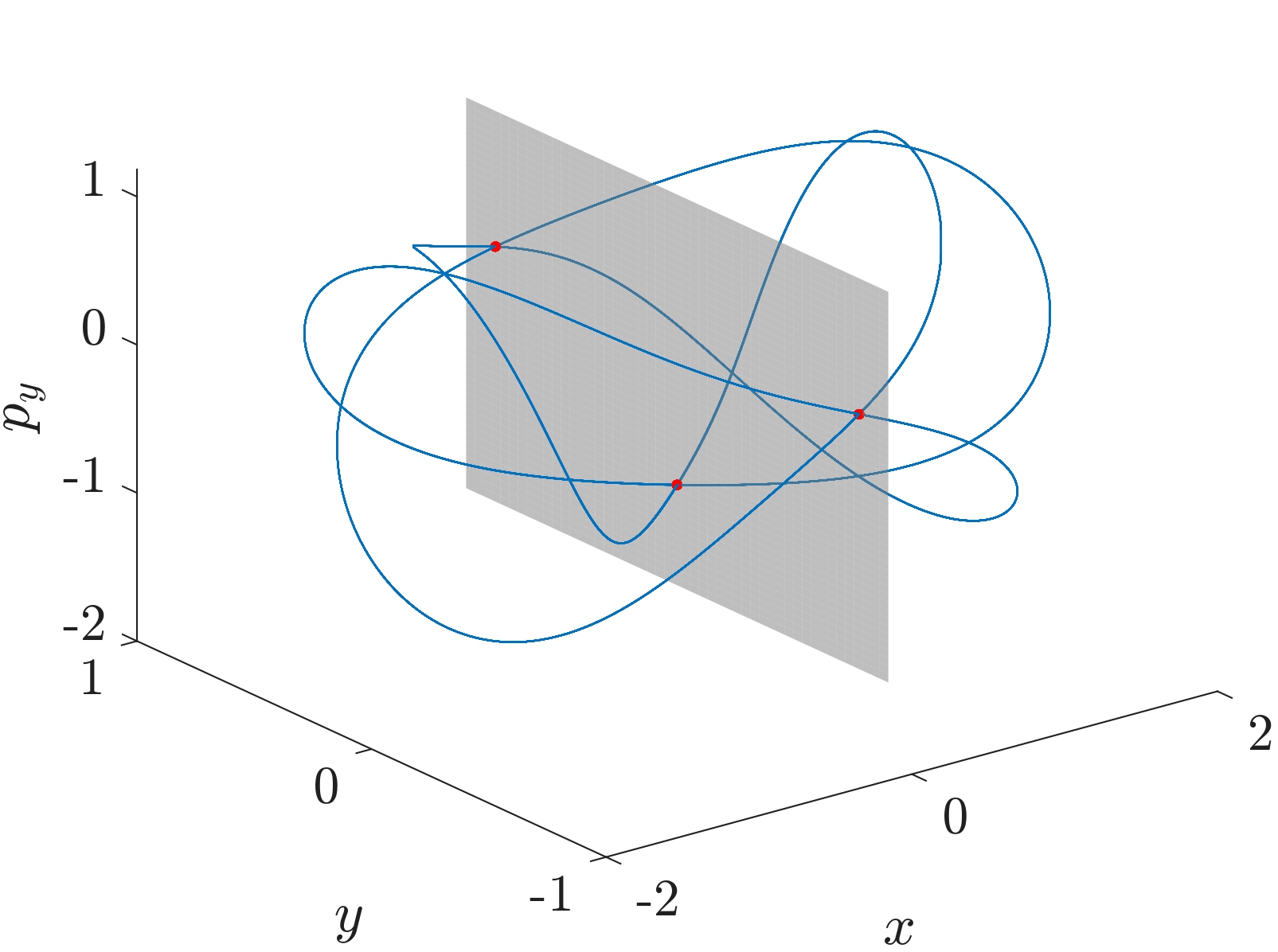}
        \caption{}
    \end{subfigure}
    \quad
    \begin{subfigure}[h!]{0.47\textwidth}
        \centering
        \includegraphics[width=\textwidth]{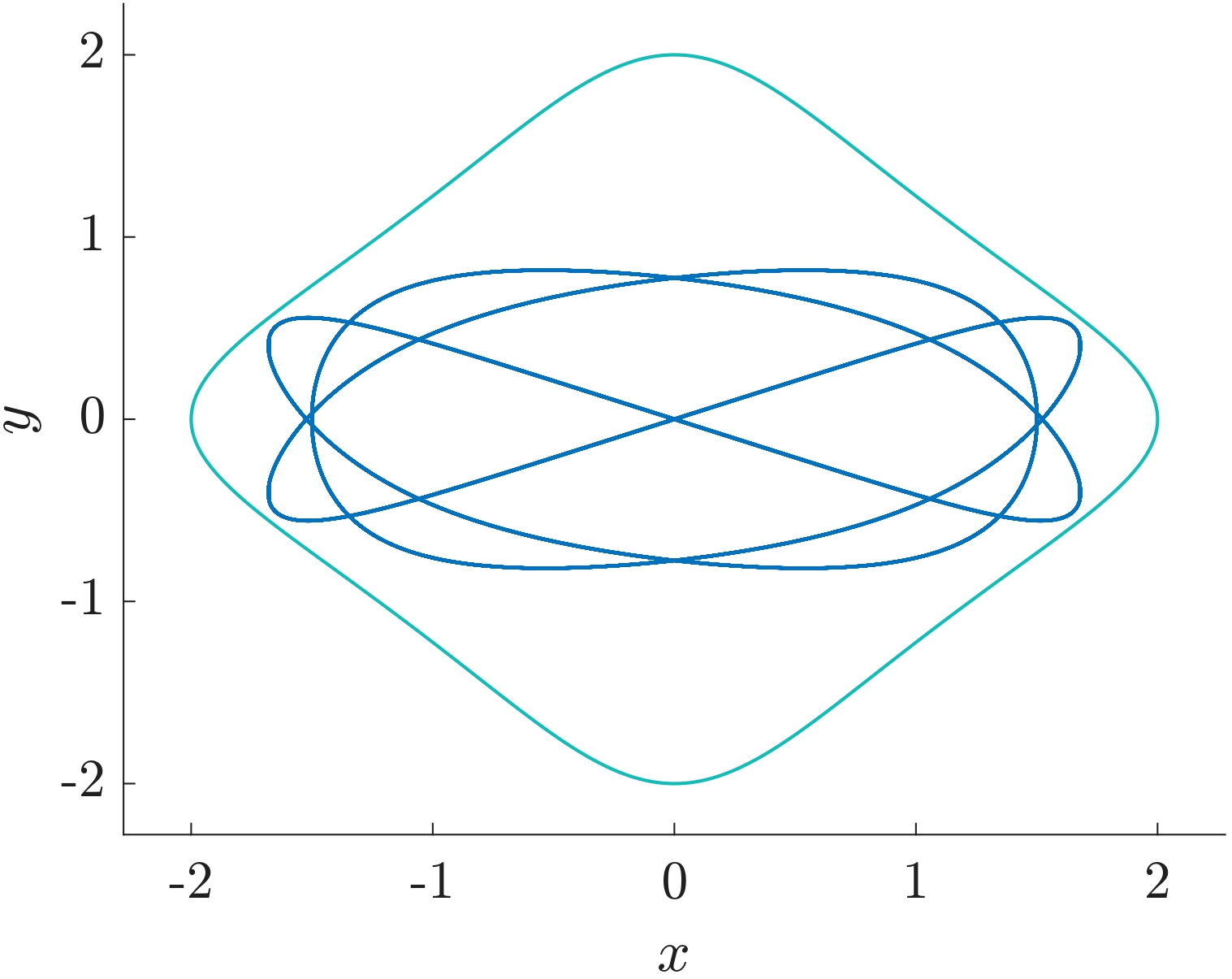}
        \caption{}
    \end{subfigure}         
    \caption{(Top) Magnification of the phase space \textbf{(a)} where a periodic orbit $\gamma_2$, with values $(\alpha=1/2,E=2)$ was found through the intersection of symmetry lines \cref{mapa_rojo}. The corresponding initial conditions are given in \cref{II}. In \textbf{(b)} the time series $y(t)$, the orbit in phase space as the Poincaré section $x=0$ (gray plane) \textbf{(c)}, and finally in the level curve on the plane $(x,y)$ (configuration space) \textbf{(d)} are shown.}
    \label{p2}
\end{figure*}

\begin{figure*}[h!]
    \centering    
    \begin{subfigure}[h!]{0.47\textwidth}
        \centering
        \includegraphics[width=\textwidth]{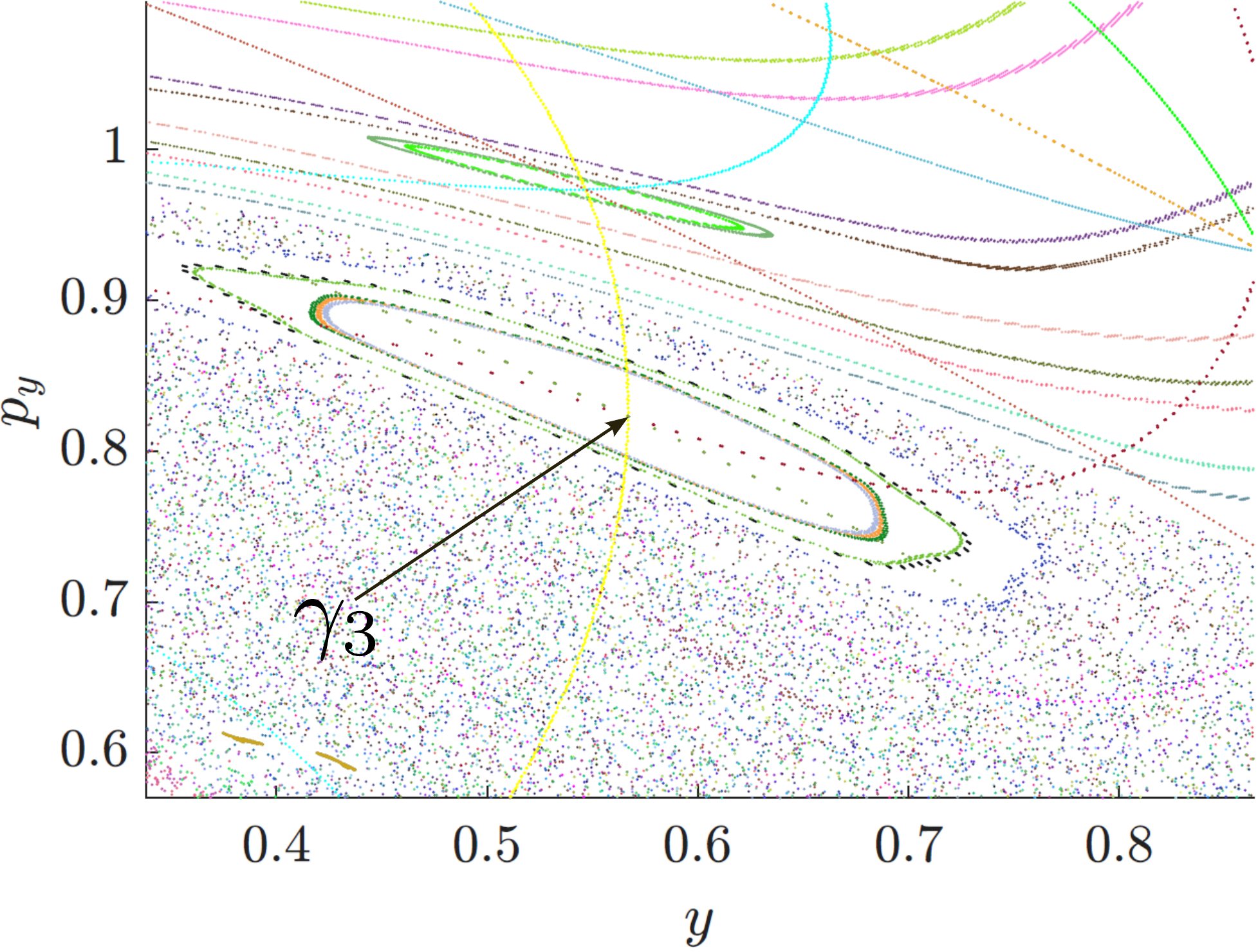} 
        \caption{}
    \end{subfigure}
    \quad
    \begin{subfigure}[h!]{0.47\textwidth}
        \centering
        \includegraphics[width=\textwidth]{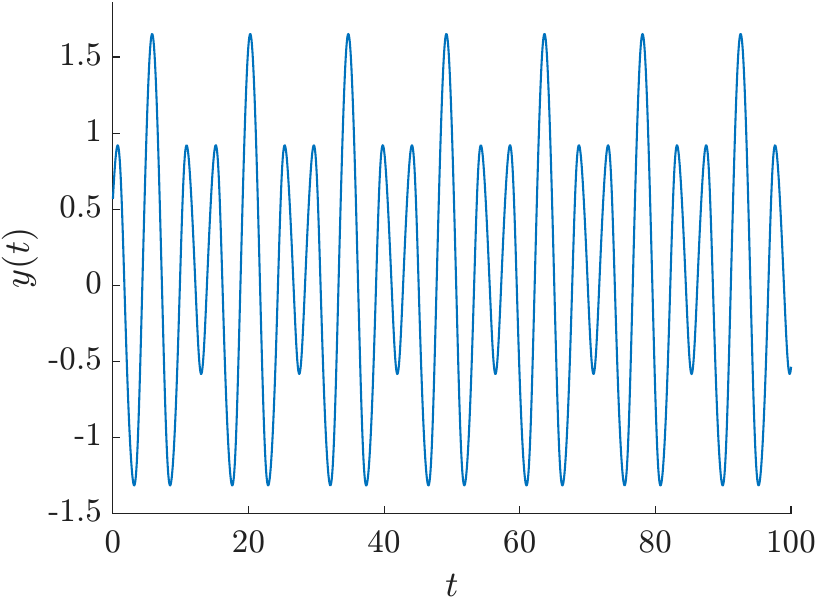}
        \caption{}
    \end{subfigure}
    \quad
    \begin{subfigure}[h!]{0.47\textwidth}
        \centering
        \includegraphics[width=\textwidth]{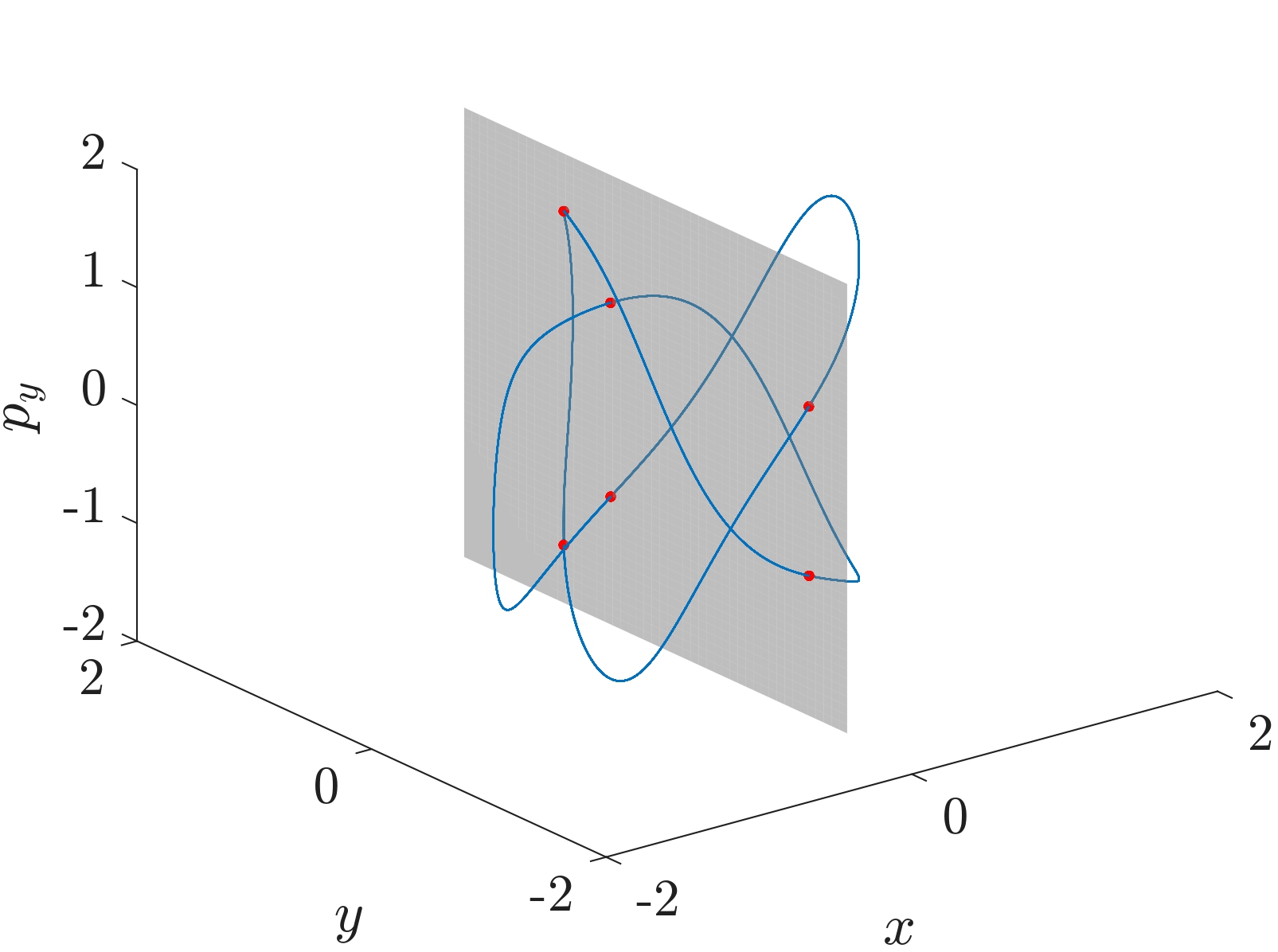}
        \caption{}
    \end{subfigure}
    \quad
    \begin{subfigure}[h!]{0.47\textwidth}
        \centering
        \includegraphics[width=\textwidth]{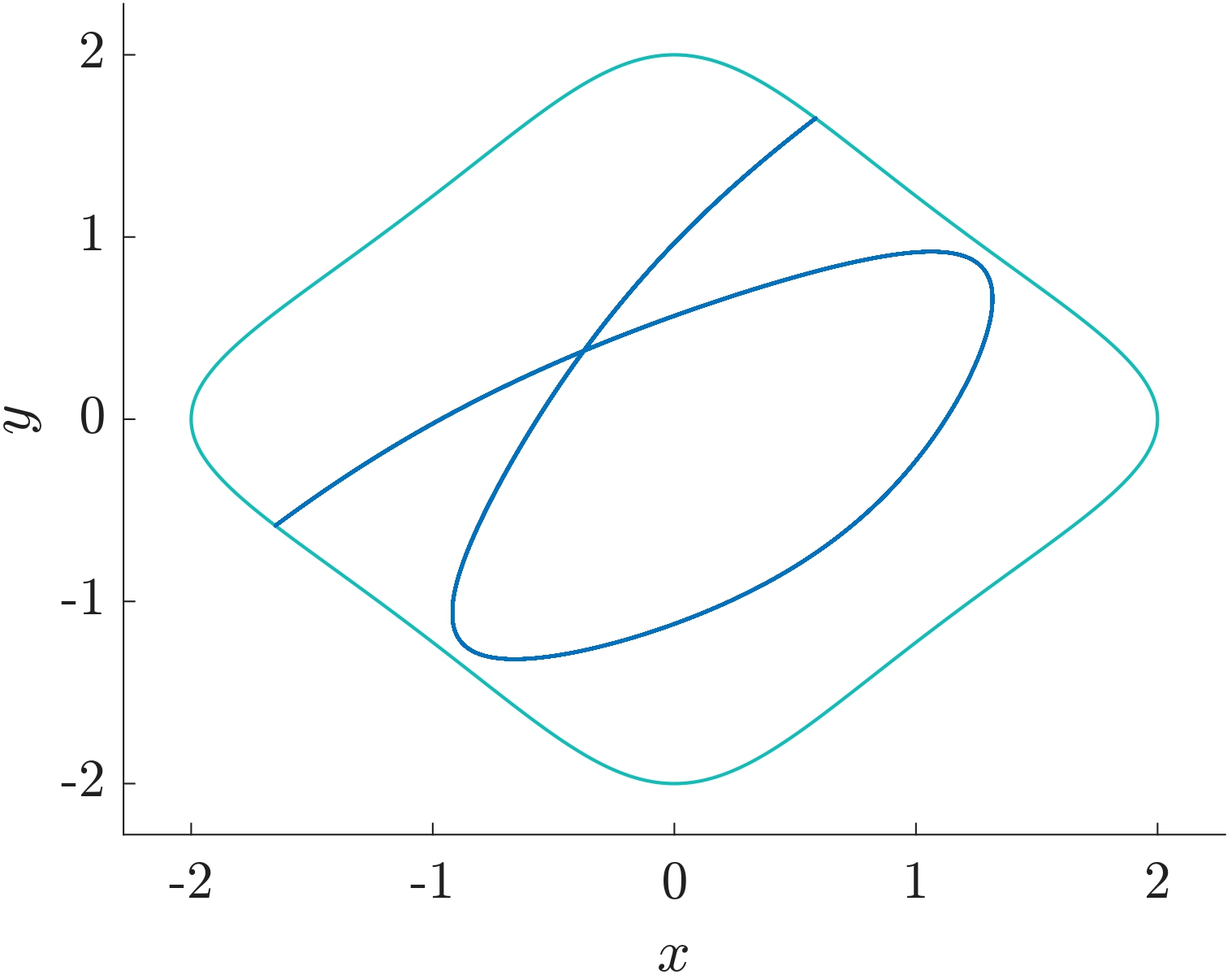}
        \caption{}
    \end{subfigure} 
    \caption{(Top) Magnification of the phase space \textbf{(a)} where a periodic orbit $\gamma_3$, with values $(\alpha=1/2,E=2)$ was found through the intersection of symmetry lines \cref{mapa_rojo}. The corresponding initial conditions are given in \cref{II}. In \textbf{(b)} the time series $y(t)$, the orbit in phase space as the Poincaré section $x=0$ (gray plane) \textbf{(c)}, and finally in the level curve on the plane $(x,y)$ (configuration space) \textbf{(d)} are shown.}
    \label{p3}
\end{figure*}

\begin{figure*}[h!]
    \centering    
    \begin{subfigure}[h!]{0.47\textwidth}
        \centering
        \includegraphics[width=\textwidth]{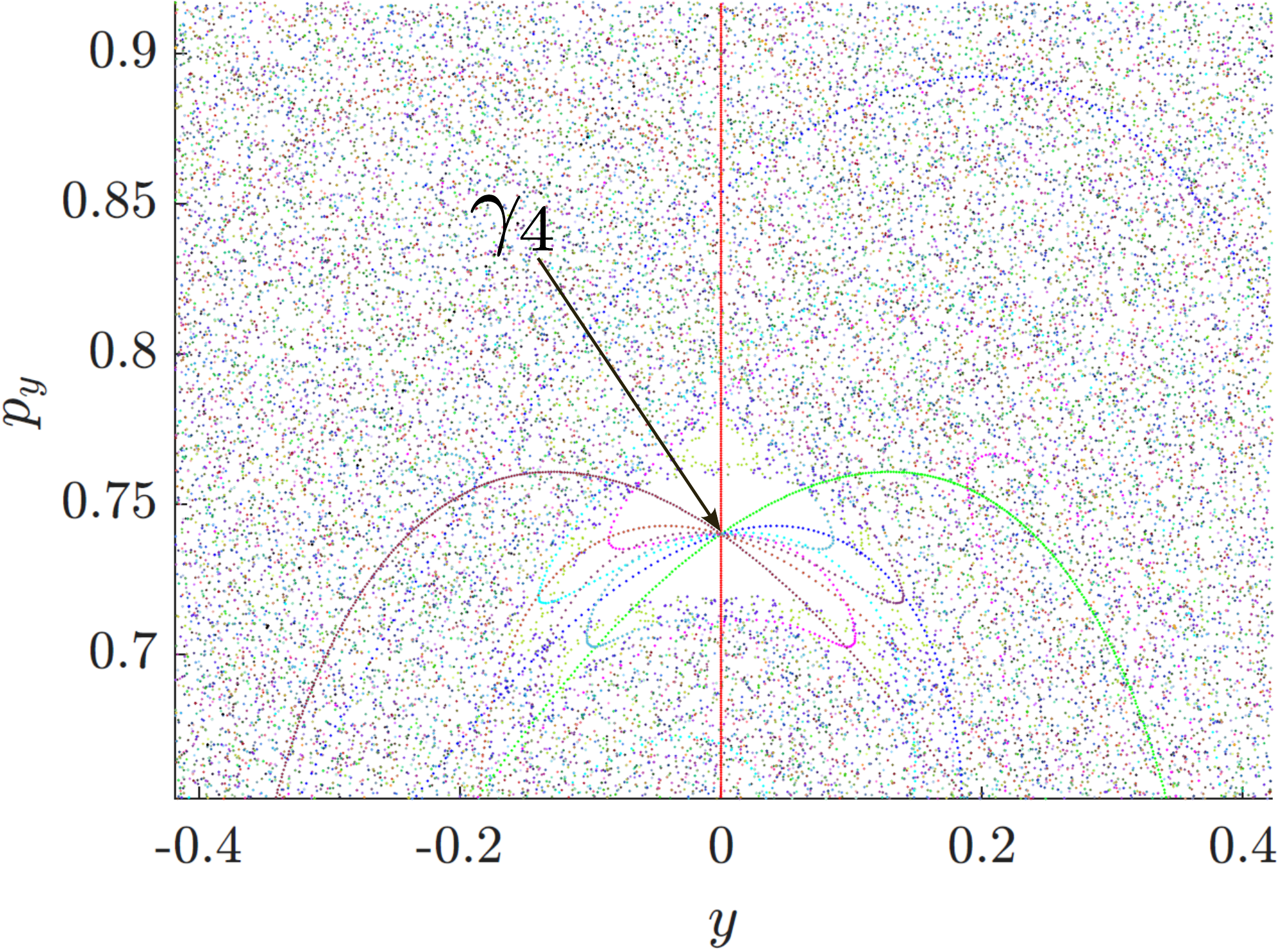} 
        \caption{}
    \end{subfigure}
    \quad
    \begin{subfigure}[h!]{0.47\textwidth}
        \centering
        \includegraphics[width=\textwidth]{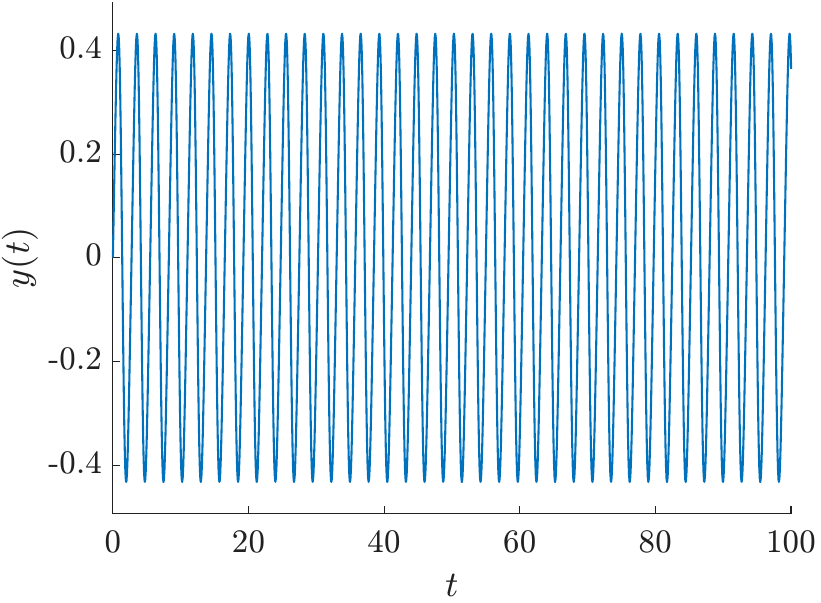}
        \caption{}
    \end{subfigure}
    \quad
    \begin{subfigure}[h!]{0.47\textwidth}
        \centering
        \includegraphics[width=\textwidth]{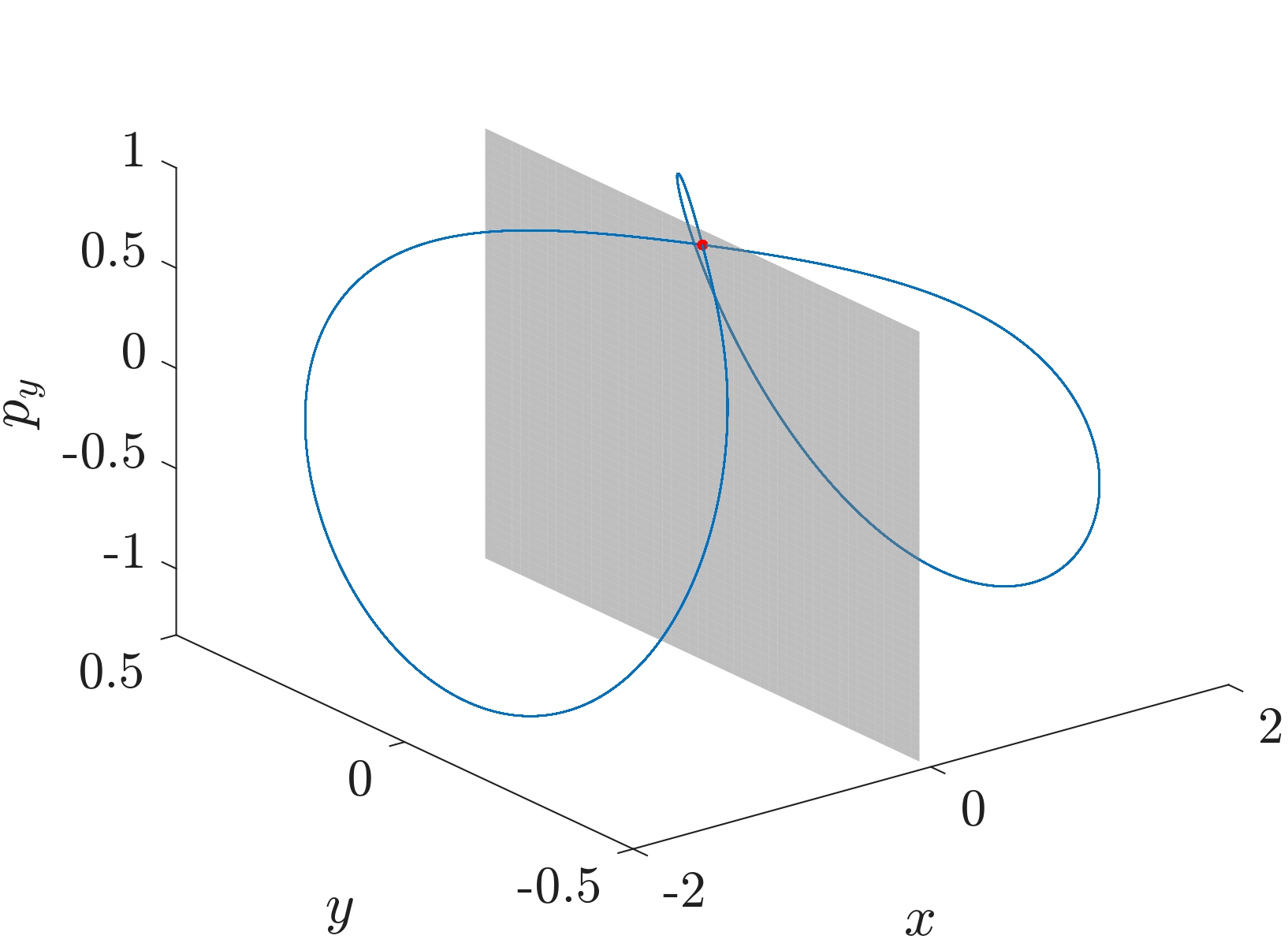}
        \caption{}
    \end{subfigure}
    \quad
    \begin{subfigure}[h!]{0.47\textwidth}
        \centering
        \includegraphics[width=\textwidth]{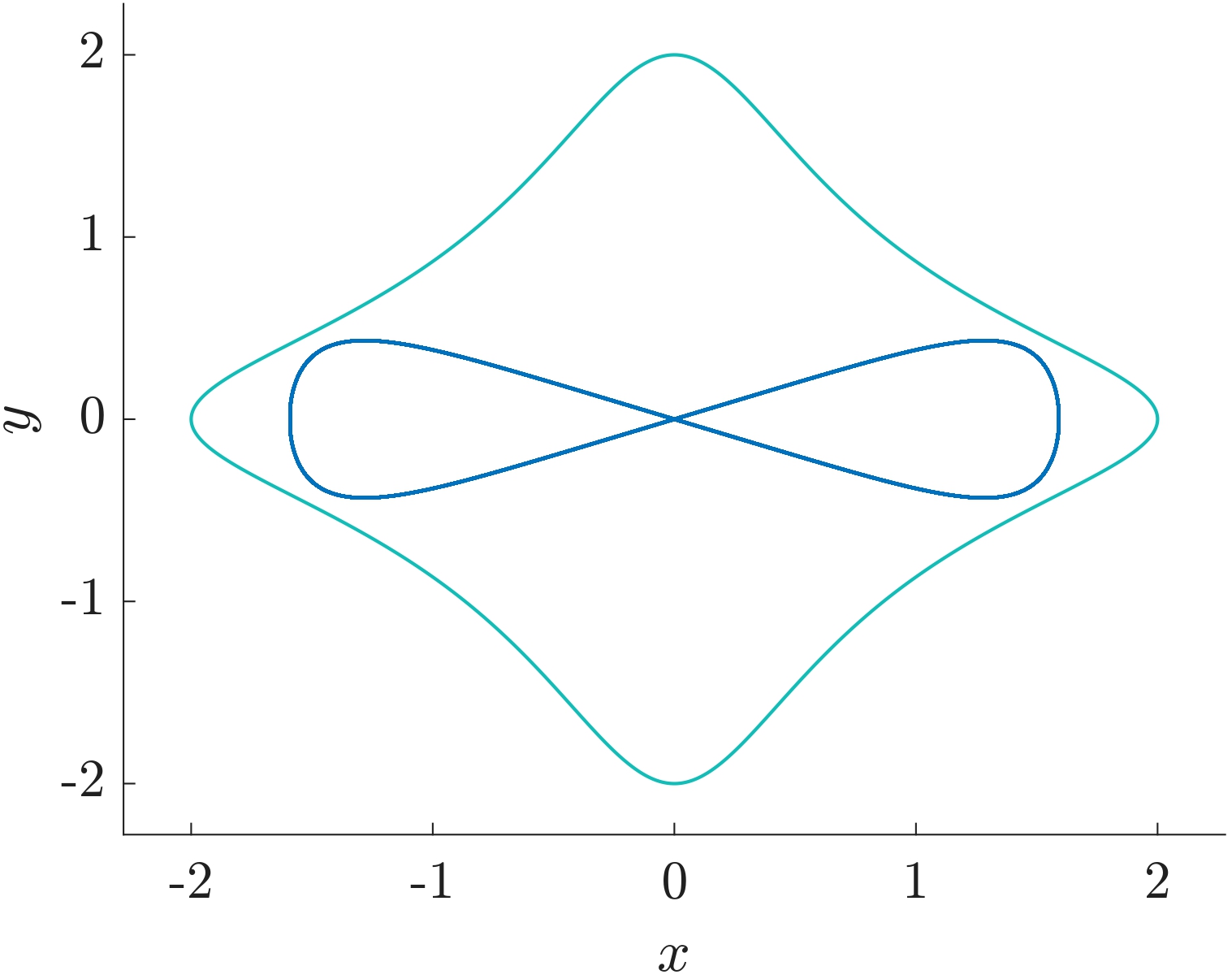}
        \caption{}
    \end{subfigure} 
    \caption{(Top) Magnification of the phase space \textbf{(a)} where a periodic orbit $\gamma_4$, with values $(\alpha=3/2,E=2)$ was found through the intersection of symmetry lines \cref{mapa_naranja}. The corresponding initial conditions are given in \cref{III}. In \textbf{(b)} the time series $y(t)$, the orbit in phase space as the Poincaré section $x=0$ (gray plane) \textbf{(c)}, and finally in the level curve on the plane $(x,y)$ (configuration space) \textbf{(d)} are shown.}
    \label{p4}
\end{figure*}

\begin{figure*}[h!]
    \centering    
    \begin{subfigure}[h!]{0.47\textwidth}
        \centering
        \includegraphics[width=\textwidth]{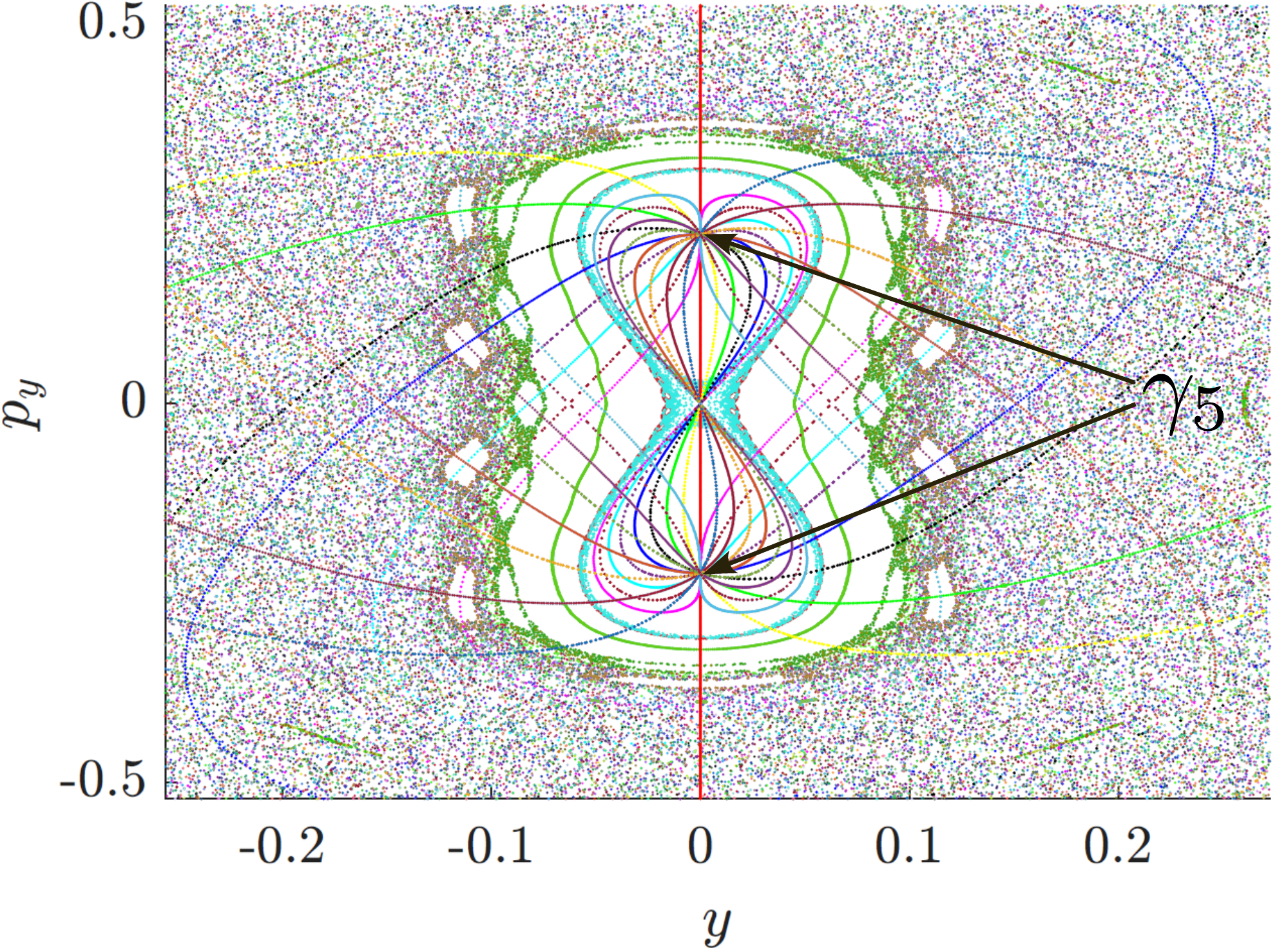} 
        \caption{}
        \label{hiperbolico_mapa}
    \end{subfigure}
    \quad
    \begin{subfigure}[h!]{0.47\textwidth}
        \centering
        \includegraphics[width=\textwidth]{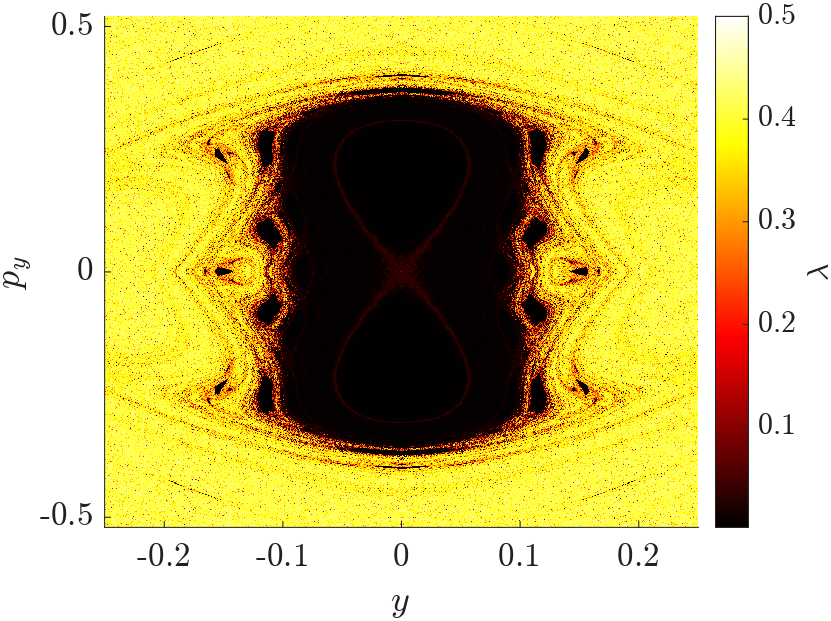}
        \caption{}
        \label{magnificacion}
    \end{subfigure}
    \quad
    \begin{subfigure}[h!]{0.47\textwidth}
        \centering
        \includegraphics[width=\textwidth]{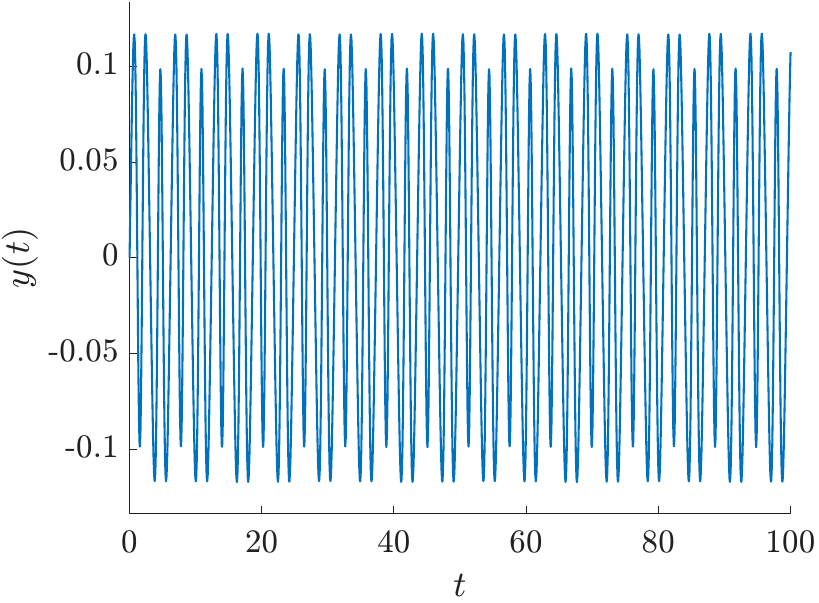}
        \caption{}
    \end{subfigure}
    \quad
    \begin{subfigure}[h!]{0.47\textwidth}
        \centering
        \includegraphics[width=\textwidth]{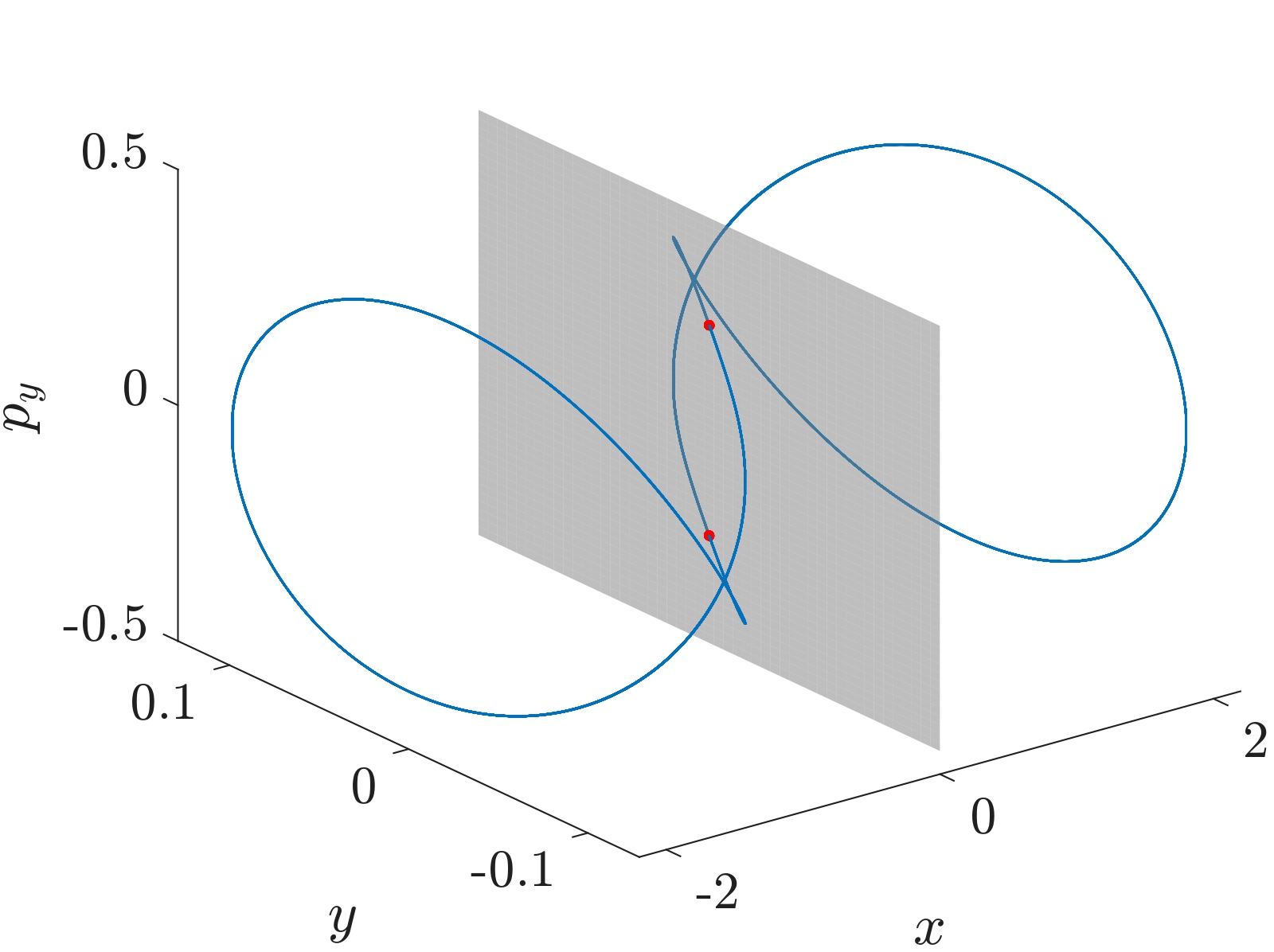}
        \caption{}
    \end{subfigure} 
    \begin{subfigure}[h!]{0.47\textwidth}
        \centering
        \includegraphics[width=\textwidth]{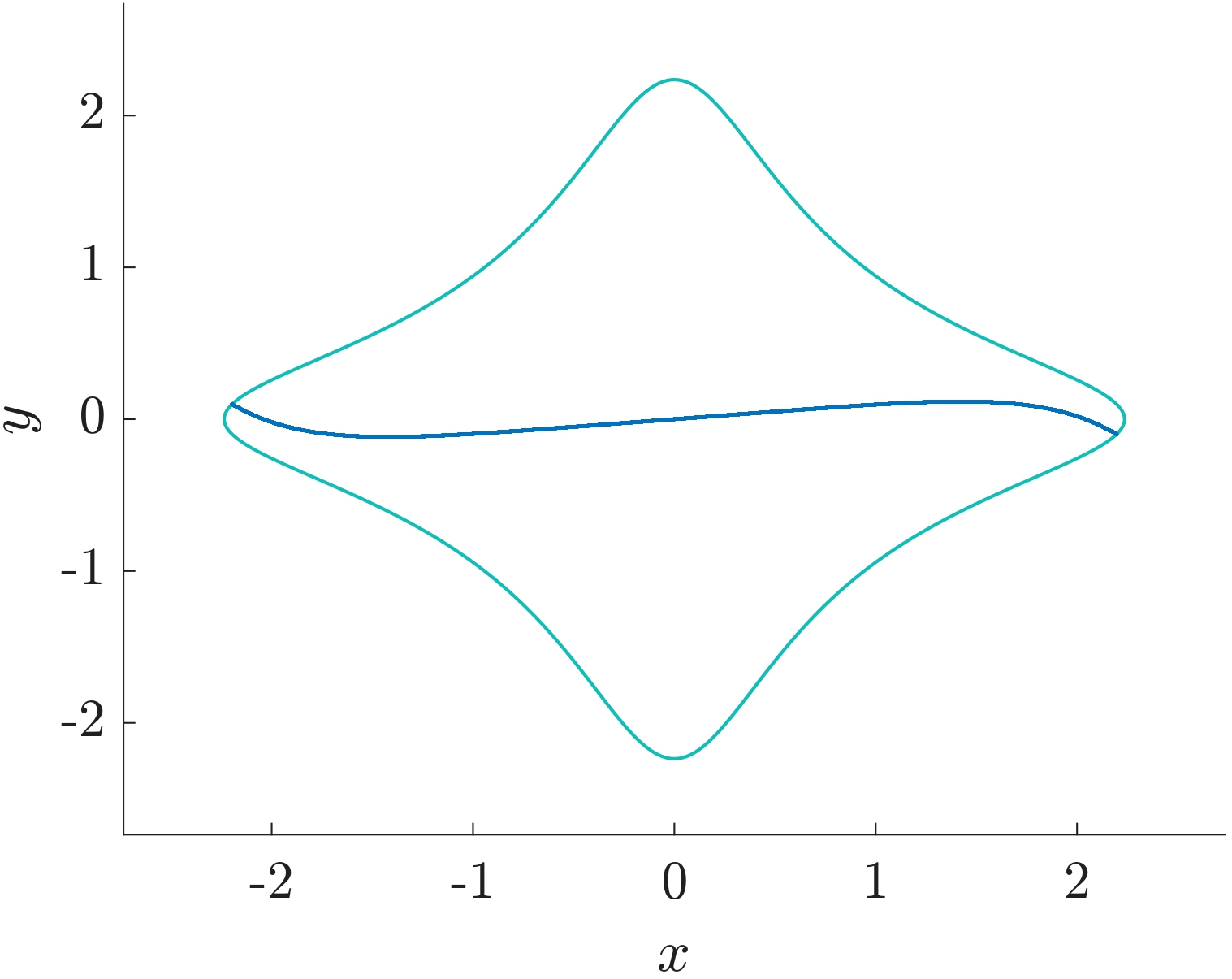}
        \caption{}
    \end{subfigure} 
    \caption{(Top) Magnification of the phase space \textbf{(a)} where a periodic orbit $\gamma_5$, with values $(\alpha=7/4,E=5/2)$ was found through the intersection of symmetry lines \cref{mapa_amarillo}. The corresponding initial conditions are given in \cref{IV}. In \textbf{(c)} the time series $y(t)$, the orbit in phase space as the Poincaré section $x=0$ (gray plane) \textbf{(d)}, in the level curve on the plane $(x,y)$ (configuration space) \textbf{(e)}, and finally in \textbf{(b)} Magnification of chaos measured with the largest Lyapunov exponent \cref{exp_amarillo} in the phase space are shown.}
    \label{p5}
\end{figure*}

\begin{figure*}[h!]
    \centering    
    \begin{subfigure}[h!]{0.47\textwidth}
        \centering
        \includegraphics[width=\textwidth]{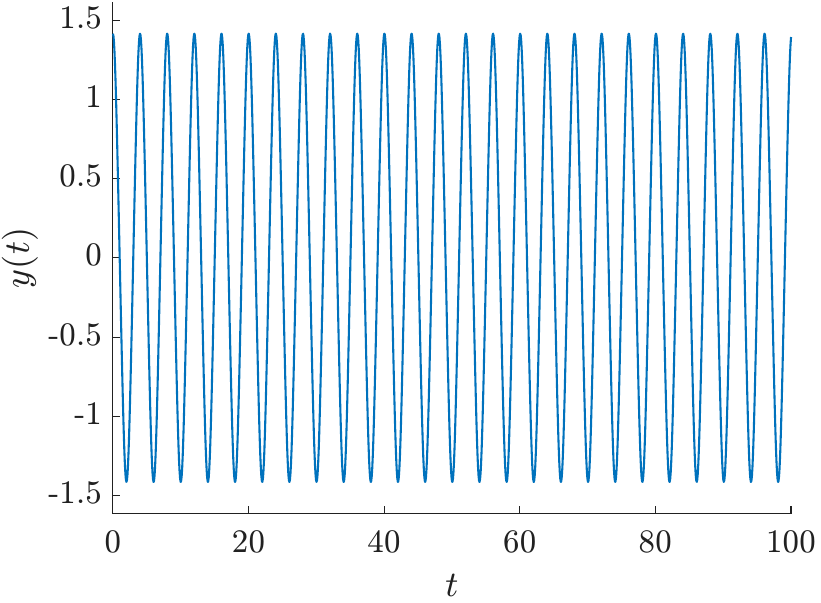} 
        \caption{}
    \end{subfigure}
    \quad
    \begin{subfigure}[h!]{0.47\textwidth}
        \centering
        \includegraphics[width=\textwidth]{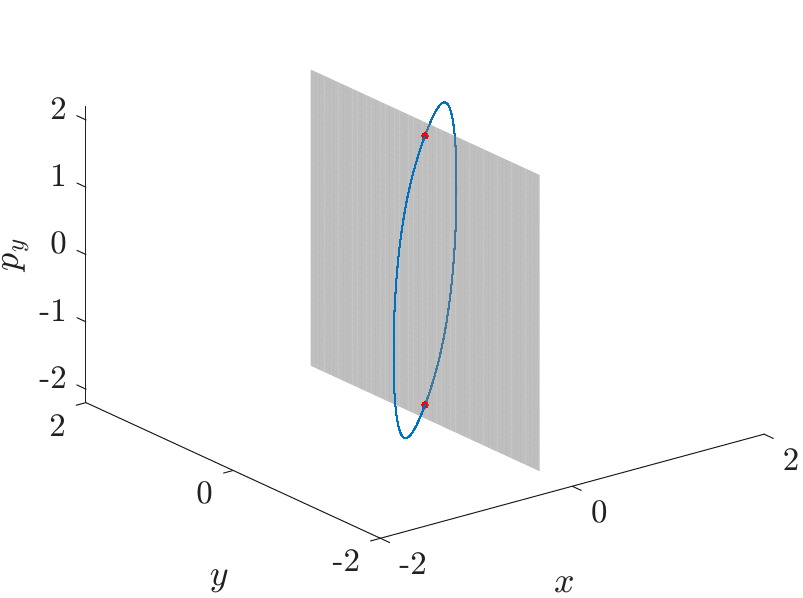}
        \caption{}
    \end{subfigure}
    \caption{Periodic orbit found analytically using the averaging theory. In \textbf{(a)} the time series $y(t)$, the orbit in phase space as the Poincaré section $x=0$ (gray plane) \textbf{(b)} with initial conditions given by \cref{lidia} for $(\alpha = 1/2, E = 2)$.}
    \label{analitico2}
\end{figure*}

\begin{table*}[h!]
\centering
\caption{\small Initial conditions obtained numerically through the intersection of symmetry lines for region \textbf{(II)}}
\begin{tabular}{cc} \noalign{\hrule height 0.8pt} \noalign{\vskip 3pt} 
 Orbit & $(x,p_x,y,p_y)$ 
 \vspace{3pt}
 \\  \hline \noalign{\vskip 5pt}
 Periodic $\gamma_2$ & $(0,\pm 1.824267370933208,\pm 0.775082651974562,\pm 0.267012063316477)$ \\[10pt] 
 Periodic $\gamma_3$ & $(0,\pm 1.732410402862078,\pm 0.566564382090614,\pm 0.823261196098501)$ \\[10pt] 
 Quasi-periodic & $(0,\pm 1.984313483298443,\pm 1/4,0)$ \\[10pt] 
 Chaotic & $(0,\pm 1.732050807568877,\pm 1,0)$\\[5pt]
\noalign{\hrule height 0.8pt}
\end{tabular}
\label{II}
\end{table*}

\begin{table*}[h!]
\centering
\caption{\small Initial conditions obtained numerically through the intersection of symmetry lines for region \textbf{(III)}}
\begin{tabular}{cc} \noalign{\hrule height 0.8pt} \noalign{\vskip 3pt} 
 Orbit & $(x,p_x,y,p_y)$ 
 \vspace{3pt}
 \\  \hline \noalign{\vskip 5pt}
 Periodic $\gamma_4$ & $(0,\pm 1.858075702823795,\pm 0.000205410560644,\pm 0.739969350974999)$ \\[10pt] 
 Quasi-periodic & $(0,\pm 1.996089927833914,\pm 1/8,0)$ \\[10pt] 
 Chaotic & $(0,\pm 1.732050807568877,\pm 1,0)$\\[5pt]
\noalign{\hrule height 0.8pt}
\end{tabular}
\label{III}
\end{table*}

\begin{table*}[h!]
\centering
\caption{\small Initial conditions obtained numerically through the intersection of symmetry lines for region \textbf{(IV)}}
\begin{tabular}{cc} \noalign{\hrule height 0.8pt} \noalign{\vskip 3pt} 
 Orbit & $(x,p_x,y,p_y)$ 
 \vspace{3pt}
 \\  \hline \noalign{\vskip 5pt}
 Periodic $\gamma_5$ & $(0,\pm 2.224886739747574, \pm 0.000198491183221,\pm 0.223335970897352)$ \\[10pt] 
 Quasi-periodic & $(0,\pm 2.235978532991764,\pm 1/50,0)$ \\[10pt] 
 Chaotic & $(0,\pm 2,\pm 1,0)$\\[5pt]
\noalign{\hrule height 0.8pt}
\end{tabular}
\label{IV}
\end{table*}

\clearpage
%----------------------------------------------------------------------
\section{Perturbative analysis of the Contopoulos Hamiltonian}

This appendix presents the perturbative treatment of the Contopoulos Hamiltonian 
that complements Sec.~V. 
Here we show explicitly how the quartic interaction lifts the degeneracies of the isotropic oscillator, 
providing the analytic basis for the numerical spectral results reported in the main text.

\vspace{0.2cm}

The Contopoulos Hamiltonian can be viewed as a two–dimensional isotropic oscillator perturbed by the quartic coupling,
\begin{equation}
    H = H_{0} + \alpha\, V, \qquad 
    H_{0}=\tfrac{1}{2}(p_x^2+p_y^2+x^2+y^2), \quad 
    V=x^{2}y^{2},
\end{equation}
with $\hbar=m=\omega=1$. The unperturbed spectrum is
\begin{equation}
    E^{(\alpha=0)}_{n_x,n_y} = n_x+n_y+1,\qquad n_x,n_y\geq 0,
\end{equation}
so that all states with $N=n_x+n_y$ form a degenerate multiplet of dimension $N+1$.

Because the perturbation $V$ couples states differing by $\Delta n_{x,y}=0,\pm 2$, the degeneracies are lifted within each $N$–manifold. The relevant matrix elements factorize as
\begin{equation}
   \langle n_x',n_y'| V |n_x,n_y\rangle = 
   \langle n_x'|x^2|n_x\rangle \,\langle n_y'|y^2|n_y\rangle,
\end{equation}
with
\begin{equation}
\begin{aligned}
   \langle n'|x^2|n\rangle & = \tfrac{1}{2}\Big[\sqrt{n(n-1)}\,\delta_{n',n-2} + (2n+1)\,\delta_{n',n} 
\\ &   + \sqrt{(n+1)(n+2)}\,\delta_{n',n+2}\Big].
\end{aligned}
\end{equation}
As a benchmark, the ground state is shifted to $E_0\simeq 1+\alpha/4$. The first excited doublet $(N=1)$ remains degenerate at first order with correction $E\simeq 2+3\alpha/4$. The first nontrivial splitting occurs at $N=2$, where the basis $\{|2,0\rangle,|1,1\rangle,|0,2\rangle\}$ yields the perturbation matrix
\begin{equation}
    V^{(N=2)}=
    \begin{pmatrix}
       5/4 & 0 & 1/2 \\
       0 & 9/4 & 0 \\
       1/2 & 0 & 5/4
    \end{pmatrix}.
\end{equation}
Its eigenvalues $\{9/4,\,7/4,\,3/4\}$ produce distinct first–order shifts
\begin{equation}
    E = 3+\alpha\Big\{\tfrac{9}{4},\,\tfrac{7}{4},\,\tfrac{3}{4}\Big\},
\end{equation}
corresponding respectively to the states $|1,1\rangle$, the symmetric combination $(|2,0\rangle+|0,2\rangle)/\sqrt{2}$, and the antisymmetric combination $(|2,0\rangle-|0,2\rangle)/\sqrt{2}$. Thus the quartic coupling completely removes the isotropic degeneracy of the $N=2$ manifold.

This procedure extends to higher $N$: each degenerate oscillator multiplet of dimension $N+1$ is split into distinct levels at order $\mathcal{O}(\alpha)$, with selection rules determined by $\Delta n_{x,y}=\pm 2$. The resulting analytic pattern explains the numerical spectra in the Contopoulos Hamiltonian, where oscillator degeneracies are lifted, in contrast to the purely quartic Yang--Mills case, which is non–degenerate from the outset.

%\appendix
%----------------------------------------------------------------------
\section{Details of the Variational Calculations}
\label{app:variational}

Here we collect the full technical details of the variational treatment of the Contopoulos Hamiltonian.  

For the Gaussian trial function $\psi_1(x,y)$, the variational energy can be written in closed form as
\begin{align}
E_0^{\rm var,1}
&= \dfrac{3\Big(27 \sqrt[3]{6}\,\alpha^2 + 3 \sqrt[3]{2}\,3^{5/6}\,\alpha G + 12\,3^{2/3} \,\alpha \sqrt[3]{S}\Big)}
{2 \Big(\sqrt[3]{2}\, S^{2/3} + 2 \sqrt[3]{3}\Big)^2 \sqrt[3]{S}} \nonumber\\[4pt]
&\quad + \ \dfrac{3\Big(4 \sqrt[6]{3}\, G\, \sqrt[3]{S} + 4\,2^{2/3}\, S^{2/3}\Big)}
{2 \Big(\sqrt[3]{2}\, S^{2/3} + 2 \sqrt[3]{3}\Big)^2 \sqrt[3]{S}},
\end{align}
with
\[
 G = \sqrt{27\alpha^2-4}, \qquad S = 9\alpha+\sqrt{3}\,G.
\]

\begin{figure}[h!]
    \centering
    \includegraphics[width=8cm]{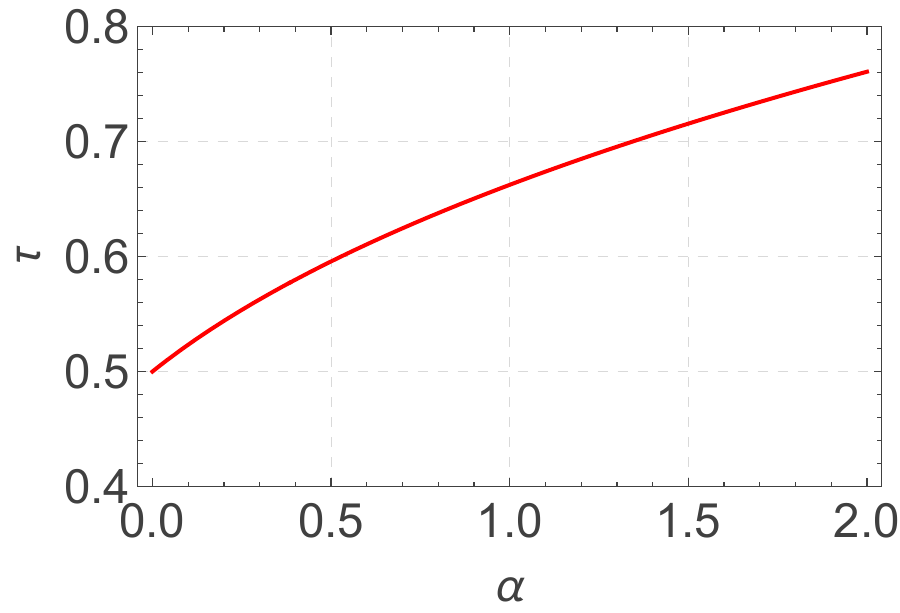}
    \caption{Optimal Gaussian variational parameter $\tau$ as a function of $\alpha$.}
    \label{fig:tau_gaussian}
\end{figure}

The corresponding optimal parameter $\tau$ is displayed in Fig.~\ref{fig:tau_gaussian}.

For the two-parameter trial function $\psi_2(x,y)$ of Eq.~\eqref{psi2C}, the energy functional reads
\begin{multline}
E[\psi_2]  =  
\frac{\sqrt{\tau _2}}{8 \,\tau _1^{5/2} \left(\tfrac{\tau _2}{\tau _1}\right)^{5/2} 
K_0\!\left(\tfrac{\tau _1^2}{\tau _2}\right)} \Bigg[
4 \tau _1 \!\left(-\alpha  \tau _1+\tau _2^2+\tau _2\right) 
K_1\!\left(\tfrac{\tau _1^2}{\tau _2}\right)\\[6pt]
+ \ 2 \Big(\alpha  \!\left(2 \tau _1^2+\tau _2\right) 
+ 2 \tau _1 (\tau _2-1)\tau _2 \Big) 
K_0\!\left(\tfrac{\tau _1^2}{\tau _2}\right)
\Bigg] \ ,
\end{multline}
where $K_q(z)$ denotes the modified Bessel function of the second kind.  

\begin{figure}[h!]
    \centering
    \begin{subfigure}{0.48\textwidth}
        \centering
        \includegraphics[width=\linewidth]{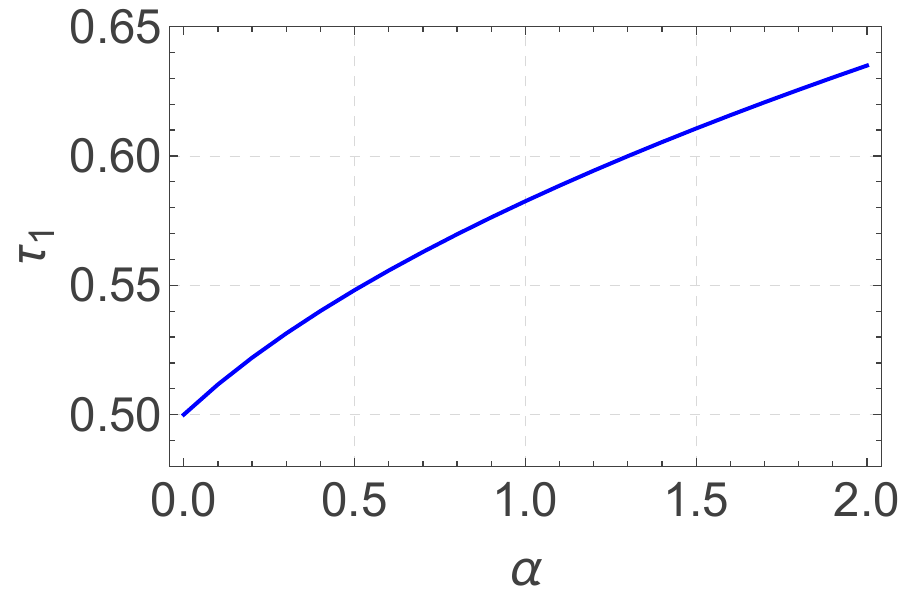}
        \caption{ $\tau_1$ \textit{vs} $\alpha$.}
        \label{fig:tau1}
    \end{subfigure}
    \hfill
    \begin{subfigure}{0.48\textwidth}
        \centering
        \includegraphics[width=\linewidth]{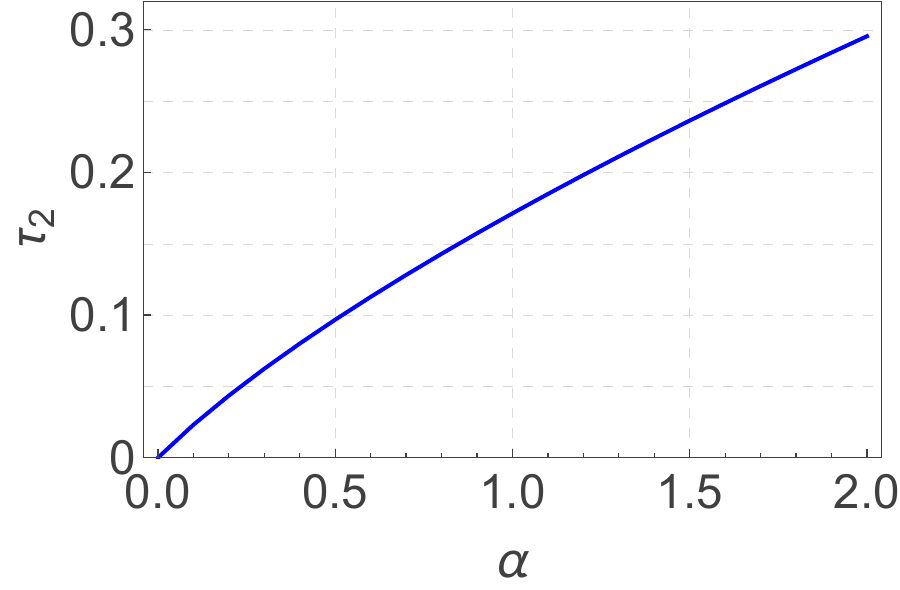}
        \caption{$\tau_2$ \textit{vs} $\alpha$}
        \label{fig:tau2}
    \end{subfigure}
    \caption{Optimal parameters $\tau_1$ and $\tau_2$ for the two--parameter trial function $\psi_2$ (\ref{psi2C}).}
    \label{fig:taus}
\end{figure}

The resulting optimal parameters $(\tau_1,\tau_2)$ are displayed in Figs.~\ref{fig:tau1}, \ref{fig:tau2}.

\section{Lagrange-Mesh method results}
\label{appx:LMM}

This appendix provides  tables \ref{tLMC} and \ref{tLMYM} showing the energy values 
obtained using the Lagrange-Mesh method for the Contopoulos~(\ref{H_cuantico}) 
and Yang-Mills~(\ref{YMV}) potentials, respectively, as a function of the 
$\alpha$ parameter.

\begin{table*}[h!]
\begin{center}
\caption{Energies $E_n$ for the lowest states $n=0,1,\ldots,6$ of the Contopoulos Hamiltonian as 
a function of the parameter $\alpha$ by using the Lagrange-Mesh method with
$N=40-50$ mesh points.}
\label{tLMC}
\begin{tabular}{cccccccc}
\noalign{\hrule height 0.8pt}
\noalign{\vskip 3pt}
$\alpha$ & $E_0$ & $E_1$ & $E_2$ & $E_3$ & $E_4$ & $E_5$ & $E_6$ \\
\hline
\noalign{\vskip 5pt}
0.0& 1.00000000000& 2.00000000000& 3.00000000000& 3.00000000000& 3.00000000000& 4.00000000000& 4.00000000000\\
0.2& 1.04433382954& 2.12476568151& 3.12890917577& 3.26198472185& 3.35407114429& 4.22617457880& 4.59187047411\\
0.4& 1.08128234293& 2.22119562728& 3.23268378127& 3.44107624580& 3.61201574033& 4.39251725908& 4.99022949877\\
0.6& 1.11371041824& 2.30262313666& 3.32229679150& 3.58346097504& 3.82382764059& 4.53009972958& 5.30658598558\\
0.8& 1.14295410151& 2.37426569844& 3.40234103907& 3.70408818682& 4.00686262357& 4.64975879672& 5.57438372985\\
1.0& 1.16978310508& 2.43885478520& 3.47531695102& 3.81001591377& 4.16972466146& 4.75685592209& 5.80913150191\\
1.2& 1.19469370869& 2.49804175418& 3.54277856837& 3.90521465460& 4.31744144275& 4.85451680817& 6.01951599506\\
1.4& 1.21802982597& 2.55291793629& 3.60577491643& 3.99216789819& 4.45325643162& 4.94475423375& 6.14475900897\\
1.6& 1.24004219065& 2.60425012809& 3.66505575410& 4.07254521720& 4.57940702207& 5.02895489523& 6.23841163006\\
1.8& 1.26092065905& 2.65260189394& 3.72117999732& 4.14752961057& 4.69751356810& 5.10812228975& 6.32630546648\\
2.0& 1.28081328782& 2.69840199160& 3.77457803145& 4.21799366117& 4.80879388237& 5.18301034842& 6.40935314939\\
\noalign{\hrule height 0.8pt}
\hline
\end{tabular}
\end{center}
\end{table*}

\begin{table*}[h!]
\begin{center}
\caption{The Lagrange--Mesh energies $E_n$ of the lowest states $n=0,1,\ldots,6$ as a function of the parameter $\alpha$ for the Yang--Mills Mechanical Hamiltonian (oscillator term removed). Results are obtained with $N=80-100$ mesh points.}
\label{tLMYM}
\begin{tabular}{cccccccc} 
\noalign{\hrule height 0.8pt} 
\noalign{\vskip 3pt} 
$\alpha$ & $E_0$ & $E_1$ & $E_2$ & $E_3$ & $E_4$ & $E_5$ & $E_6$ \\
\hline 
\noalign{\vskip 5pt} 
0.2&  0.40827290201&  0.87629766872&  1.12586992362&  1.29491829990&  1.50804697528&  1.75093632342&  1.83647628758\\
0.4&  0.51439162334&  1.10406587879&  1.41850721621&  1.63149482394&  1.90002012838&  2.20604153090&  2.31381513236\\
0.6&  0.58883141749&  1.26383993617&  1.62378541356&  1.86759536161&  2.17498010210&  2.52528716009&  2.64865713665\\
0.8&  0.64809283414&  1.39103584116&  1.78720710113&  2.05555467147&  2.39387535497&  2.77943816173&  2.91522439082\\
1.0&  0.69813684211&  1.49844793563&  1.92521048847&  2.21427914574&  2.57872405418&  2.99405899721&  3.14033027841\\
1.2&  0.74188109773&  1.59233853927&  2.04584142305&  2.35302270878&  2.74030321374&  3.18166245003&  3.33709888042\\
1.4&  0.78099796533&  1.67629713588&  2.15371168463&  2.47708959502&  2.88479008409&  3.34942069212&  3.51305275694\\
1.6&  0.81654580402&  1.75259533743&  2.25173984723&  2.58983659980&  3.01609395056&  3.50187264684&  3.67295257517\\
1.8&  0.84924185886&  1.82277260487&  2.34190381497&  2.69353860780&  3.13686411768&  3.64209432154&  3.82002461722\\
2.0&  0.87959730309&  1.88792609628&  2.42561321990&  2.78981690606&  3.24898871774&  3.77227795521&  3.95656822139\\
\noalign{\hrule height 0.8pt}
\label{}
\end{tabular}
\end{center}
\end{table*}

\end{document}